\documentclass{iopjournal}

\usepackage{amsmath}
\usepackage{bm}
\usepackage{subfig}

\usepackage[natbib, sorting=none, maxnames=2, style=chem-angew]{biblatex}
\AtEveryBibitem{%
  \clearfield{issn} 
  \clearfield{doi} 
  \clearfield{note}

  \ifentrytype{online}{}{
    \clearfield{url}
  }
}
\newcommand{\p}[2]{\frac{\partial#1}{\partial#2}}
\newcommand{\Bc}[2]{\nabla{#1}\times\nabla{#2}}
\newcommand{\oo}[1]{\frac{1}{#1}}
\newcommand{\jac}[0]{\mathcal{J}}
\newcommand{\dive}[1]{\nabla\cdot{#1}}
\newcommand{\curl}[1]{\nabla\times{#1}}
\newcommand{\gradperp}{\nabla_\perp}
\newcommand{\gradpar}{\nabla_\parallel}
\newcommand{\ttt}[1]{\texttt{#1}}
\newcommand{\dS}[0]{\delta\mathcal{S}}
\newcommand\numberthis{\addtocounter{equation}{1}\tag{\theequation}}

\newcommand{\BB}{\bm B}
\newcommand{\bA}{\bm A}
\newcommand{\bb}{\bm b}
\newcommand{\rr}{\bm r}
\newcommand{\JJ}{\bm J}

\newcommand{\ab}{\bar{\alpha}}

\begin{document}

\articletype{Paper} 

\title{Shear Alfvén Waves in Chaotic Magnetic Fields}

\author{M. R. Thomas$^{1, *}$\orcid{0009-0006-9974-1782}, Z. Qu$^2$\orcid{0000-0003-4628-6983} and M. J. Hole$^{1}$\orcid{0000-0002-9550-8776}}

\affil{$^1$Mathematical Sciences Institute, The Australian National University, Canberra, ACT 2601, Australia.}

\affil{$^2$School of Physical and Mathematical Sciences, Nanyang Technological University, 639798, Singapore.}

\affil{$^*$Author to whom any correspondence should be addressed.}

\email{matthew.thomas1@anu.edu.au}


\begin{abstract}
The shear Alfvén spectrum is computed in the presence of symmetry breaking perturbations that introduce chaotic magnetic field trajectories.
Quadratic flux minimised surfaces allow the creation of pseudo straight field line coordinates in the chaotic region.
With these coordinates, the reduced ideal MHD equations are cast into an eigenvalue problem and solved numerically.
The spectrum is computed with varying perturbation strength, showing how shear Alfvén waves change as increasing number of flux surfaces are destroyed.
Solutions on specific flux surfaces are shown to remain relatively unchanged while the flux surface remains intact, and retain some original features at large perturbations where the flux surface is destroyed.
\end{abstract}

\section{Introduction}

Shear Alfvén waves are a low frequency plasma wave propagating along the magnetic field.
These waves are of particular interest in fusion plasmas due to energetic particle drive leading to instability.
Discrete shear Alfvén waves pose the greatest risk, as they can exist in gaps in the frequency spectrum, experiencing minimal continuum damping \cite{heidbrink_basic_2008}.
The most typical example being the toroidal Alfvén eigenmode (TAE) \cite{cheng_lown_1986}, residing inside gaps created by poloidal mode coupling due to toroidal geometry.
The excitation of TAEs has been demonstrated in plasma experiments showing a significant fast ion loss, decreasing confinement and causing damage to the first wall \cite{duong_loss_1993}. 

Resonant magnetic perturbation (RMP) coils have become a standard feature of modern tokamaks \cite{evans_resonant_2015}.
Originally created to suppress edge localised modes, RMP coils introduce a symmetry breaking magnetic field that can introduce magnetic island chains and chaotic magnetic field trajectories.
Experimental evidence \cite{chu_observation_2018, kramer_mitigation_2016, gonzalez-martin_active_2023}, has shown that RMP coils also effect the growth rate of TAEs.
This means that RMP coils may prove a powerful method for TAE suppression if phased to suppress. However, if resonant, RMP fields will amplify TAE drive.

Numerical simulations compared with these experiments \cite{kramer_mitigation_2016, gonzalez-martin_active_2023} have shown that RMP coils can modify the energetic particle distribution and therefore change the drive.
Underlying this work is the assumption that the structure and frequency of shear Alfvén waves remain unchanged from the unperturbed state.
However, if this assumption is not valid, then direct modification of a TAE or the continuous spectrum by RMP coils may be a contributing factor to the modified growth rates.

Typically, the shear Alfvén continuum is theoretically studied in tokamaks by assuming perfect toroidal symmetry and perfect nested flux surfaces.
Under these circumstances it is possible to construct straight field line coordinates, or treat the magnetic field lines as a Hamiltonian system with action-angle coordinates \cite{morrison_magnetic_2000}.
Using these coordinates, much of the complexity in the dynamics are shifted into the coordinates allowing for simple representation and computation of shear Alfvén waves.

One of the simplest symmetry breaking perturbations is a single resonant magnetic island chain.
Under such perturbation, the remaining helical symmetry can be exploited to construct straight field line coordinates \cite{cook_analytical_2015}.
These coordinates have been employed both analytically \cite{biancalani_continuous_2010} and numerically for tokamak \cite{qu_shear_2023} and stellarator \cite{konies_numerical_2022, konies_shear_2024} geometries.
A single island chain has revealed a rich spectrum of shear Alfvén waves including frequency gaps and the so called magnetic island Alfvén Eigenmode (MiAE).

However, these coordinates are not sufficient for studying the more general case of a TAE interacting with an RMP field.
First, there is a coordinate singularity at the magnetic island separatrix, meaning computations must be done entirely inside \textit{or} outside the magnetic island chain, whereas a TAE will exist in both regions.
Second, these coordinates do not generalise to cases with multiple islands and, the focus of this work, overlapping island chains where the magnetic field becomes chaotic.

In this case, straight field line coordinates are not constructable as the Hamiltonian system describing the field line evolution is non-integrable. 
Despite this, some structure remains in the magnetic field; for example the Kolomogorov-Arnold-Moser (KAM) \cite{meiss_symplectic_1992} theorem tells us that for sufficiently small perturbations some flux surfaces persist.
Taking advantage of the residual structures, the construction of action-angle coordinates for an integrable system can be generalised to pseudo action-angle coordinates in the non-integrable case.

Originally proposed by Dewar \textit{et al}. \cite{dewar_almost_1994}, then expanded by Hudson and Suzuki \cite{hudson_chaotic_2014}, the invariant action surfaces of the integrable system are generalised to surfaces which are quadratic flux minimising (QFM).
With these surfaces, a pseudo magnetic field can be produced, allowing the construction of an approximately straight coordinate system for non-integrable systems.
These coordinates simplify the dynamics of shear Alfvén waves allowing more feasible numerical computation.

In this work, we employ the pseudo straight field line QFM coordinates to compute the shear Alfvén continuum in cylindrical geometry under a symmetry breaking perturbation.
First, we outline the geometry and the magnetic field, followed by the computation of the QFM surfaces in section \ref{sec:theory}.
Section \ref{sec:numerical_method} covers the numerical method we employ, including the conversion to the QFM coordinates and selection of QFM surfaces.
We next compute the spectrum in a chaotic magnetic field, and show how the resulting continuum evolves as the perturbation increases in section \ref{sec:chaotic_spectrum}.
This is followed by a direct discussion of the effects the magnetic island chains have on the chaotic spectrum in section \ref{sec:single_island}.
Finally, we have some concluding remarks and a discussion on future work in section \ref{sec:conclusion}.


\section{Theory}\label{sec:theory}

We consider the cylindrical limit of a circular cross section tokamak, with equilibrium magnetic field 
\begin{align}\label{eqn:B_unperturbed}
    \BB_0 = \Bc{\psi}{\theta} - \oo{q(\psi)}\Bc{\psi}{\varphi},
\end{align}
where $q$ is field line helicity; the number of of toroidal circuits of the field per poloidal circuit. 
This is written in terms of the straight field line coordinates, $(\psi, \theta, \varphi)$, where the radial coordinate, $\psi$, is the toroidal flux, $\theta$ is the poloidal angle and $\varphi$ is the toroidal angle.
In these coordinates the flux surfaces are nested cylinders and the geometric and magnetic axis align at $\psi=0$.
The metric tensor and Jacobian of these coordinates are given in appendix \ref{app:cyl_metric}.

\begin{figure}[ht]
    \centering
    \includegraphics[width=0.6\linewidth]{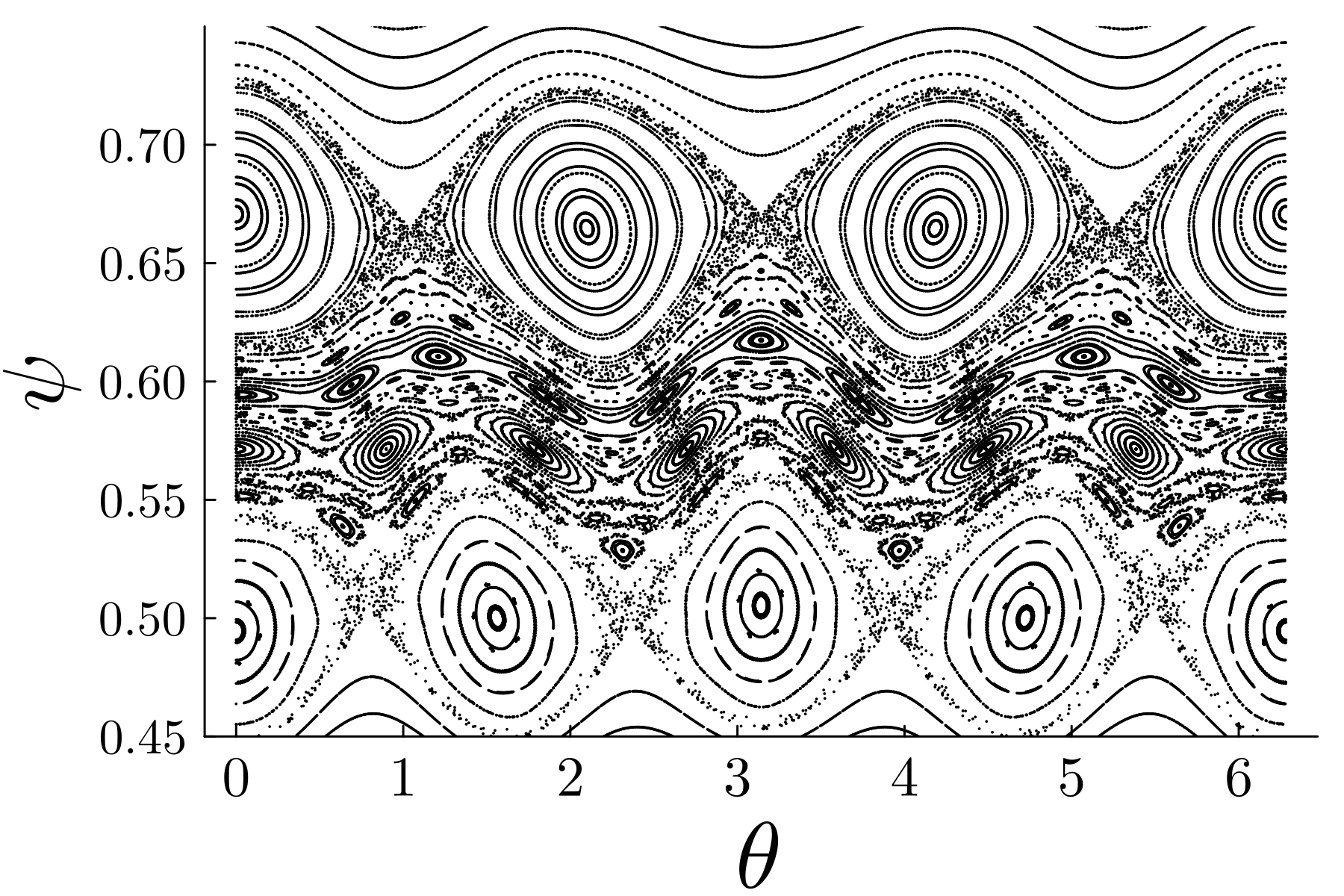}
    \caption{Poincaré plot of the magnetic field including the perturbation from equation \ref{eqn:B_perturbation}, with $k=1.3\times10^{-3}$.}
    \label{fig:base_poincare}
\end{figure}

To model the broken symmetry and create a chaotic field, we add a perturbation of the form,
\begin{align}\label{eqn:B_perturbation}
    \BB_1 = k\left[\sin(3\theta-2\varphi) + \sin(4\theta-3\varphi)\right]\Bc{\theta}{\varphi},
\end{align}
where $k$ is the amplitude of the perturbation.
Magnetic island chains will then form on the resonant flux surfaces, where the helicity of the perturbations match the helicity of the equilibrium field. 
We take a $q$-profile of the form
\begin{align}
    q(\psi) = \oo{1-\psi/2},
\end{align}
resulting in a $(4, -3)$ island chain at $\psi=1/2$ and a $(3, -2)$ island chain at $\psi=2/3$.
For large enough $k$, the two island chains will overlap leading to a chaotic region.
A Poincaré map with this perturbation for $k=1.3\times10^{-3}$ is shown in figure \ref{fig:base_poincare}.

To compute the shear Alfvén spectrum, we assume zero beta and constant equilibrium current profile, such that the linearised ideal MHD equation can be reduced to \cite{berk_continuum_1992, rosenbluth_excitation_1975},
\begin{align}\label{eqn:SAW_gov}
    -\dive{\left(\frac{\omega^2}{v_A^2}\gradperp\Phi\right)} = \BB\cdot\nabla\left(\frac{1}{B^2}\left[\dive{\left(B^2\gradperp\left(\frac{\gradpar\Phi}{B}\right)\right)}\right]\right) - \dive{\left(\frac{\mu_0\bb\cdot\JJ}{B}[\curl(\bb(\gradpar\Phi))]_\perp\right)},
\end{align}
where $\omega$ is the wave frequency, $v_A=B/\sqrt{\mu_0\rho}$ is the Alfvén velocity, $\rho$ is the mass density, taken to be uniform, $B=|\BB|$ is the magnitude of the magnetic field, $\Phi$ is the perturbation to the electrostatic potential, $\bb=\BB/B$ is the normalised magnetic field vector, $\JJ$ is the equilibrium current, $\gradpar=\bb\cdot\nabla$ is the gradient operator parallel to the magnetic field and $\gradperp=\nabla - \bb\gradpar$ is the gradient operator perpendicular to the magnetic field.

Once the perturbation is added, the $(\psi, \theta, \varphi)$ coordinates are no longer straight field line coordinates as the system is no longer integrable. 

\subsection{Quadratic-Flux-Minimising Surfaces}\label{sec:qfm_theory}

To produce approximate straight field line coordinates, we follow the work of Hudson and Suzuki \cite{hudson_chaotic_2014}, employing the specific algorithm and notation given in the appendix of Helander \textit{et al}. \cite{helander_heat_2022}.
The pseudo field lines defining the QFM surfaces are found by extremising the constrained-area action integral,
\begin{align}
    \mathcal{S} = \oint \bA\cdot d\bm l + \nu\left[\oo{2\pi a}\oint \theta\nabla\varphi\cdot d\bm l - b\pi -\alpha\right],
\end{align}
noting that $\varphi$ is playing the role of time in the Hamiltonian system. The integers $a, b$\footnote[1]{Typically $q, p$ are used for this, but to avoid confusion with the $q$-profile we take $a, b$ instead.} define the rational surface $q(\psi)=a/b$, $\nu$ is a Lagrange multiplier and $\bA$ is the vector potential.
The parameter $\alpha$ is chosen to be between $0$ and $2\pi/a$ and will be used in the construction of the pseudo straight field lines coordinates below.

Considering variations in $\psi(\varphi), \theta(\varphi)$ and $\nu$ gives three equations,
\begin{subequations}
\begin{align}
    \frac{\dS}{\delta \psi} &= \dot{\theta} \jac B^\varphi - \jac B^\theta,\\
    \frac{\dS}{\delta\theta} &= \jac B^\psi - \dot{\psi} \jac B^\varphi +\nu/(2\pi a),\\
    \frac{\dS}{\delta\nu} &= \oo{2\pi a}\int_0^{2\pi a} \theta d\varphi - b\pi - \alpha.
\end{align}
\end{subequations}
Extrema are found by determining zeros of these equations. 
For our chosen $q$ profile, it is possible to eliminate $\psi$ from above, reducing the degrees of freedom, however we will consider a more general case without this substitution.
Our variables are expanded as a trial curve in the form,
\begin{subequations}
\begin{align}
    \psi(\varphi) &= \psi_0^c + \sum_{n=1}^{aN}\left[\psi_n^c\cos(n\varphi/a) + \psi_n^s\sin(n\varphi/a)\right],\\
    \theta(\varphi) &= \theta_0^c + b\varphi/a + \sum_{n=1}^{aN}\left[\theta_n^c\cos(n\varphi/a) + \theta_n^s\sin(n\varphi/a)\right],
\end{align}
\end{subequations}
where $N$ is the Fourier resolution.
This particular trial curve automatically enforces the periodicity, $\psi(\varphi+2\pi a) = \psi(\varphi)$ and $\theta(\varphi+2\pi a) = \theta(\varphi) + 2\pi b$, while also allowing efficient numerical integration with the fast Fourier transform.
Projecting the gradient equations onto $\sin$ and $\cos$ gives $2aN+1$ equations for the coefficients of both $\psi$ and $\theta$. Combining these with the final equation for $\nu$ yields a total of $4aN+3$ equations for the unknown coefficients in the trial curve.
Once solved, the coefficients are used to construct a pseudo straight field line poloidal angle,
\begin{align}
    \vartheta = \alpha + b\varphi/a.
\end{align}
We then vary the chosen value for $\alpha$, building up the QFM surface from each of the pseudo field lines. This process is then repeated with different values of $a, b$ to build up a set of QFM surfaces.
We can then write our original cylindrical coordinates in terms of the new QFM coordinates, $(s, \vartheta, \zeta)$, via
\begin{subequations}\label{eqn:tor_to_qfm}
\begin{align}
    \psi(s, \vartheta, \zeta) &= \sum_{m=0, n=0}^{M, N} \psi^c_{m, n}(s)\cos(m\vartheta - n\zeta),\\ 
    \theta(s, \vartheta, \zeta) &= \sum_{m=0, n=0}^{M, N} \theta^s_{m, n}(s)\sin(m\vartheta - n\zeta),\\
    \varphi(s, \vartheta, \zeta) &= \zeta,
\end{align}
\end{subequations}
using interpolation for values in-between the surfaces. 

\begin{figure}[ht]
    \centering
    \subfloat[]{\includegraphics[width=0.45\textwidth]{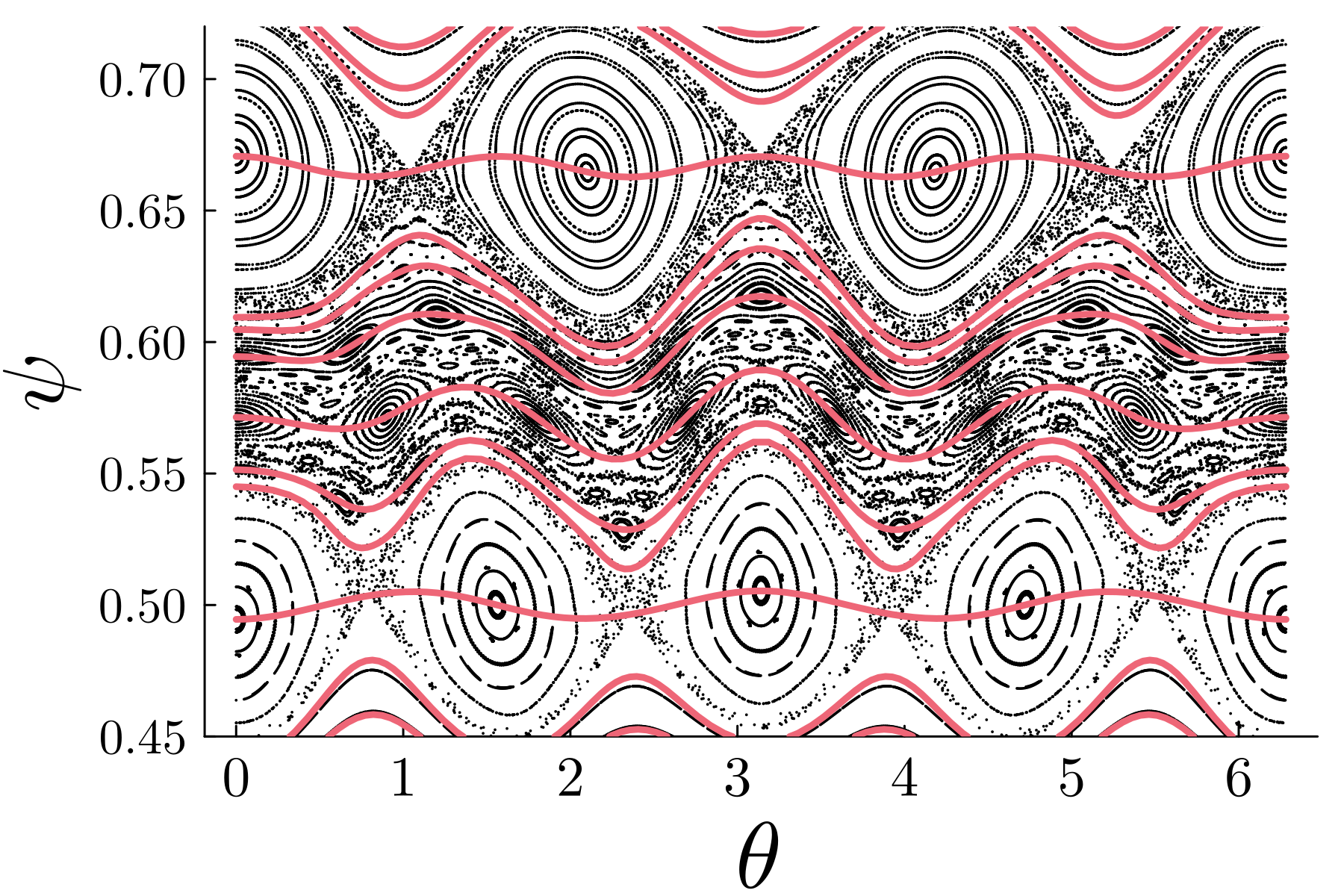}\label{subfig:qfm_surfs}}
    \qquad
    \subfloat[]{\includegraphics[width=0.45\textwidth]{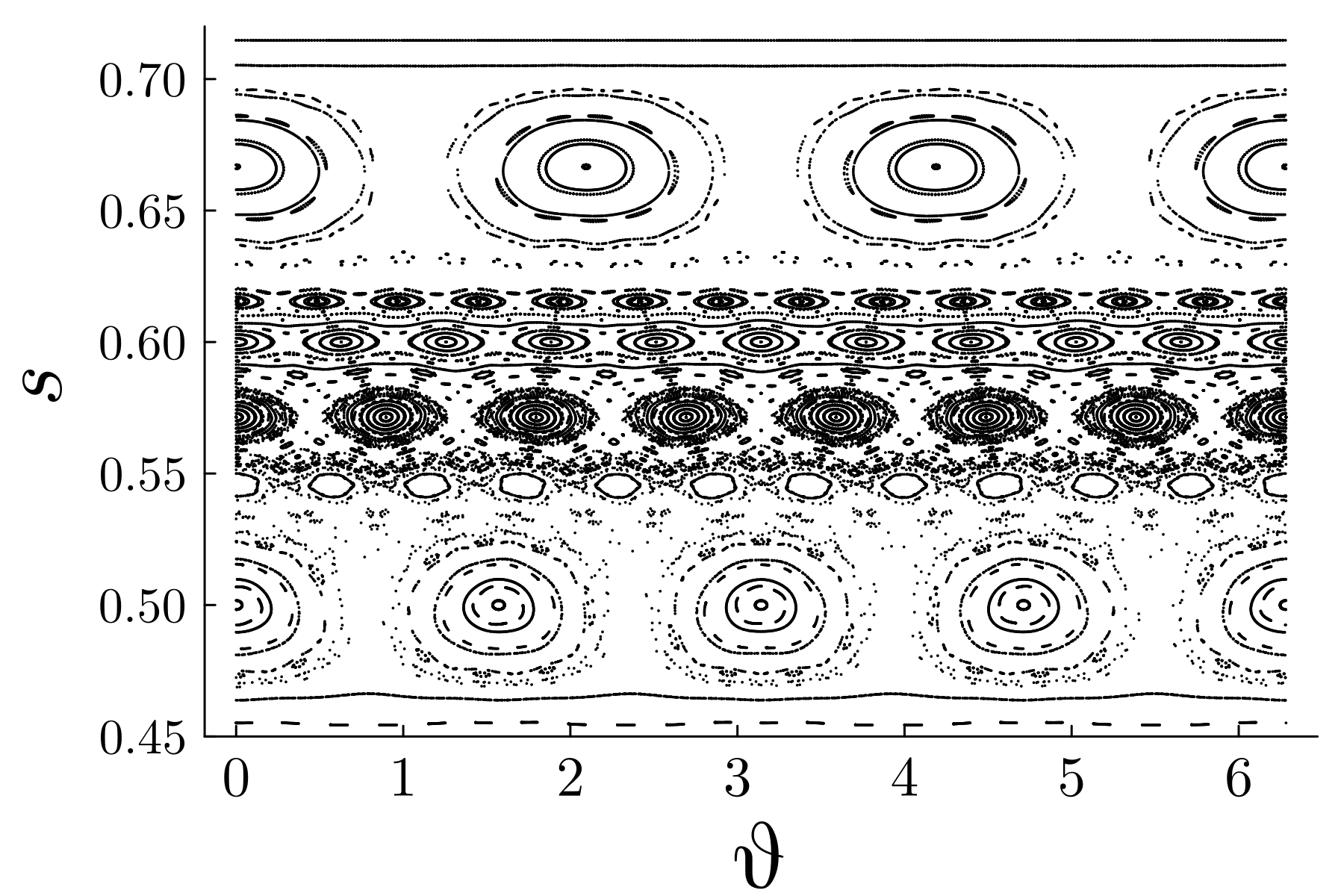}\label{subfig:qfm_poincare}}
    \caption{Quadratic-Flux-Minimising (QFM) surfaces on Poincaré map with $k=1.3\times10^{-3}$ (a) and Poincaré map with the same magnetic field, in QFM coordinates (b).}
    \label{fig:qfm_poincare}
\end{figure}

An example of QFM surfaces are shown in figure \ref{subfig:qfm_surfs}. 
These surfaces have been overlaid on a Poincaré plot with $k=1.3\times10^{-3}$ showing that the QFM surfaces cling to the residual structures. 
In particular, surfaces pass through the $X$ and $O$ points of each of the island chains.

In figure \ref{subfig:qfm_poincare} we have plotted the same perturbed magnetic field in the QFM coordinates constructed from the QFM surfaces.
Here we can see the structure imposed by the new coordinates, as intact flux surfaces have been straightened and island chains have been organised.
This will simplify the numerical computation of shear Alfvén waves.

\section{Numerical Method}\label{sec:numerical_method}

To numerically solve equation \ref{eqn:SAW_gov} with QFM coordinates, we employ the finite element Galerkin method \cite{braess_finite_2007}, implemented in the \ttt{Julia} package \ttt{ChaoticShearAlfvenWaves.jl}\footnote{The package is freely available using the \ttt{Julia} package manager or at \url{https://github.com/Syndrius/ChaoticShearAlfvenWaves}}. 
We approximate the perturbation to the electric field as a linear combination of basis functions, 
\begin{align}
    \Phi = \sum_i u_i S_i,
\end{align}
where $u_i$ are the coefficients to be determined, and $S_i$ are given by the tensor product of $1$d cubic Hermite splines.
We then obtain the weak form by multiplying by a test function, $\Psi$, which is also an expansion of cubic Hermite functions, and integrating by parts, noting that each term has been manipulated to ensure the operators are self adjoint \cite{fesenyuk_ideal_2002, feher_simulation_nodate}, 
\begin{align*}\numberthis\label{eqn:weak_form}
        \omega^2\int \frac{\mu_0\rho}{B^2}\left(\nabla\Psi\right)\cdot\left(\gradperp\Phi\right) d\rr = - \int &  B^2\nabla\left(\frac{\gradpar\Psi}{B}\right)\cdot\left(\gradperp\left(\frac{\gradpar\Phi}{B}\right)\right)d\rr\\
            &- \int\frac{\mu_0 J_\parallel}{B}\oo{2}\bigg[\gradperp\Psi\cdot\left(\curl{\gradperp\Phi}\right) + \gradperp\Phi\cdot\left(\curl{\gradperp\Psi}\right)\bigg] d\rr.
\end{align*}
This is then converted into matrix form with matrix elements given by
\begin{subequations}\label{eqn:matrix_elements}
\begin{align}
    P_{ij} &= \int \Psi_i \hat{P} \Phi_j d\rr,\\
    Q_{ij} &= \int \Psi_i \hat{Q} \Phi_j d\rr,
\end{align}
\end{subequations}
where the integration is done via Gaussian quadrature.
The two operators, $\hat{P}$ and $\hat{Q}$ are the differential operators defining the right and left hand side of equation \ref{eqn:weak_form} respectively.
Further details and expressions for the two operators, $\hat{P}, \hat{Q}$, are given in appendix \ref{app:weakform}.
These two operators are given in terms of the metric tensor, allowing the spectrum to be computed in arbitrary coordinates.

To specifically compute these in QFM coordinates, we must transform the metric tensor and magnetic field from the original cylindrical coordinates, $(\psi, \theta, \varphi)$, into the new QFM coordinates, $(s, \vartheta, \zeta)$.
From equation \ref{eqn:tor_to_qfm}, we compute the Jacobian matrix of the transformation,
\begin{align}
    \jac_\mu^{i} = \p{x^i}{x^\mu} = \p{(\psi, \theta, \varphi)}{(s, \vartheta, \zeta)},
\end{align}
where Greek indices are used to denote summation over the new QFM coordinates, and Latin indices for the old cylindrical coordinates.
We then compute the new metric tensor and covariant magnetic field,
\begin{subequations}
\begin{align}
    g_{\mu\nu} &= \jac^i_\mu \jac^j_\nu g_{ij},\\
    B^\mu &= (\jac_\mu^i)^{-1} B^i.
\end{align}
\end{subequations}
Additionally, we need the derivatives of these terms, requiring some care to ensure the derivatives are with respect to the new coordinates, for example, the derivative of the metric tensor is,
\begin{align}
    \partial_\sigma g_{\mu\nu} = \jac^j_\nu g_{ij}\partial_\sigma\jac^i_\mu + \jac^i_\mu g_{ij}\partial_\sigma\jac^j_\nu + \jac^i_\mu \jac^j_\nu \partial_\sigma g_{ij}.
\end{align}

We then iterate through the grid of $(s, \vartheta, \zeta)$ values, at each point we use the QFM surfaces to determine the $(\psi^c_{m, n}, \theta^s_{m, n})$ coefficients in equation \ref{eqn:tor_to_qfm}, interpolating values and their derivatives between surfaces with $5$th order B-splines.
Summing over the coefficients, we compute $(\psi, \theta, \varphi)$ and the Jacobian matrix.
With the original coordinates, we can determine the original cylindrical metric and the magnetic field given by equations \ref{eqn:B_unperturbed} and \ref{eqn:B_perturbation}, which are then converted to the new coordinates using the Jacobian matrix.
Finally, the matrix elements are computed with the new metric tensor and new magnetic field.

These matrix elements are collated into two matrices, allowing us to write this equation as a generalised eigenvalue problem,
\begin{align}
    \bm P \bm u = \omega^2 \bm Q \bm u,
\end{align}
where $\bm u$ is the vector of unknown coefficients, and $\omega^2$ is the eigenvalue.

Directly solving for the entire spectrum is impractical for even moderate grid sizes, instead a shift-and-invert transformation is applied \cite{saad_numerical_2011}. 
A target frequency, $\sigma$, is selected and the eigenvalue problem is transformed into
\begin{align}
    (\bm P - \sigma \bm Q)^{-1} \bm Q \bm u = \oo{\omega^2-\sigma}\bm u.
\end{align}
Under this transformation, the eigenvalues closest to $\sigma$ are now the largest eigenvalues of the system and many eigenvalue solvers are able to efficiently find the largest eigenvalues \cite{saad_numerical_2011}.
To build up a larger portion of the spectrum, multiple shift-and-invert transformations can be taken.

Solutions of the new system are found with the Krylov-Schur algorithm \cite{stewart_krylov--schur_2002},  inside the Scalable Library for Eigenvalue Problem Computations \cite{hernandez_slepc_2005} (SLEPc).
This library is an extension of the Portable, Extensible Toolkit for Scientific Computation \cite{balay_petsctao_2025} (PETSc) library, allowing the matrices to be stored and solved in parallel.

\subsection{Choosing QFM Surfaces}

The selection of QFM surfaces, which has some degree of freedom, is the final consideration for the QFM coordinates. 
Our method starts by taking a small set of surfaces that roughly divide up the domain.
These are chosen to be low order rationals, i.e. small values for $a, b$, such that the algorithm outlined in section \ref{sec:qfm_theory} is efficient.
With this first guess of surfaces, we compute the squared radial component of the magnetic field, $(B^s)^2$, which approximates the quadratic flux.
We can then view this over regions of our domain, and for $s$ values where $(B^s)^2$ is large, we add extra surfaces. Typically, this is done by determining the mediant, $(a_1+a_2)/(b_1+b_2)$, of the two surrounding surfaces. 

\begin{figure}[ht]
    \centering
    \includegraphics[width=0.6\linewidth]{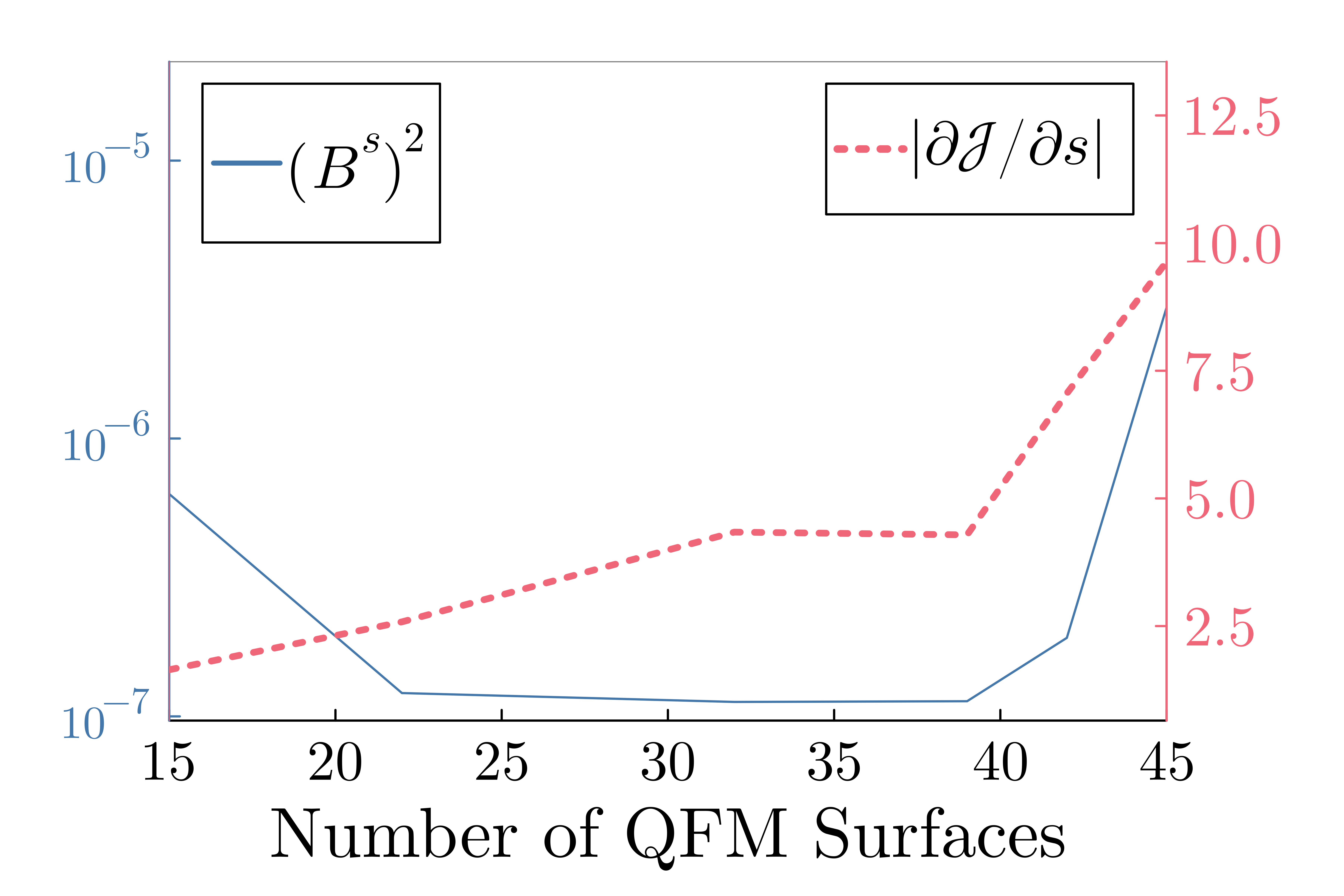}
    \caption{Key metrics vs number of QFM surfaces. Chosen number of surfaces is based on minimising $(B^s)^2$ while maintaining a smooth Jacobian.}
    \label{fig:number_of_surfaces}
\end{figure}

While repeating this process, we track both the quadratic flux and the derivative of the Jacobian of the coordinate transformation, $\partial\jac/\partial s$, as shown in figure \ref{fig:number_of_surfaces}.
Additional surfaces tend to increase Jacobian derivative which can cause instabilities in the matrix construction and for very large values can even prevent solutions of the eigenvalue problem.
Figure \ref{fig:number_of_surfaces} also shows that it is possible for additional surfaces to increase rather than decrease the quadratic flux.

It is important to note that the number of surfaces axis label in figure \ref{fig:number_of_surfaces} is somewhat misleading.
In regions far from the chaotic regions, an arbitrary number of surfaces can be added without significant changes to the Jacobian, however, these make negligible difference to the quadratic flux.
Adding more surfaces inside the chaotic region, in particular, surfaces with large $(a, b)$ values or surfaces close to the nearly intact separatrixes and the nested flux surfaces of the main island structures, is much more volatile.

These surfaces are typically more deformed, resulting in problematic interpolation between surfaces.
By tracking the quadratic flux and Jacobian derivative we are able to manually remove surfaces that cause issues, leaving us with a smooth coordinate system.

\section{Chaotic Spectrum}\label{sec:chaotic_spectrum}

Combining the quadratic flux minimised (QFM) coordinates with our numerical method, we are now equipped to solve the shear Alfvén wave governing equation (equation \ref{eqn:SAW_gov}).
We will investigate the behaviour of solutions as the perturbation amplitude, $k$, increases.

To give a broad overview of the change with increasing $k$, we first consider the continuum, constructed from all solutions.
Each eigenfunction, $\Phi$, is decomposed into Fourier harmonics,
\begin{align}
    \Phi(s, \vartheta, \zeta) = \sum_{m, n} \Phi(s)e^{i(m\vartheta + n\zeta)},
\end{align}
and the radial location of the peak is then paired with the normalised eigenvalue, $\omega$.
The continuum is then split into branches, based on the Fourier mode numbers, $(m, n)$.

Additionally, as representations of the general behaviour of individual solutions, we will highlight the change of two specific solutions.
In the unperturbed case, the first has a frequency of $\omega=0.357$, located on the flux surface with irrational rotational transform, $\iota\equiv 1/q$, of $\iota_1=1.37016...$, located at $\psi\approx0.547$.
Importantly, this flux surface will be intact until a critical perturbation of $k_{c1}=1.005\times10^{-3}$, computed by Greene's reside \cite{greene_method_1979, hudson_calculation_2006, hudson_chaotic_2014}.
The second solution exists on the flux surface with a critical perturbation of $k_{c2}=1.385\times10^{-3}$ with irrational rotational transform $\iota_2=1.43622...$, at $\psi\approx0.609$ and has a frequency of $\omega=0.173$.

\begin{figure}[htp]
    \centering
    \subfloat[$k=0$]{\includegraphics[width=0.45\textwidth]{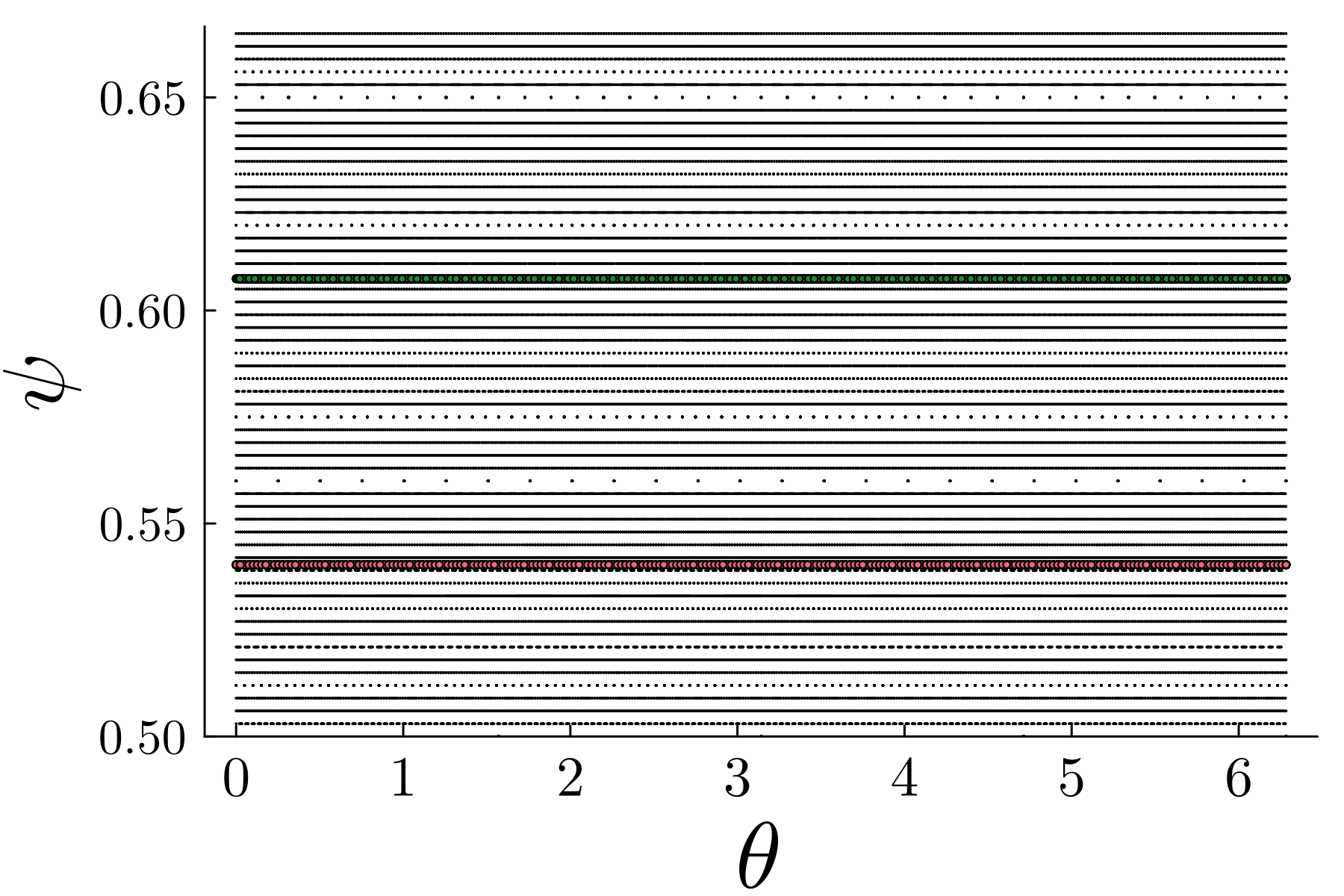}\label{subfig:k0_poincare}}
    \qquad
    \subfloat[$k=0$]{\includegraphics[width=0.45\textwidth]{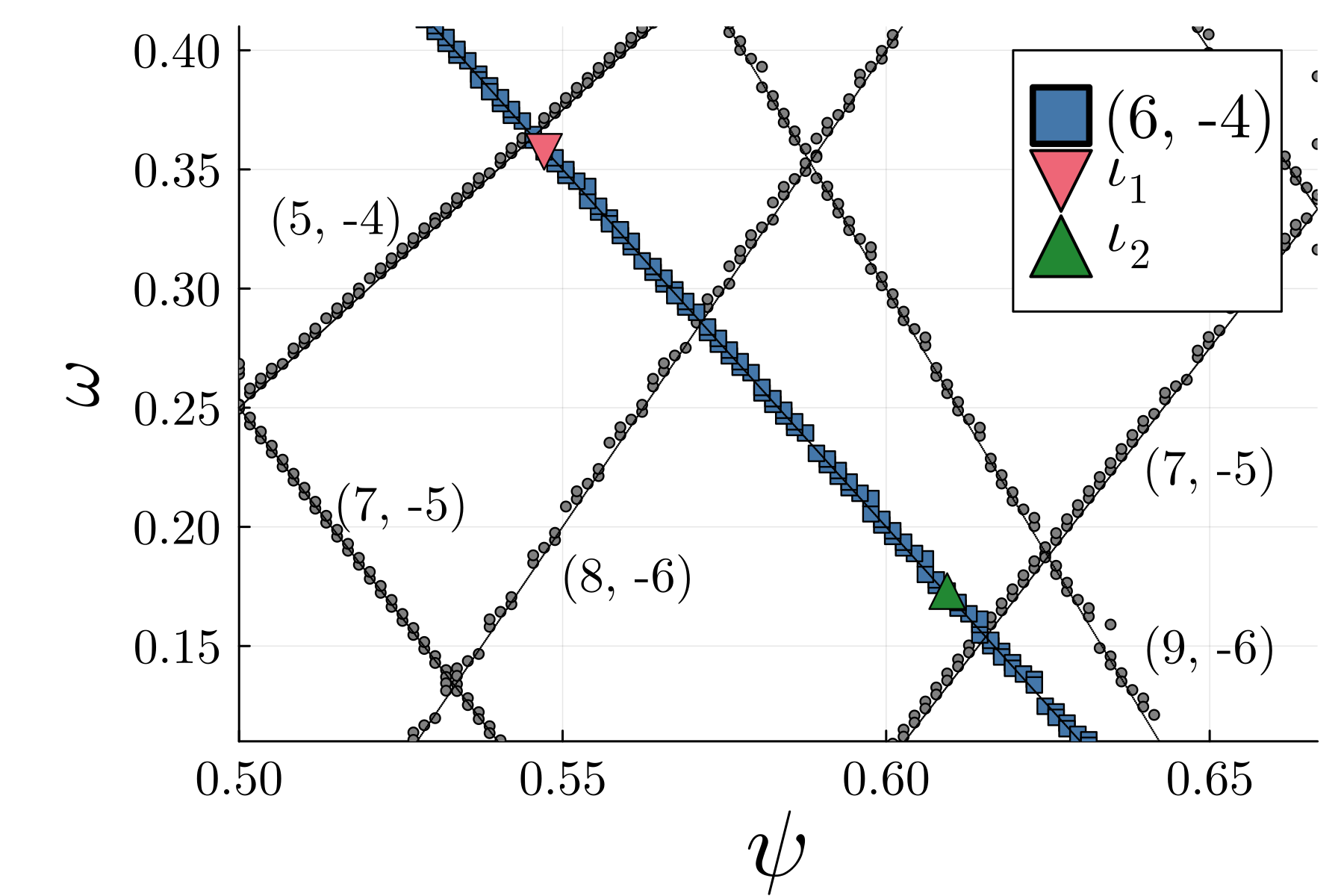}\label{subfig:k0_cont}}
    
    \subfloat[$k=0.5\times10^{-3}$]{\includegraphics[width=0.45\textwidth]{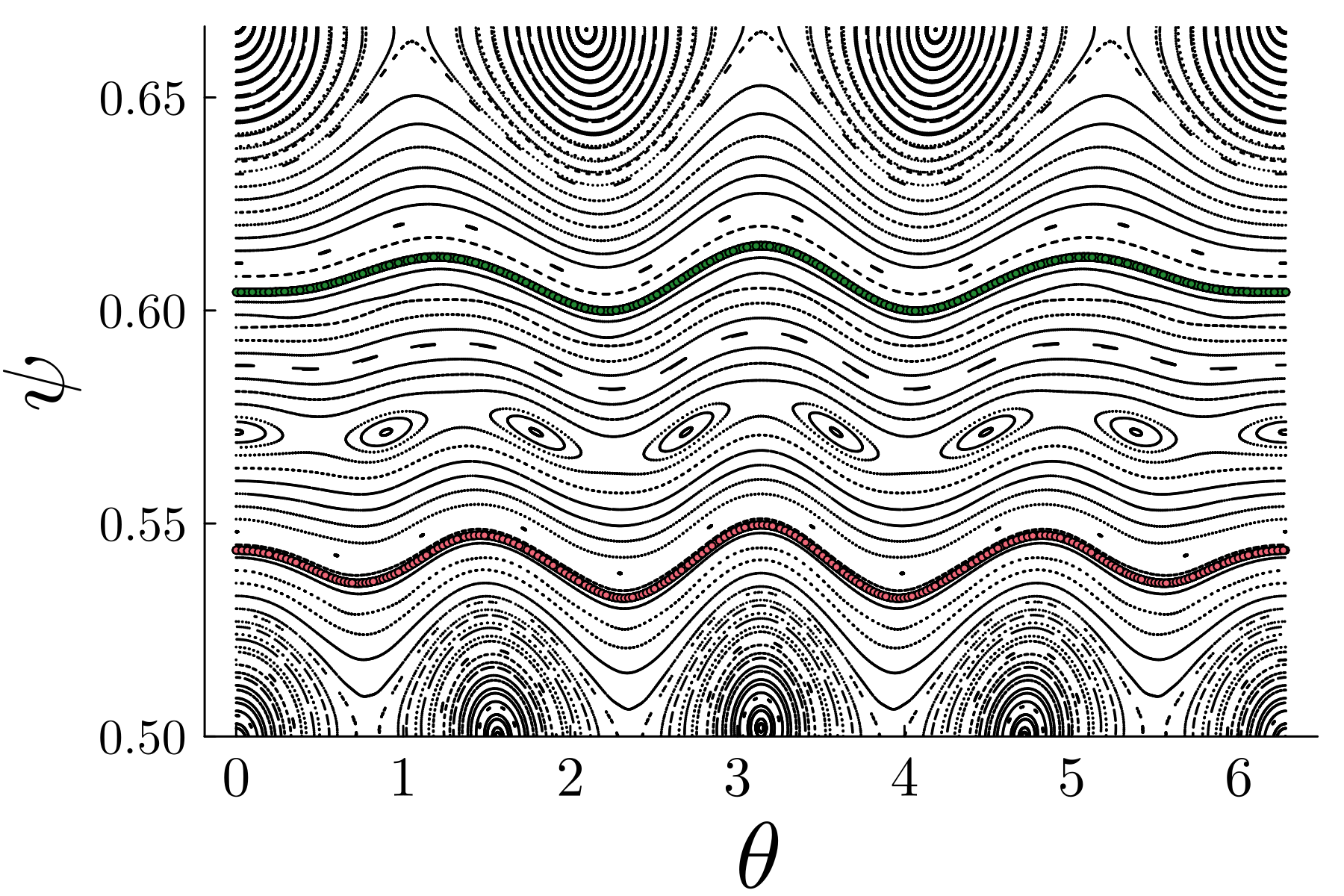}\label{subfig:k05_poincare}}
    \qquad
    \subfloat[$k=0.5\times10^{-3}$]{\includegraphics[width=0.45\textwidth]{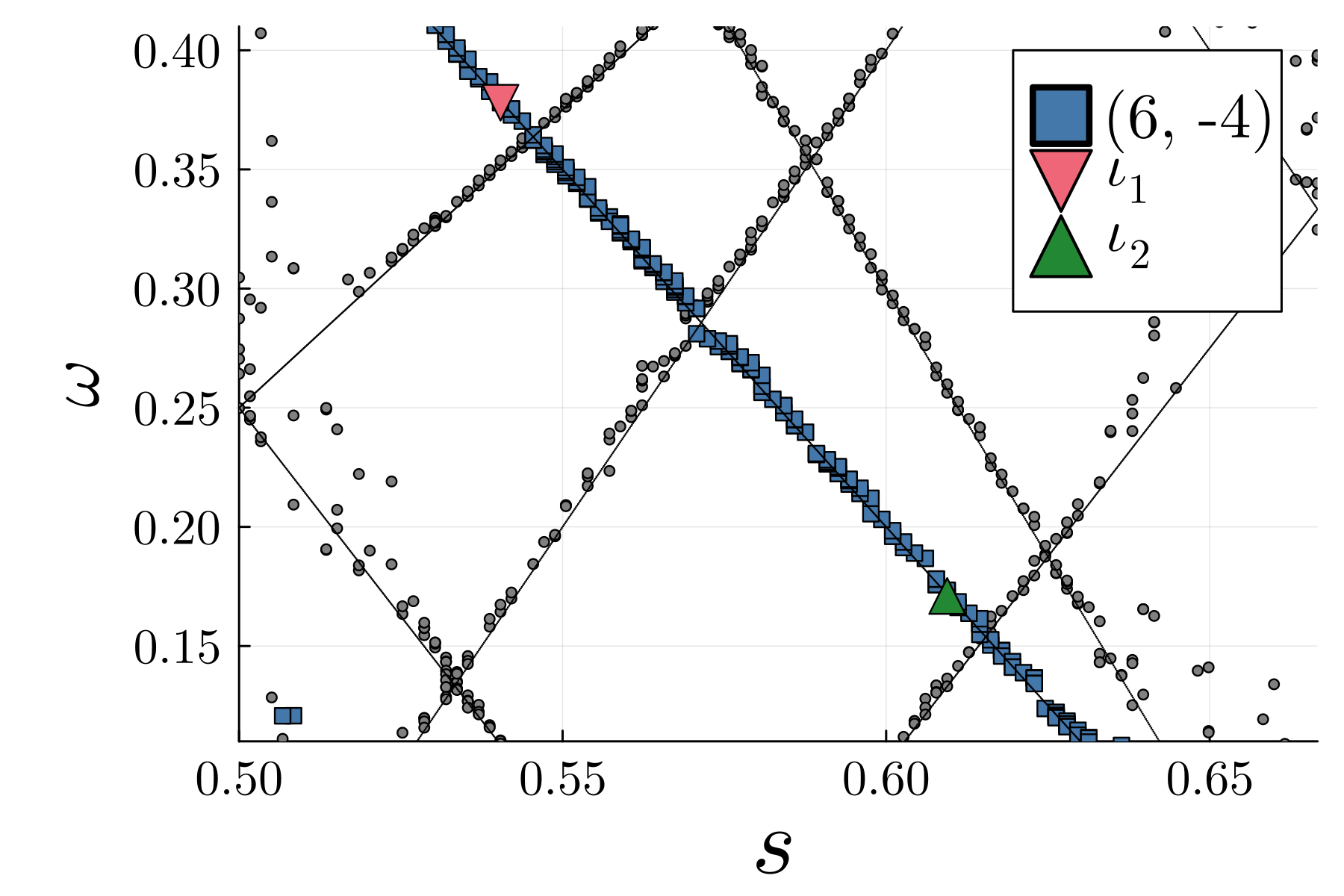}\label{subfig:k05_cont}}

    \subfloat[$k=1.2\times10^{-3}$]{\includegraphics[width=0.45\textwidth]{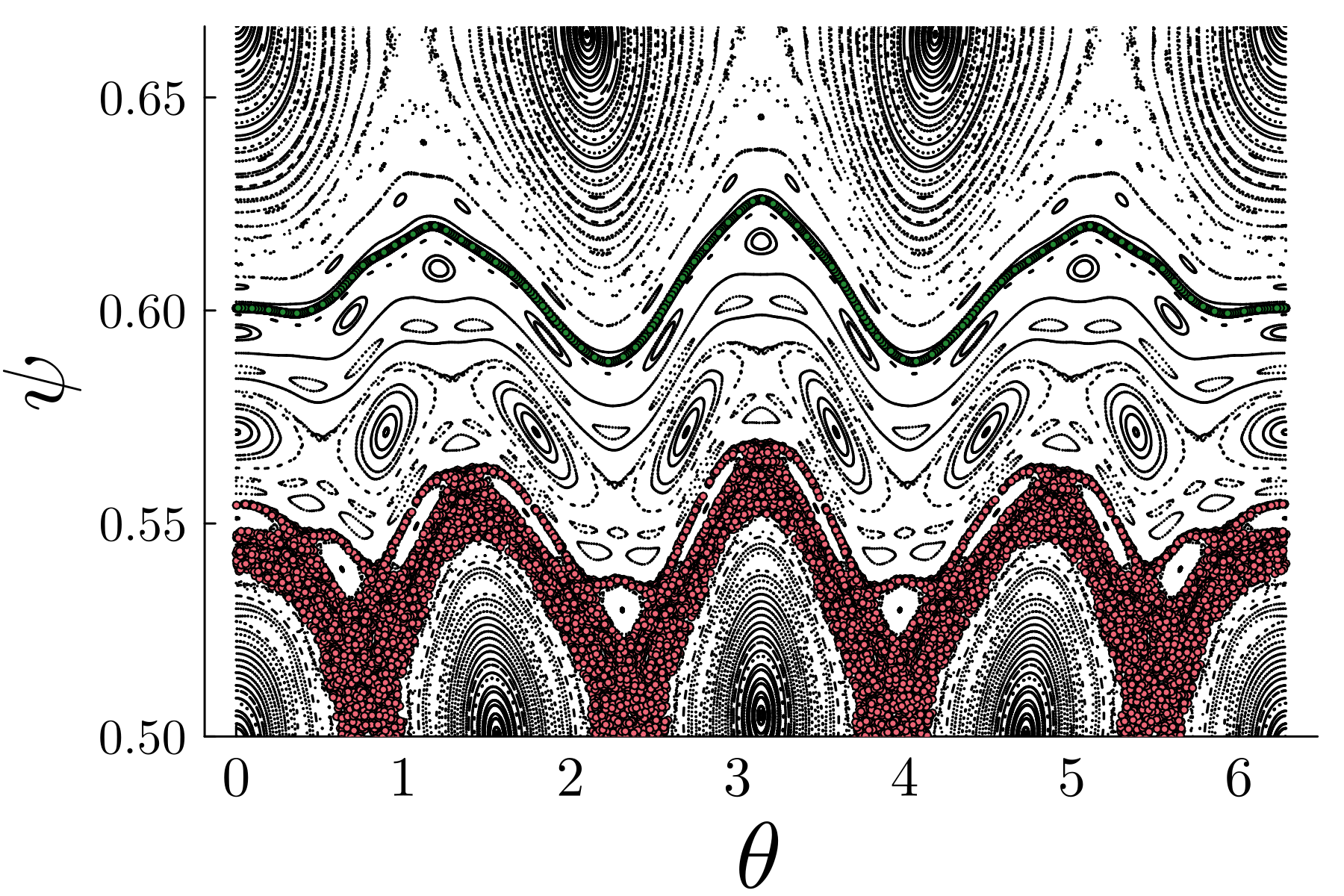}\label{subfig:k12_poincare}}
    \qquad
    \subfloat[$k=1.2\times10^{-3}$]{\includegraphics[width=0.45\textwidth]{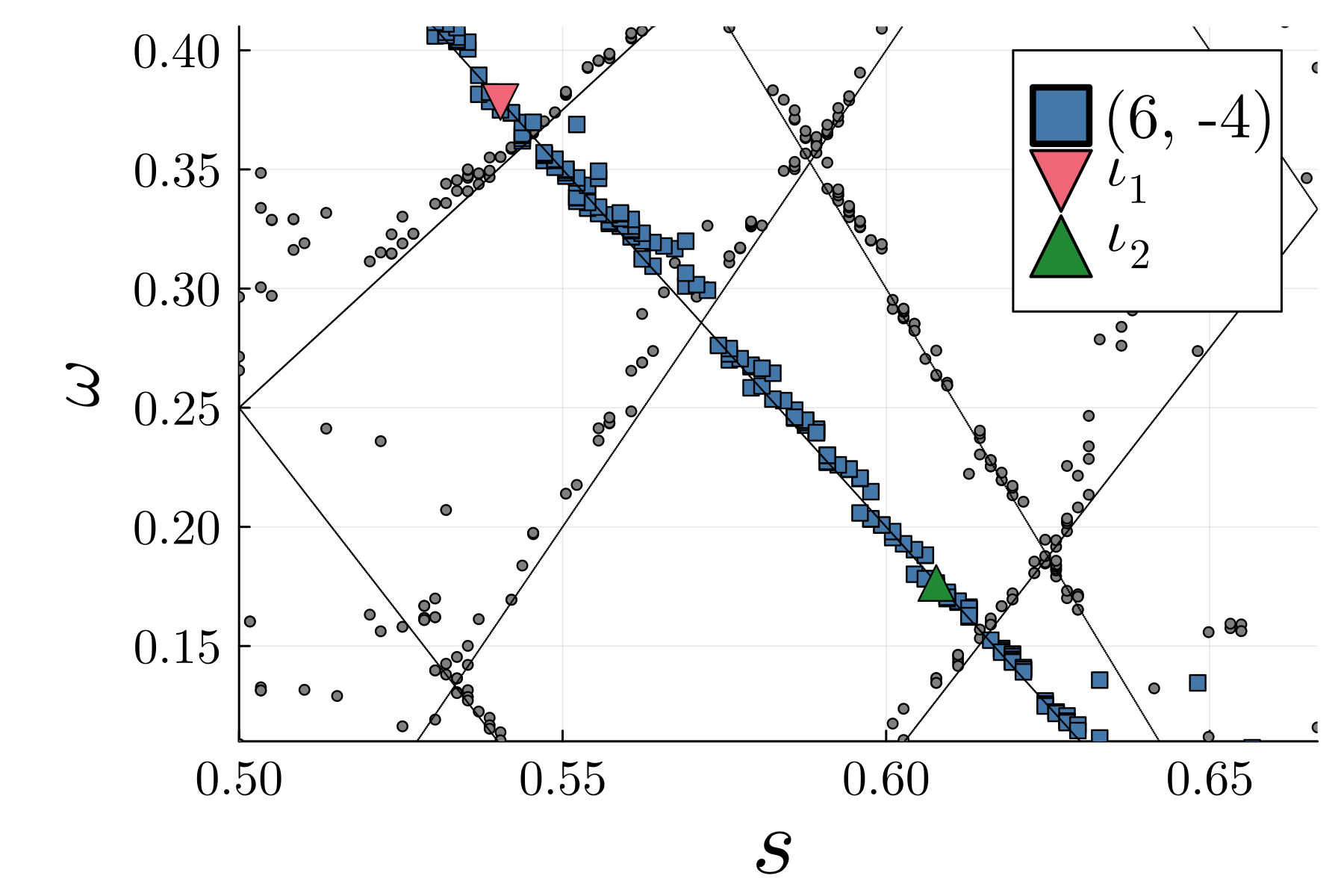}\label{subfig:k12_cont}}
    
    \subfloat[$k=1.7\times10^{-3}$]{\includegraphics[width=0.45\textwidth]{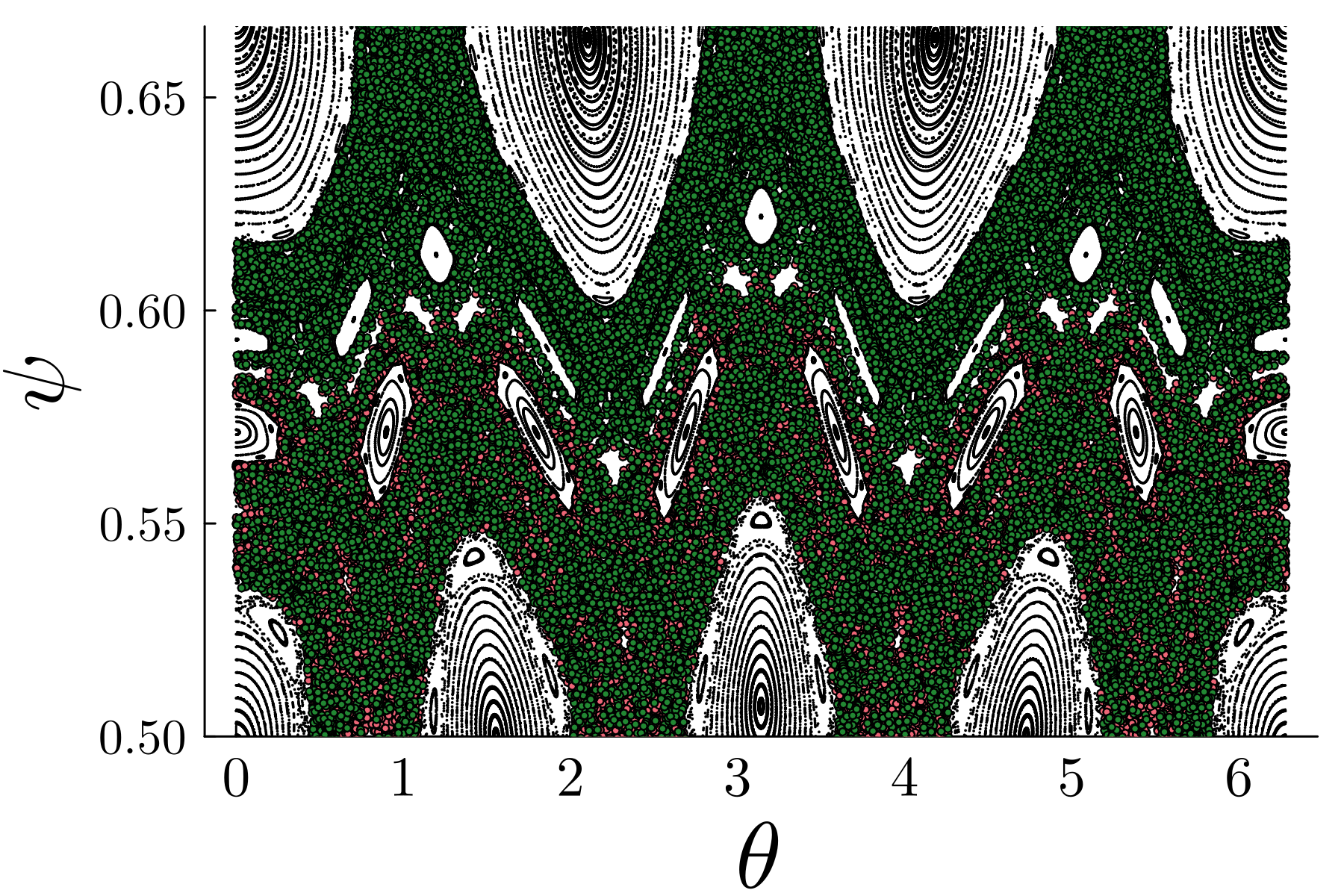}\label{subfig:k17_poincare}}
    \qquad
    \subfloat[$k=1.7\times10^{-3}$]{\includegraphics[width=0.45\textwidth]{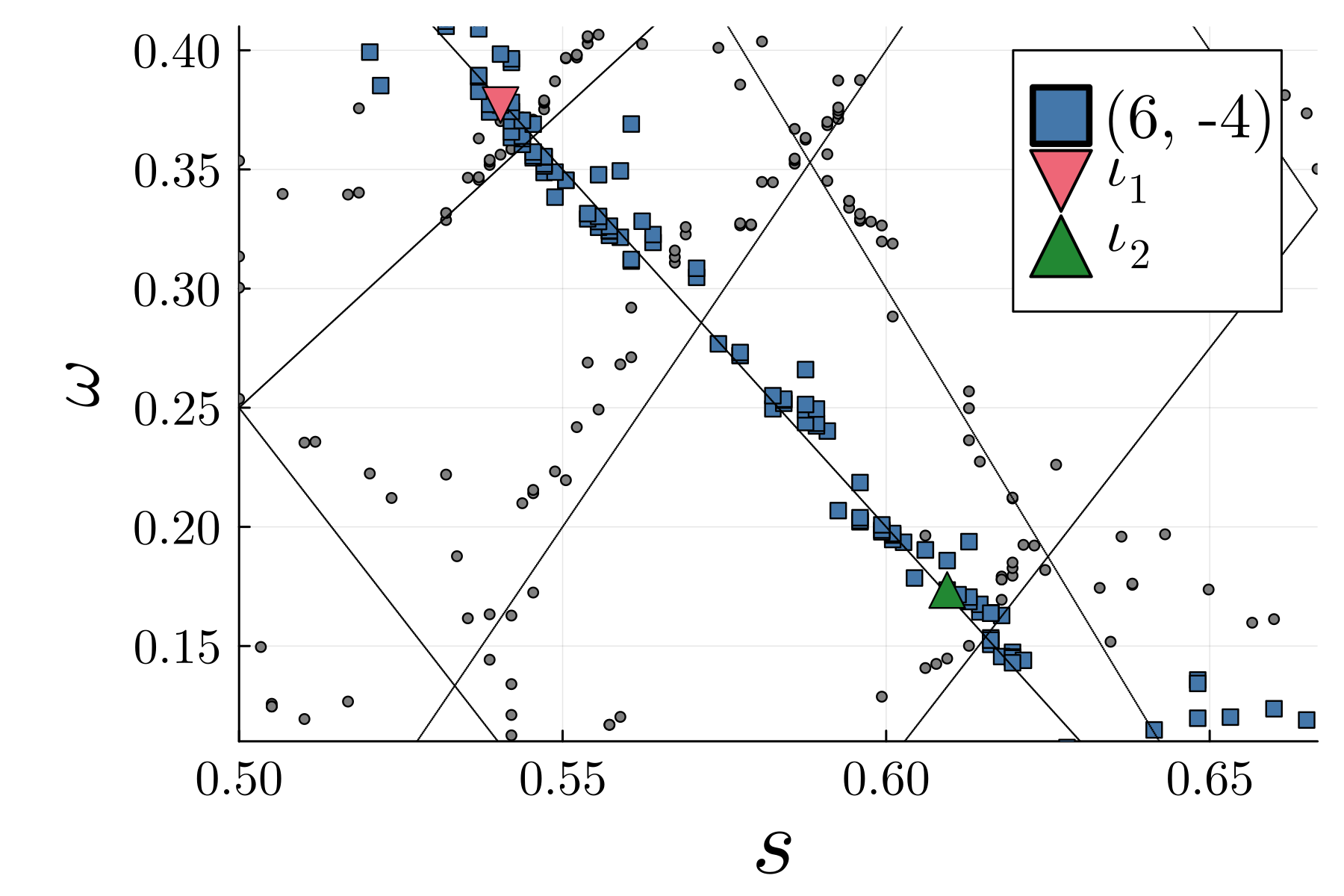}\label{subfig:k17_cont}}
    \caption{Change of the Poincaré map (left column) and corresponding continuum (right column) for increasing perturbation amplitude $k$.  Specific solutions are highlighted in the continuum plots that are localised to the flux surfaces highlighted in the Poincaré maps. Continuum plots are constructed in QFM coordinates, $(s, \vartheta, \zeta)$, while Poincaré maps are computed in original cylindrical coordinates, $(\psi, \theta, \varphi)$.}
    \label{fig:poinc_cont}
\end{figure}

We consider $4$ different amplitudes of the perturbation, $k=0, k=0.5\times10^{-3}, k=1.2\times10^{-3}$ and $k=1.7\times10^{-3}$, with a grid resolution of $N_s\times N_\vartheta \times N_\zeta = 150\times40\times30$, where roughly $2/3$ of the radial points are placed in the chaotic region.
The different amplitudes result in the Poincaré plots shown in the left column of figure \ref{fig:poinc_cont}.
Within these Poincaré plots we have highlighted the two flux surfaces, $\iota_1$ and $\iota_2$, by tracing an orbit with periodicity closely matching the irrational frequency.
With a small perturbation of $k=0.5\times10^{-3}$, shown in figure \ref{subfig:k05_poincare}, the unperturbed straight flux surfaces distort to conform to the first order island chains and there is an introduction of a second order $(7, -5)$ island chain at $\psi\approx0.57$.
However, most flux surfaces, in particular $\iota_1$ and $\iota_2$, remain intact and there are no visible chaotic trajectories.
Increasing the perturbation to a moderate $k=1.2\times10^{-3}$, figure \ref{subfig:k12_poincare}, we see higher order island chains forming and  destroyed flux surfaces, leading to chaotic trajectories, most notable for $\psi<0.55$.
Here we can see that the orbit closely following the surface $\iota_1$ is no longer constrained to a single surface, showing that this flux surface has broken.
In figure \ref{subfig:k17_poincare} with a large perturbation of $k=1.7\times10^{-3}$, we can see that all flux surfaces in this region have been destroyed and the entire region becomes chaotic.

On the right hand side of figure \ref{fig:poinc_cont}, we have shown the computed continuum, the radial peak location of each eigenvalue for each solution.
We have also plotted in black the analytical continuum for $k=0$, given by, 
\begin{align}
    \omega^2 = \frac{v_A^2}{R_0^2}\left(\frac{m}{q} + n\right).
\end{align}
We have chosen to specifically highlight the $(6, -4)$ branch and have also highlighted the two specific solutions with triangle dots.

As soon as the perturbation is non-zero, the introduction of island chains causes a significant distortion to the spectrum, shown in figure \ref{subfig:k05_cont}.
For low perturbation, many flux surfaces are still intact and the continuum in regions of intact flux surfaces remains relatively unchanged from the unperturbed case.
Increasing to a moderate perturbation, figure \ref{subfig:k12_cont}, we see larger distortion due to the larger island and formation of higher order island chains.
Despite this, the spectrum still maintains most of the same structure as the unperturbed case, even in the region between the $(4, -3)$ and $(7, -5)$ island chains, around $s=0.55$, where very few flux surfaces are still intact.
In the fully chaotic case with no intact flux surfaces over the plotted domain, the continuum is notably distorted but surprisingly, still contains some of the original structure, shown in figure \ref{subfig:k17_cont}.
To understand this, we now focus on the specific highlighted solutions.

\begin{figure}[htp]
    \centering
    \subfloat[$k=0$]{\includegraphics[width=0.45\textwidth]{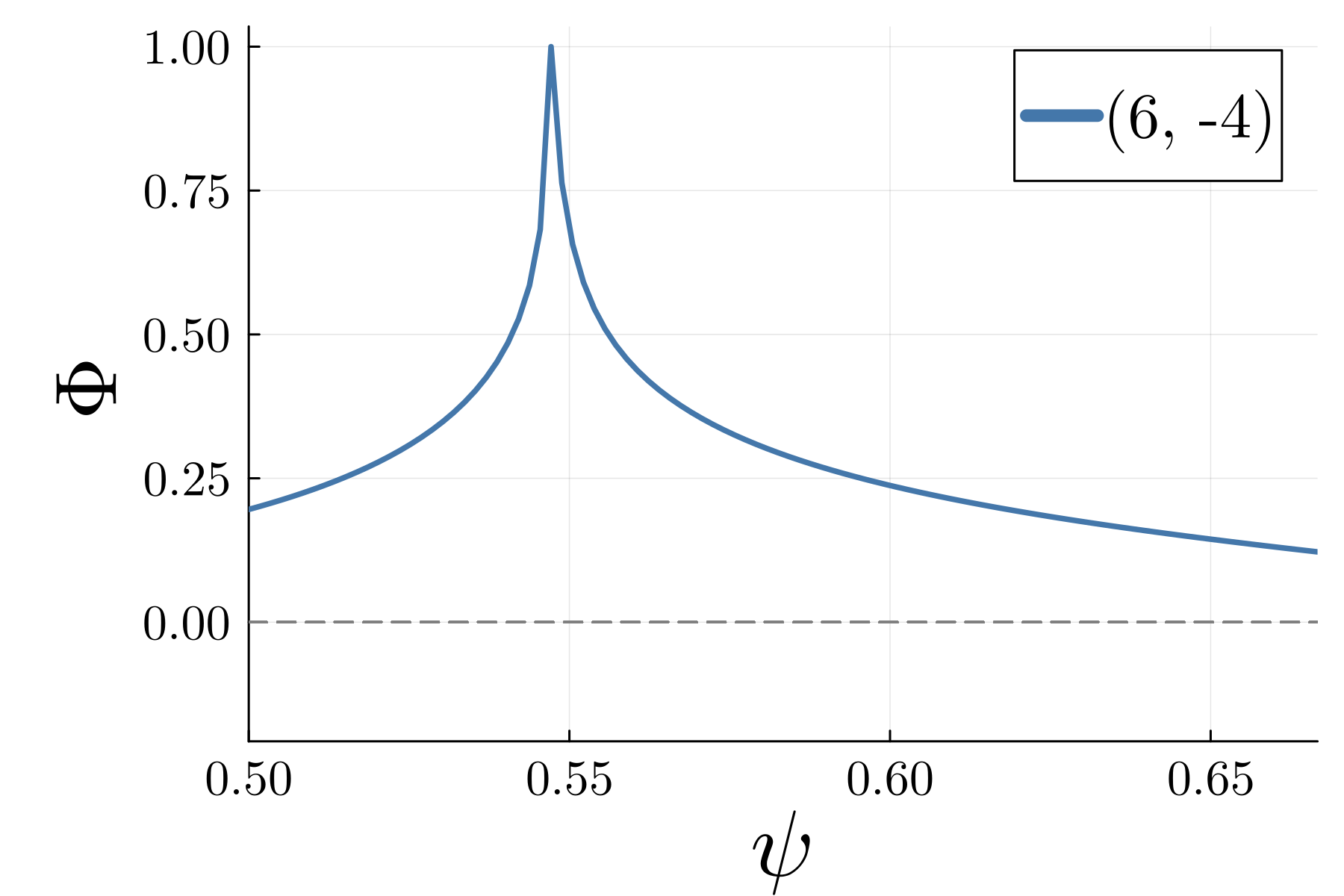}\label{subfig:k0_phi1}}
    \qquad
    \subfloat[$k=0$]{\includegraphics[width=0.45\textwidth]{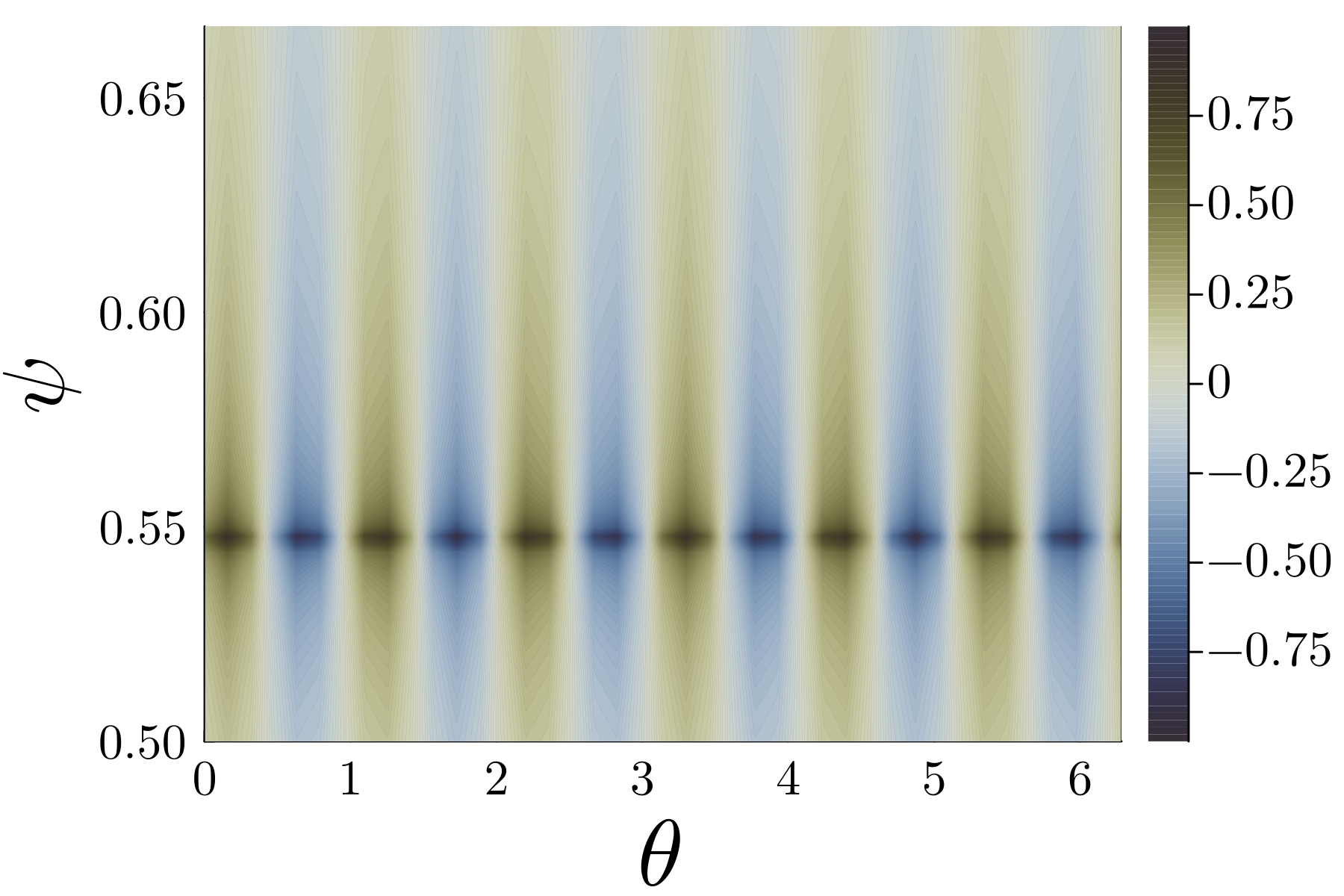}\label{subfig:k0_phi1_cont}}

    \subfloat[$k=0.5\times10^{-3}$]{\includegraphics[width=0.45\textwidth]{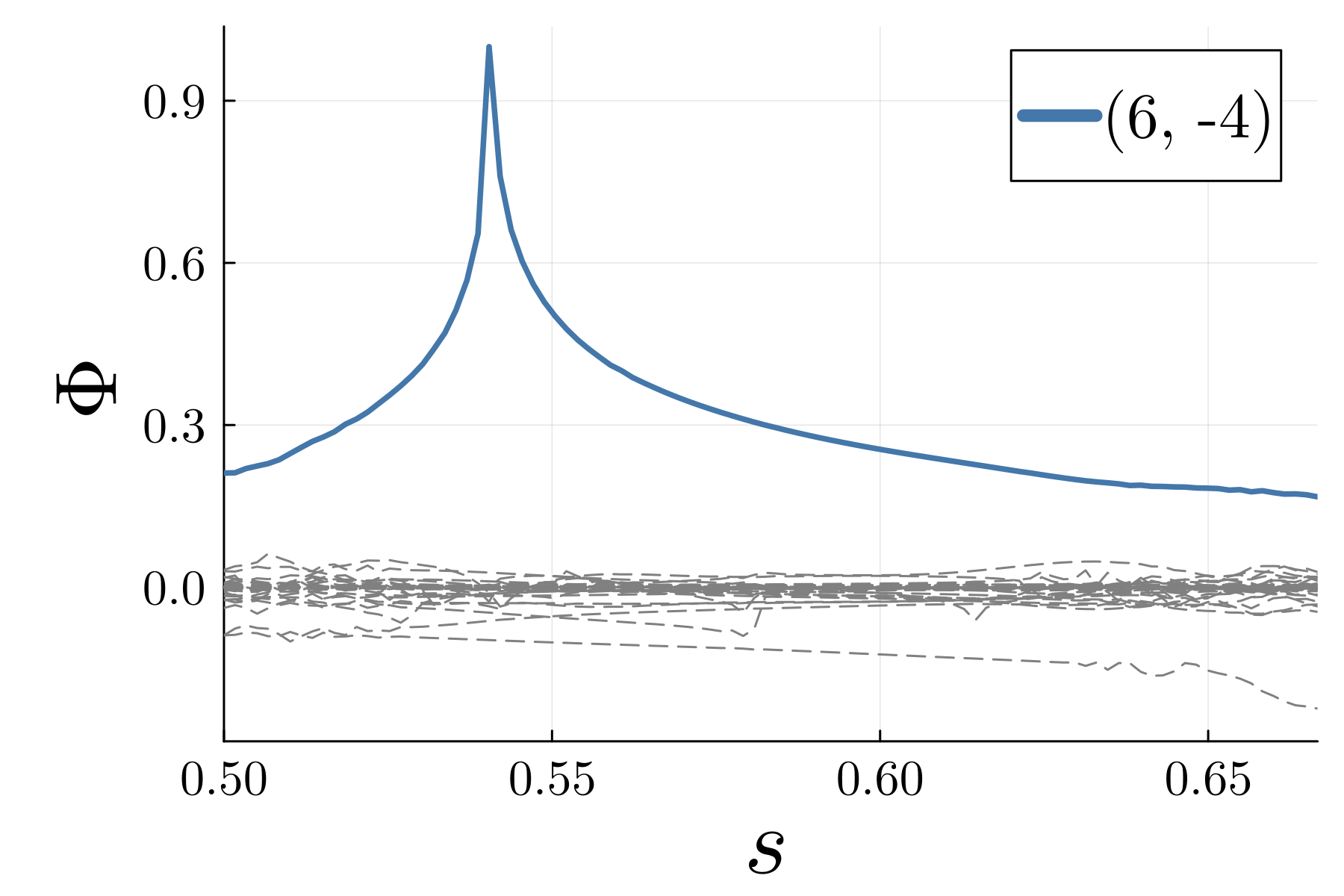}\label{subfig:k05_phi1}}
    \qquad
    \subfloat[$k=0.5\times10^{-3}$]{\includegraphics[width=0.45\textwidth]{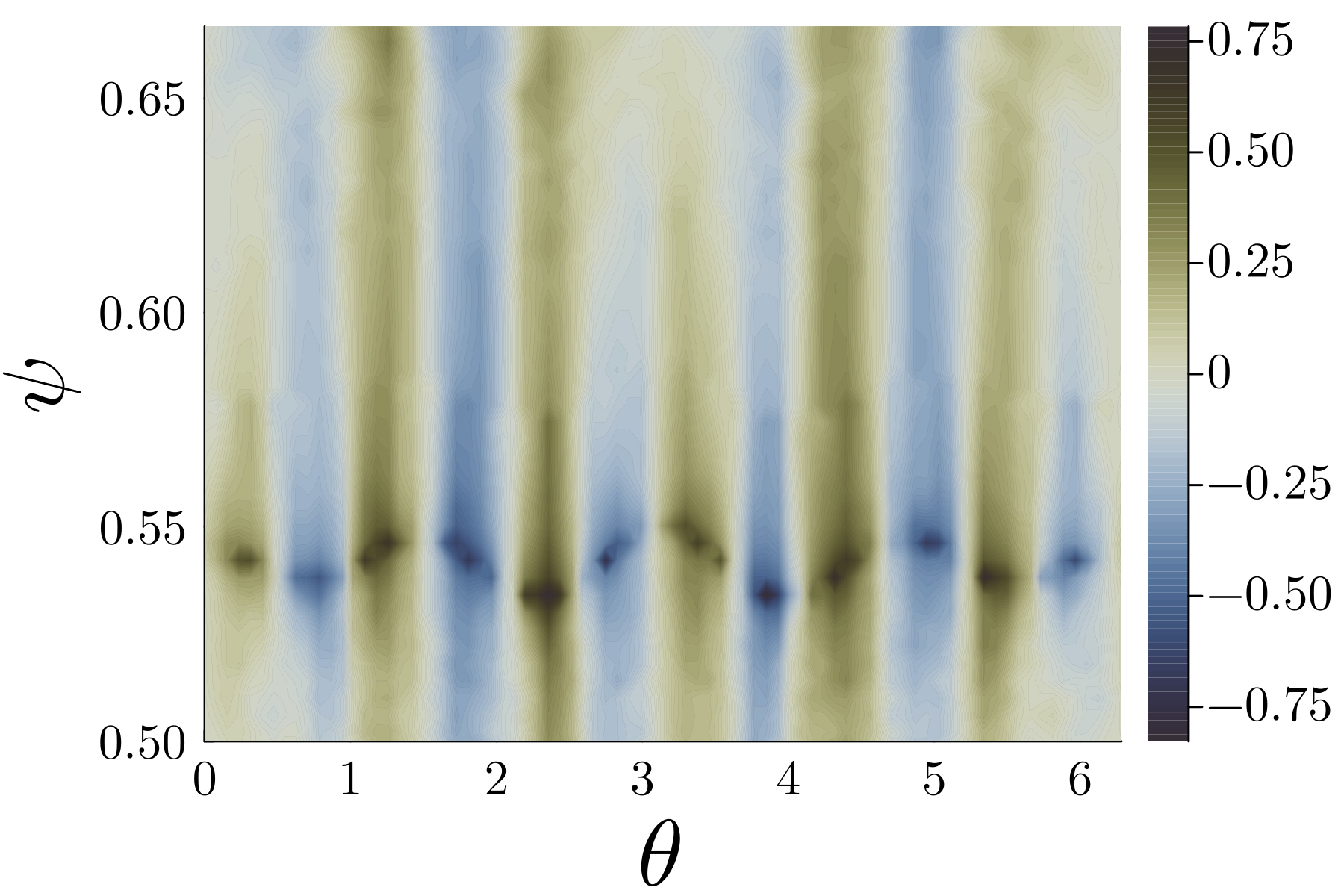}\label{subfig:k05_phi1_cont}}

    \subfloat[$k=1.2\times10^{-3}$]{\includegraphics[width=0.45\textwidth]{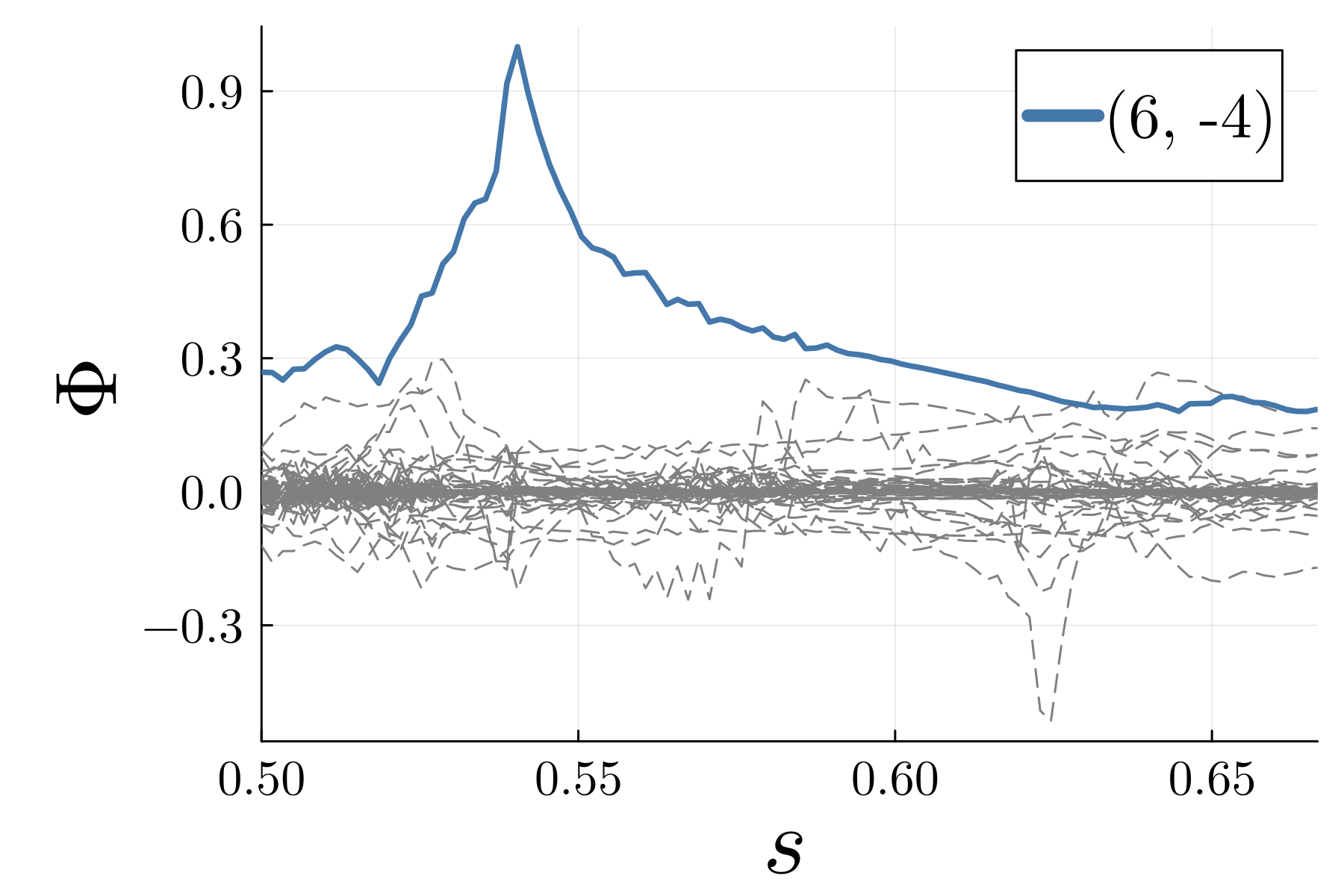}\label{subfig:k12_phi1}}
    \qquad
    \subfloat[$k=1.2\times10^{-3}$]{\includegraphics[width=0.45\textwidth]{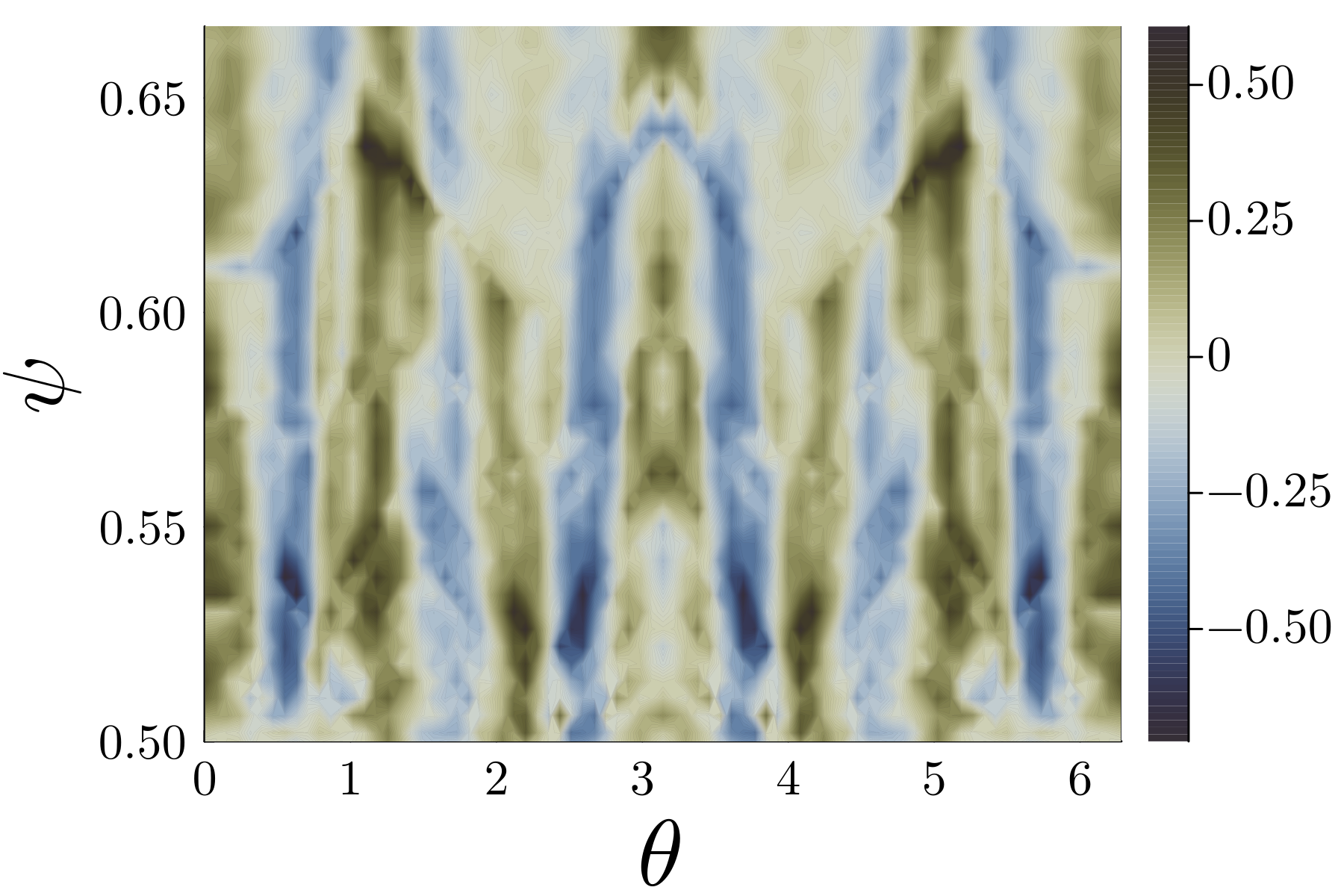}\label{subfig:k12_phi1_cont}}

    \subfloat[$k=1.7\times10^{-3}$]{\includegraphics[width=0.45\textwidth]{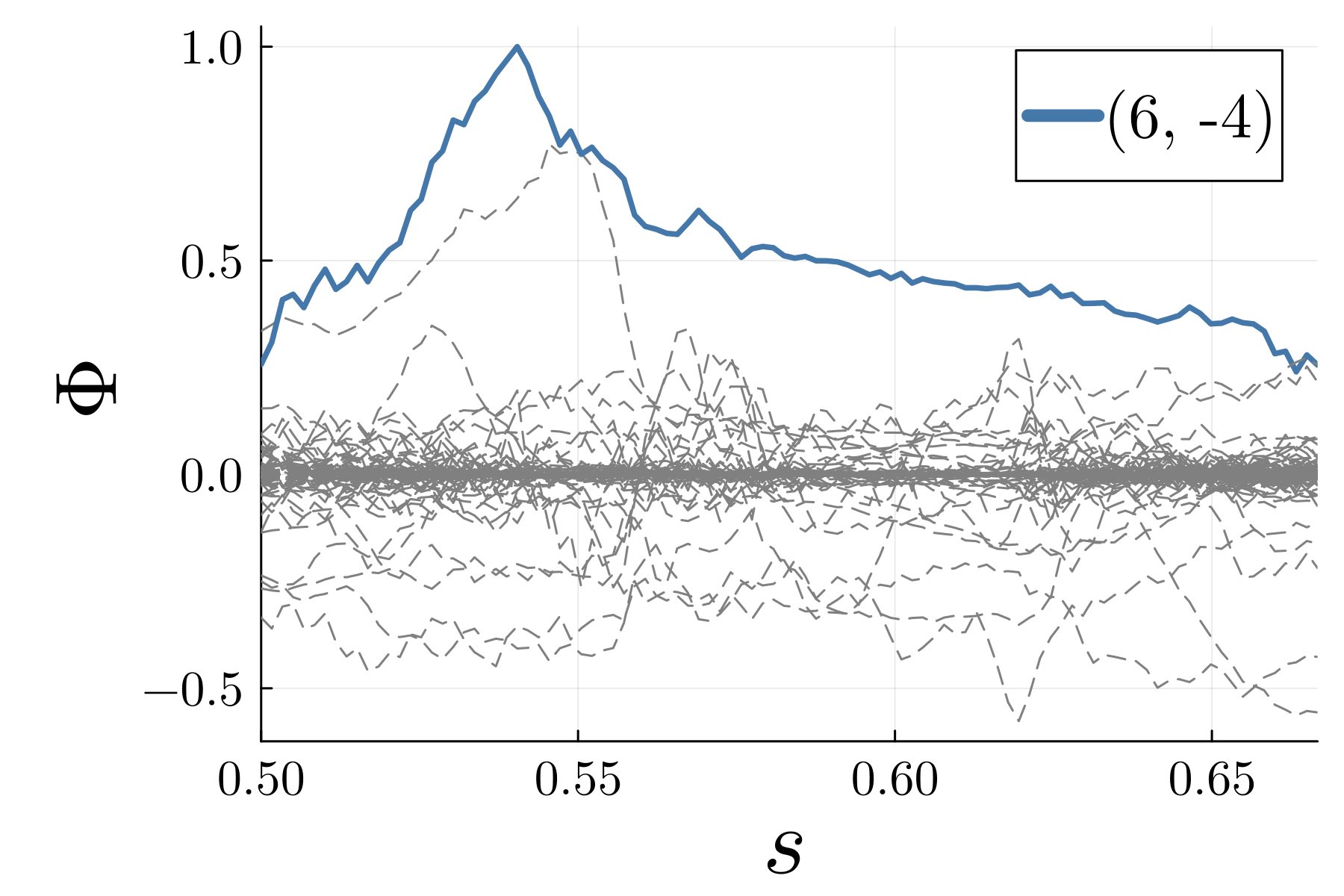}\label{subfig:k17_phi1}}
    \qquad
    \subfloat[$k=1.7\times10^{-3}$]{\includegraphics[width=0.45\textwidth]{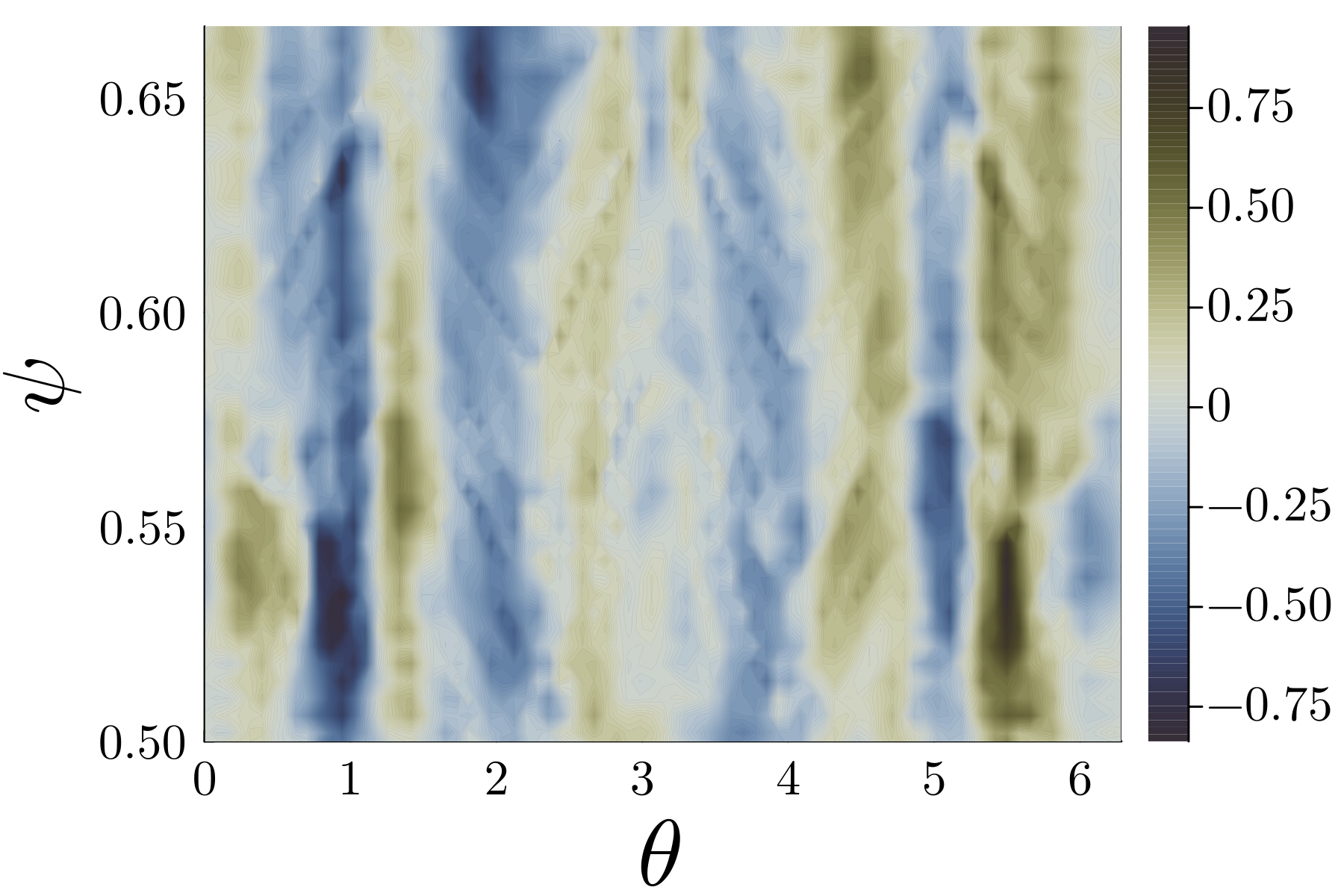}\label{subfig:k17_phi1_cont}}
    \caption{Specific solution residing on $\iota_1$ as $k$ increases. Left column shows the Fourier harmonics in QFM coordinates, $(s, \vartheta, \zeta)$. Right column shows contour plot mapped back to original cylindrical coordinates, $(\psi, \theta, \varphi)$, at $\varphi=0$.}
    \label{fig:solution_1}
\end{figure}

The first of these solutions is shown in figure \ref{fig:solution_1}.
The left hand column shows the Fourier harmonic structure of this solution in the QFM coordinates, $(s, \vartheta, \zeta)$, as $k$ increases. 
The right hand solution shows the contour plot of the potential, but here we have mapped the solution from the QFM coordinates back to the original cylindrical coordinates, $(\psi, \theta, \varphi)$.
This allows us to gain a physical understanding of what is happening to the solution as it evolves with $k$ and to compare with the Poincaré maps.

In the unperturbed case, figure \ref{subfig:k0_phi1}, we have perfect symmetry and straight field line coordinates, so the solution appears as a single harmonic with a sharp localised peak.
Increasing to $k=0.5\times10^{-3}$ in figure \ref{subfig:k05_phi1}, we can see that the harmonic structure remains mostly unchanged, with the slight increase of other harmonics. 
The contour plot, figure \ref{subfig:k05_phi1_cont}, shows the modification of the solution due to the distorted flux surfaces. 
However, because the flux surfaces are still intact, we can see that this solution remains relatively unchanged, it has just conformed to the new flux surface.
This distortion shows the importance of the QFM coordinates.
In the QFM coordinates, we see a simple representation, but in cylindrical coordinates, we see a distorted peaked structure, demanding significantly more numerical resources to accurately resolve.

At $k=1.2\times10^{-3}$, figures \ref{subfig:k12_phi1} and \ref{subfig:k12_phi1_cont}, we are above the critical threshold for the first flux surface, and we can now see a dramatic change in the structure of this solution.
We see a significant modification to the harmonic structure, however, we still see the original $(6, -4)$ harmonic as the largest with roughly the same shape.
In contrast, the contour plot only shows the smallest residue of the original structure and we can see that the solution has \textit{smeared} across the chaotic region.

Despite the flux surfaces being destroyed, the original harmonic persists, which is why the continuum plots still have much of the unperturbed structure.
This is further shown in the next plots, figures \ref{subfig:k17_phi1} and \ref{subfig:k17_phi1_cont}, where all flux surfaces are destroyed.
The contour plot is again unrecognisable, but we still see the persistence of the $(6, -4)$ harmonic, leading to similar continuum structure.

\begin{figure}[htp]
    \centering
    \subfloat[$k=0$]{\includegraphics[width=0.45\textwidth]{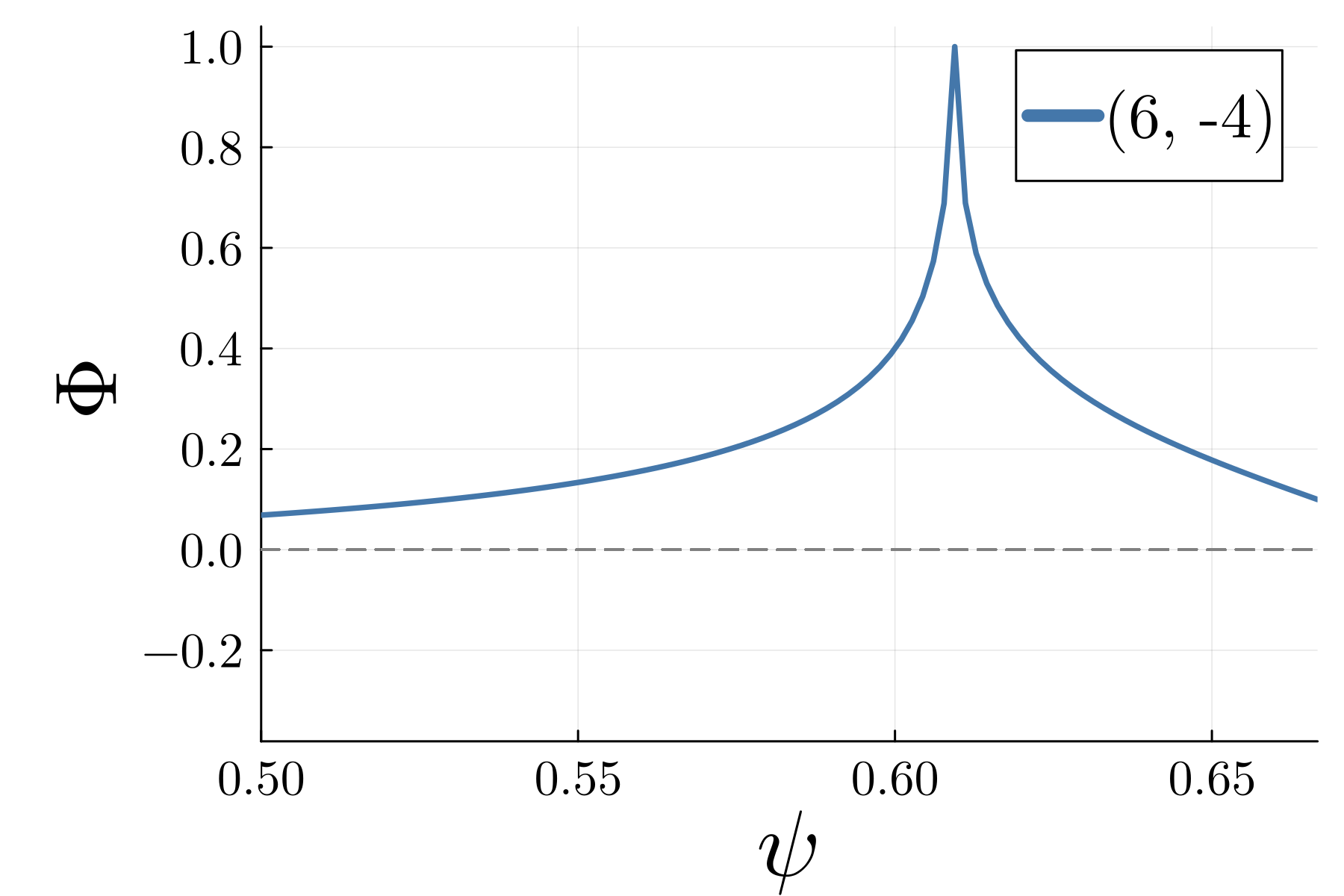}\label{subfig:k0_phi2}}
    \qquad
    \subfloat[$k=0$]{\includegraphics[width=0.45\textwidth]{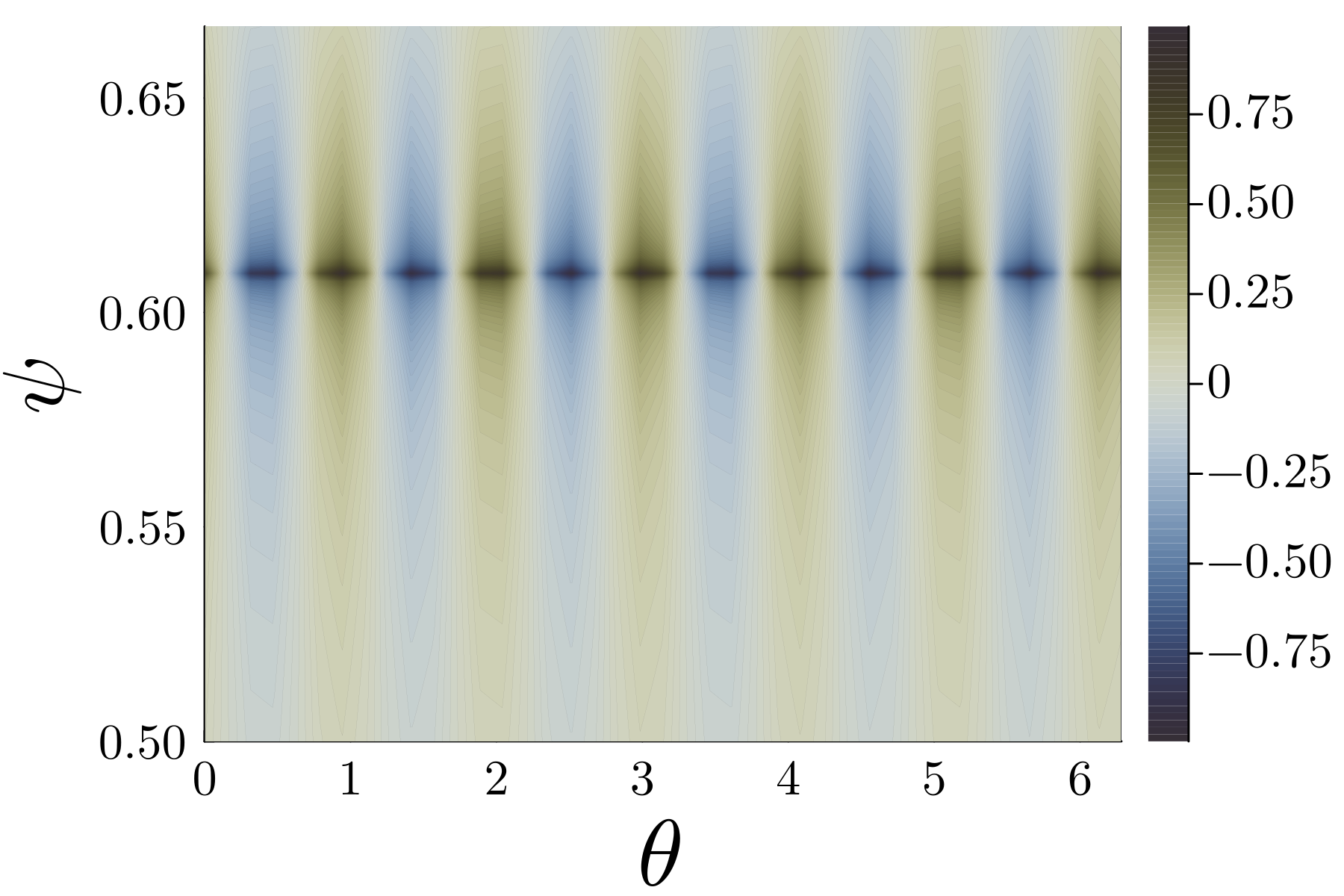}\label{subfig:k0_phi2_cont}}

    \subfloat[$k=0.5\times10^{-3}$]{\includegraphics[width=0.45\textwidth]{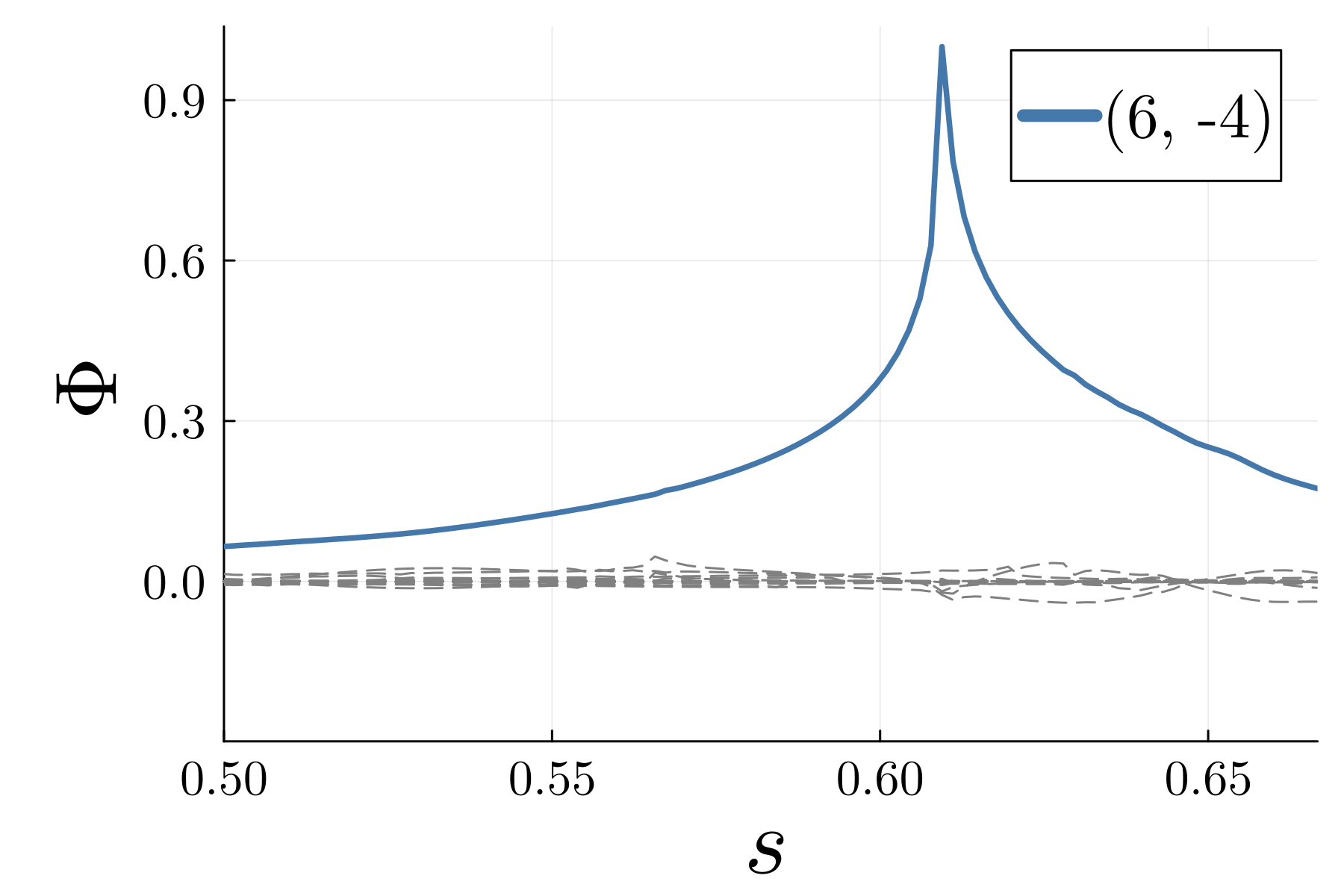}\label{subfig:k05_phi2}}
    \qquad
    \subfloat[$k=0.5\times10^{-3}$]{\includegraphics[width=0.45\textwidth]{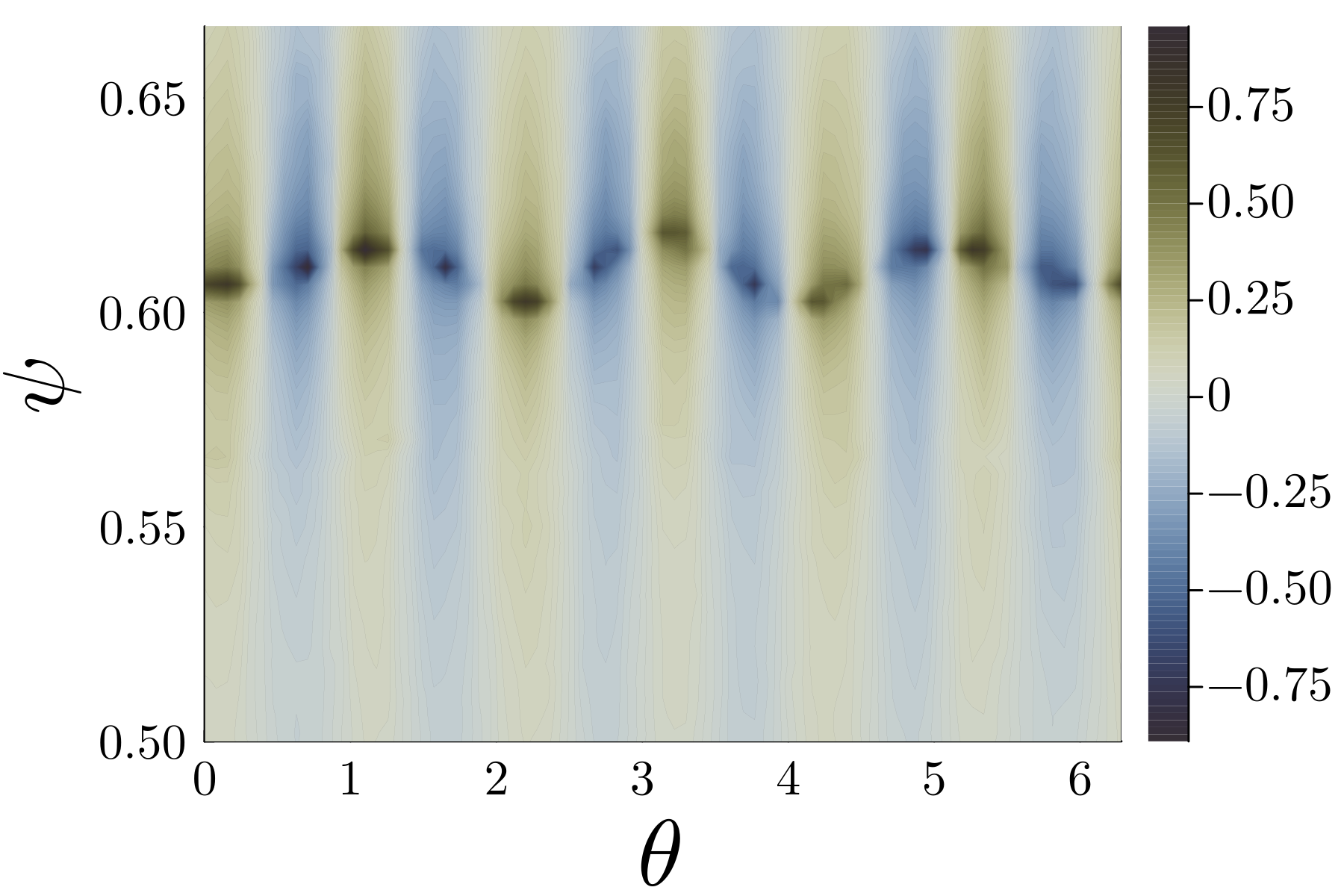}\label{subfig:k05_phi2_cont}}

    \subfloat[$k=1.2\times10^{-3}$]{\includegraphics[width=0.45\textwidth]{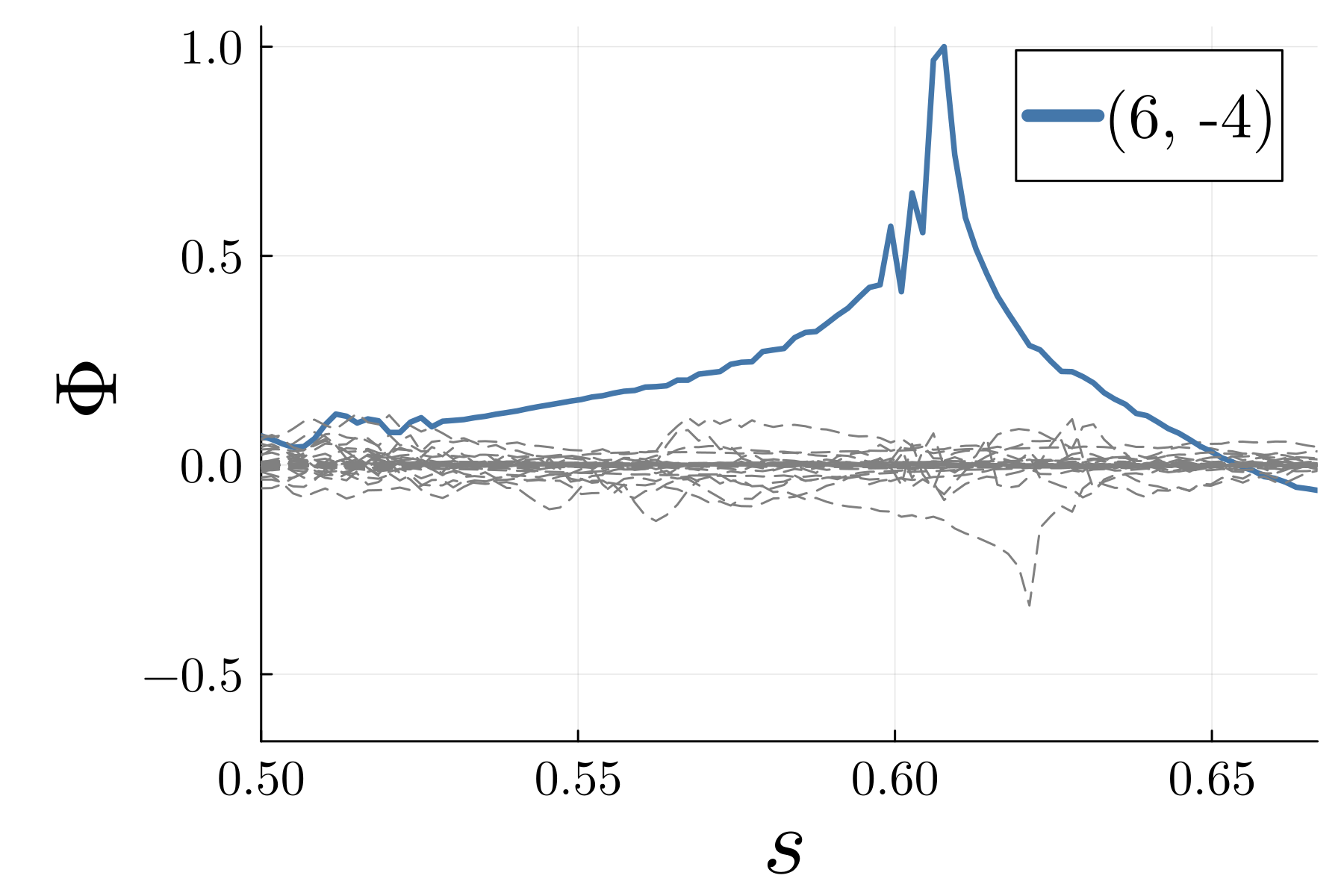}\label{subfig:k12_phi2}}
    \qquad
    \subfloat[$k=1.2\times10^{-3}$]{\includegraphics[width=0.45\textwidth]{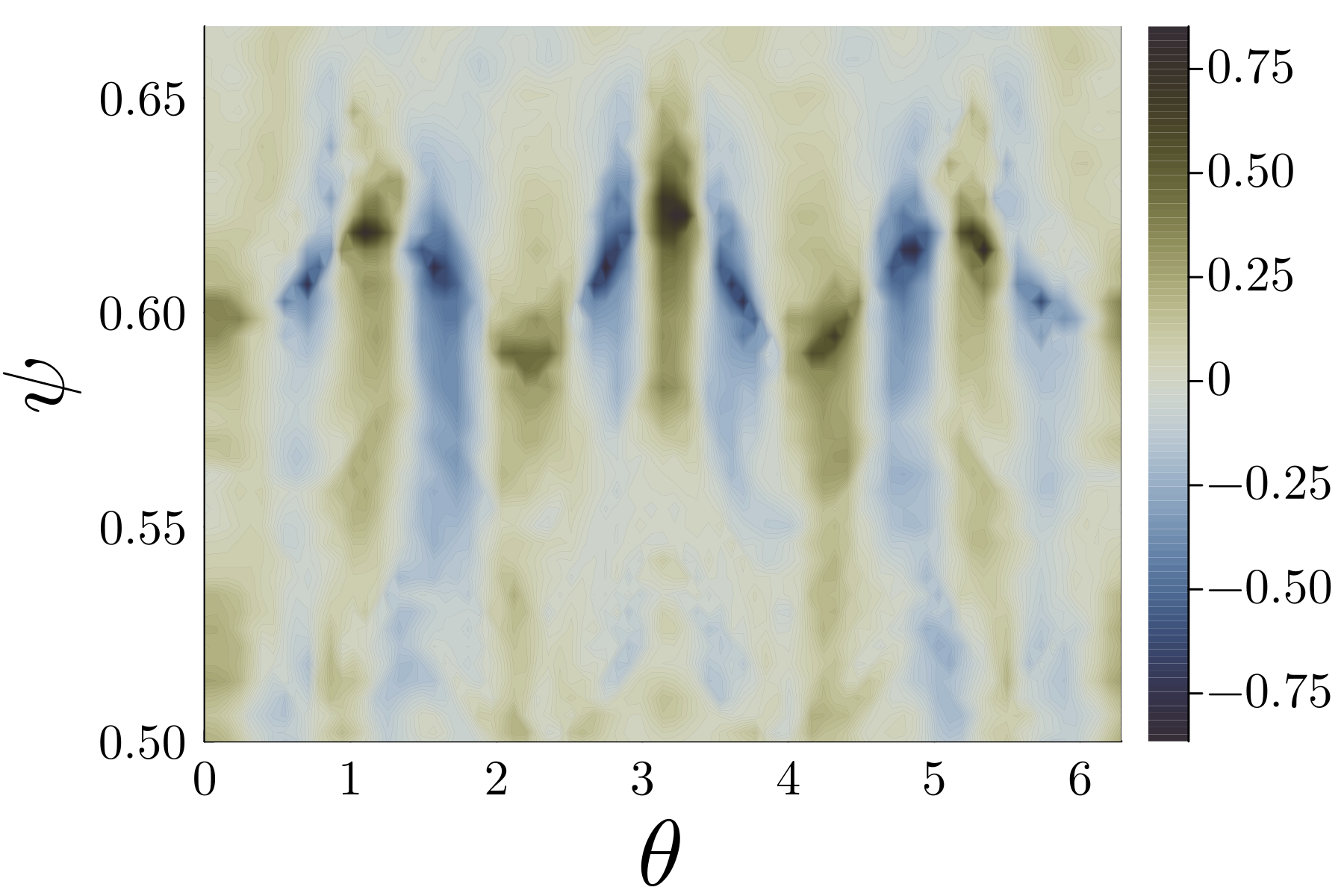}\label{subfig:k12_phi2_cont}}

    \subfloat[$k=1.7\times10^{-3}$]{\includegraphics[width=0.45\textwidth]{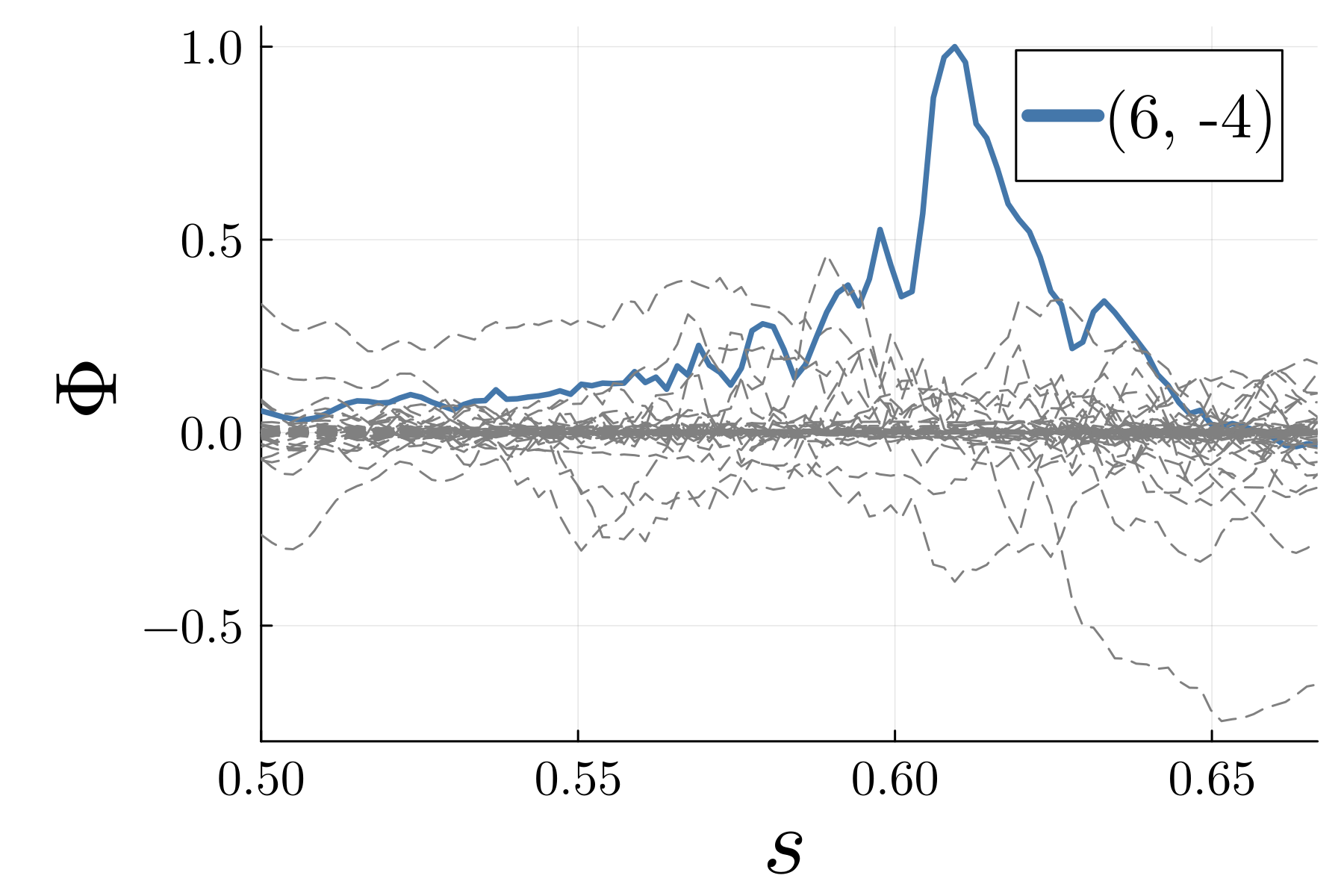}\label{subfig:k17_phi2}}
    \qquad
    \subfloat[$k=1.7\times10^{-3}$]{\includegraphics[width=0.45\textwidth]{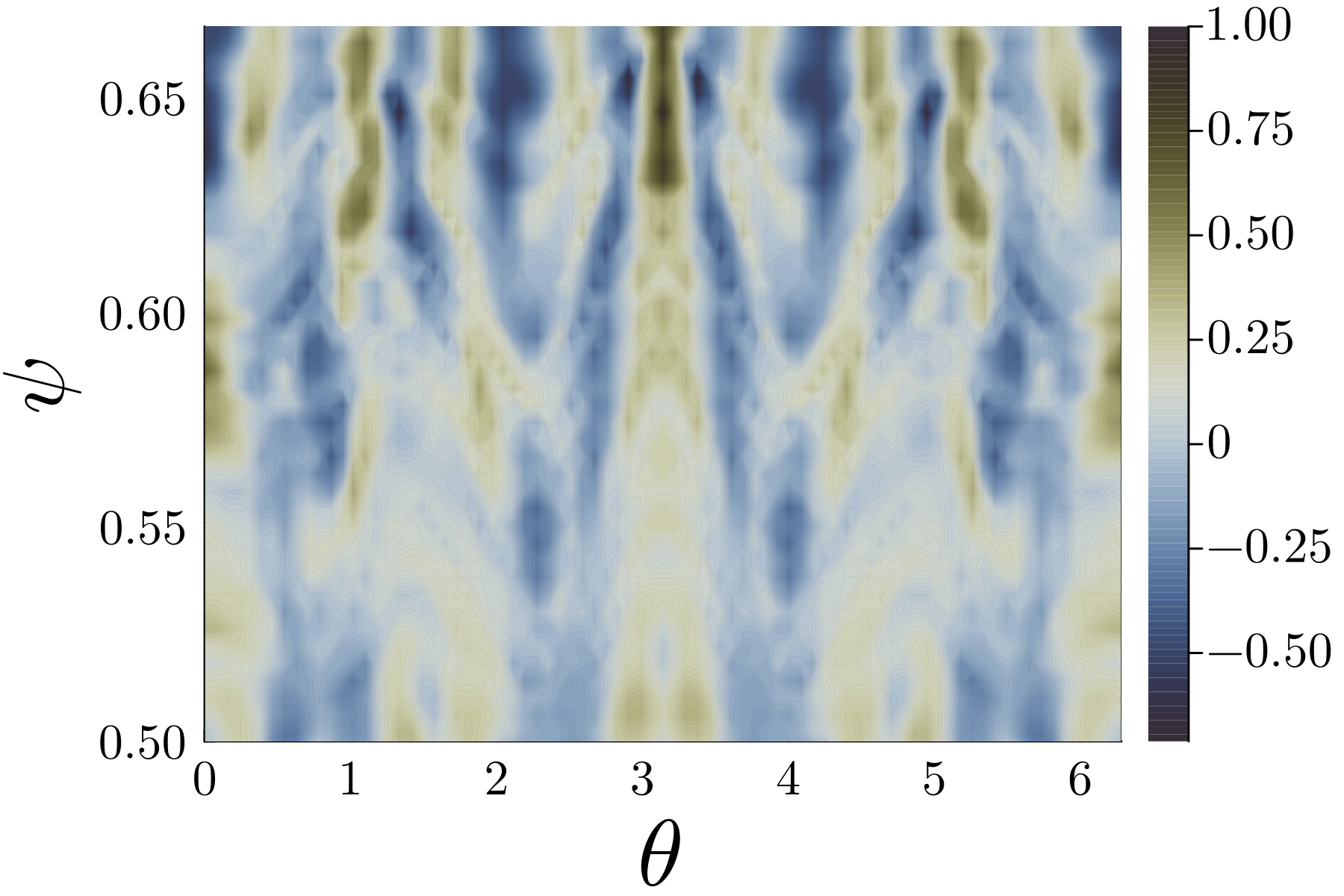}\label{subfig:k17_phi2_cont}}
    \caption{Specific solution residing on $\iota_2$ as $k$ increases. Left column shows the Fourier harmonics in QFM coordinates, $(s, \vartheta, \zeta)$. Right column shows contour plot mapped back to original cylindrical coordinates, $(\psi, \theta, \varphi)$, at $\varphi=0$.}
    \label{fig:solution_2}
\end{figure}

The flux surface for the second solution remains intact for a larger perturbation.
This is reflected in the structure shown in figure \ref{fig:solution_2}.
The unperturbed and $k=0.5\times10^{-3}$ cases show similar behaviour to our last example.
However, now that the flux surface is still intact for $k=1.2\times10^{-3}$, we see that the solution is still sharp and clearly recognisable for this moderate perturbation, shown in figures \ref{subfig:k12_phi2} and \ref{subfig:k12_phi2_cont}.
While the solution is mostly unchanged from the unperturbed case, we do see minor differences.
Because the flux surface is still intact, this is likely due to the imperfections in the numerical QFM coordinates and the destruction of neighbouring flux surfaces.
Once the perturbation exceeds the critical perturbation of this flux surface, figures \ref{subfig:k17_phi2} and \ref{subfig:k17_phi2_cont}, we again see the persistence of the original $(6, -4)$ harmonic and the smearing across the chaotic region as before.

The two solutions we have shown in detail are representative of the entire spectrum.
In particular, two key observations are reflected in the majority of solutions.
First, provided a flux surface is still intact, solutions on that surface will only have minor modification compared to the same solution in the unperturbed case.
The level of modification is typically correlated to the nearby flux surfaces, if a solution is found in region where all flux surfaces are intact, the solution will be almost unchanged from the unperturbed case.
In contrast, if the specific flux surface is in a region of broken flux surfaces we expect to see a much larger distortion.

Second, above the critical threshold where a flux surface is destroyed, we see that the original Fourier harmonic persists while gaining other harmonics.
Additionally, the original harmonic maintains a similar structure to the unperturbed case.
Surprisingly, this behaviour persists even well above the critical threshold where all flux surfaces are broken.

Physically, this is most likely due to the cantorus \cite{percival_stochastic_1979, hudson_calculation_2006} that forms when a flux surface is broken. 
A cantorus acts as a `leaky' flux surface with gaps, which allows the magnetic field to pass through.
However, it will still act as a partial barrier, with the same frequency as the original flux surface.
This persistent structure appears to be sufficient for the original Fourier harmonic to still be observed, as the magnetic field line can remain contained to a cantorus for many orbits.
The gaps however, explain why we see a smearing of the wave as it is no longer perfectly constrained to the single surface and can leak to a wider region.

\begin{figure}[ht]
    \centering
    \subfloat[]{\includegraphics[width=0.45\textwidth]{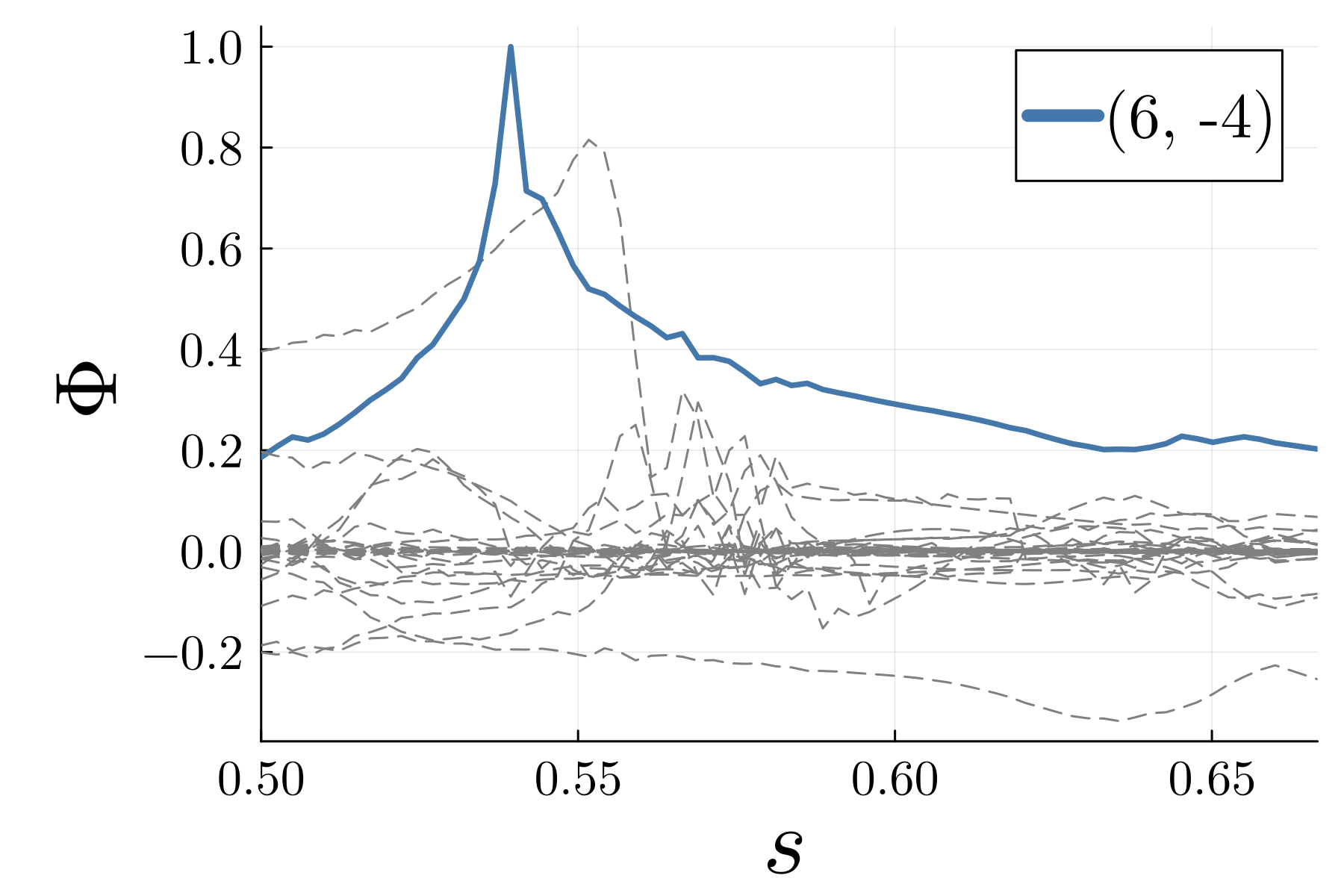}\label{subfig:low_res_1}}
    \qquad
    \subfloat[]{\includegraphics[width=0.45\textwidth]{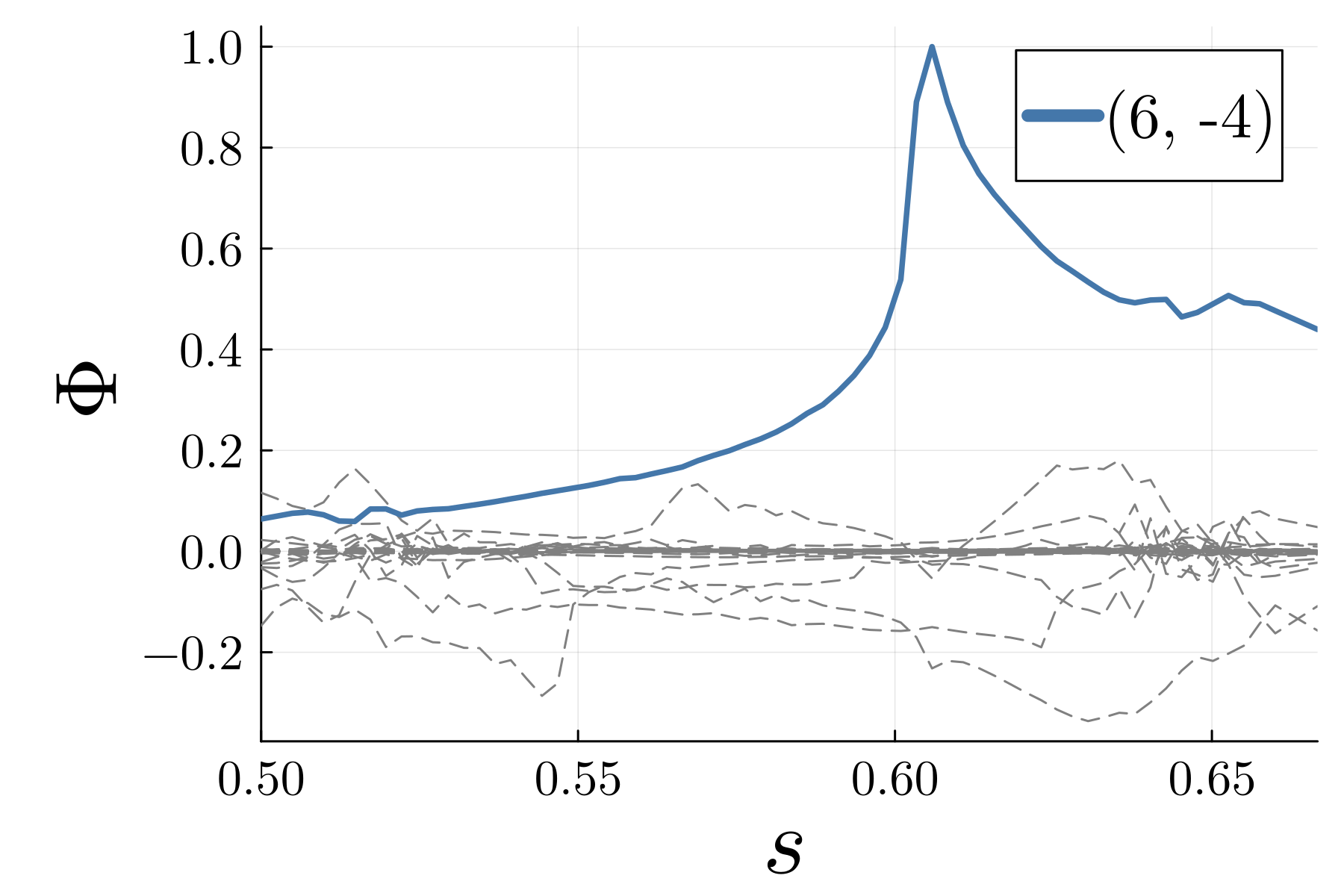}\label{subfig:low_res_2}}
    \caption{Fourier harmonics of the first (a) and second (b) solutions computed with a lower grid resolution.}
    \label{fig:low_res}
\end{figure}

Considering the same cases in a lower resolution of $100\times 20\times 15$, much of the same behaviour is observed.
As a specific example, figure \ref{fig:low_res} shows the harmonic structure of two solutions that are closest in frequency to the high resolution cases for $k=1.2\times10^{-3}$. 
Here we can see the same general properties, both have a dominant $(6, -4)$ harmonic.
The first solution, figure \ref{subfig:low_res_1}, on a broken flux surface, also includes a variety of other harmonics while the second, figure \ref{subfig:low_res_2}, existing on an intact flux surface, is predominantly this single harmonic.

While the general trends are consistent across different resolutions, we can see that the exact structure is not.
One reason for this is due to the numerical construction of the QFM coordinates.
Different perturbation strengths require a different set of QFM surfaces, resulting in a slightly different mapping between QFM, $(s, \vartheta, \zeta)$, and cylindrical, $(\psi, \theta, \varphi)$, coordinates.
Because these solutions are immersed in a continuum, determining the exact same solution across different cases is not possible without infinite resolution.

Therefore, the low resolution solutions are not the same solutions presented before, but instead are neighbouring solutions.
With a chaotic field, this minimal difference in coordinate mapping can still result in large differences in magnetic field trajectories and may explain the variation in the additional harmonics seen.

Despite both Fourier plots showing the unperturbed solutions remaining mostly unchanged for intact flux surfaces and the persistence of the original harmonic for destroyed flux surfaces, there are some notable exceptions to these trends.
Any solutions that have a strong interaction with the magnetic island chains, at either resolution, do not follow these observations.
This includes solutions localised inside the island chains, with helicity matching the islands or particularly broad structure, such as the $m=\pm 1$ solutions. 
Examples of these solutions are shown in the next section where we further investigate the effect of the magnetic island structures.


\section{Single Island Chain}\label{sec:single_island}

We now consider the simpler perturbation of a single island chain to understand how these structures effect the spectrum.
This case allows a more direct comparison with the use of pre-existing straight field line coordinates, $(\kappa, \ab, \tau)$, where $\kappa$ is the radius from the $O$-point, $\ab$ is the helical straight field line angle and $\tau$ is the extend toroidal angle along the island magnetic axis, see appendix \ref{app:island_metric} for details.
These coordinates are only defined inside or outside the magnetic island chain, with a coordinate singularity at the separatrix, given by $\kappa=1$.
Previous work employing these coordinates \cite{biancalani_continuous_2010, qu_shear_2023, konies_numerical_2022, konies_shear_2024} has revealed a rich structure in the shear Alfvén spectrum inside magnetic islands.

We will consider the simplest possible island chain, $m=1, n=-1$, such that we maximise the number of grid points for resolving the structure inside the island chain.
This is introduced in a similar way as before, with a perturbation of the form,
\begin{align}\label{eqn:single_isl_perturbation}
    \BB_1 = A\sin(\theta-\varphi)\Bc{\theta}{\varphi}.
\end{align}
We employ the same algorithm as before and again make use of QFM coordinates which will align our coordinate system with the flux surfaces outside the island.
The $q$-profile is chosen such that the island coordinates are analytical \cite{qu_shear_2023}, this is given by,
\begin{align}
    \oo{q} = \oo{q_0} - \frac{q'}{q_0^2}\left(\psi-\psi_0\right)
\end{align}
where we have taken $\psi_0=0.5, q'=1.0$ and $q_0=-m_0/n_0=1.0$. We take an island width of $w=0.1$, allowing us to compute the amplitude via,
\begin{align}
    A = \frac{q' w^2}{16 q_0^2}=6.25\times10^{-4}.
\end{align}

\begin{figure}[ht]
    \centering
    \includegraphics[width=0.5\linewidth]{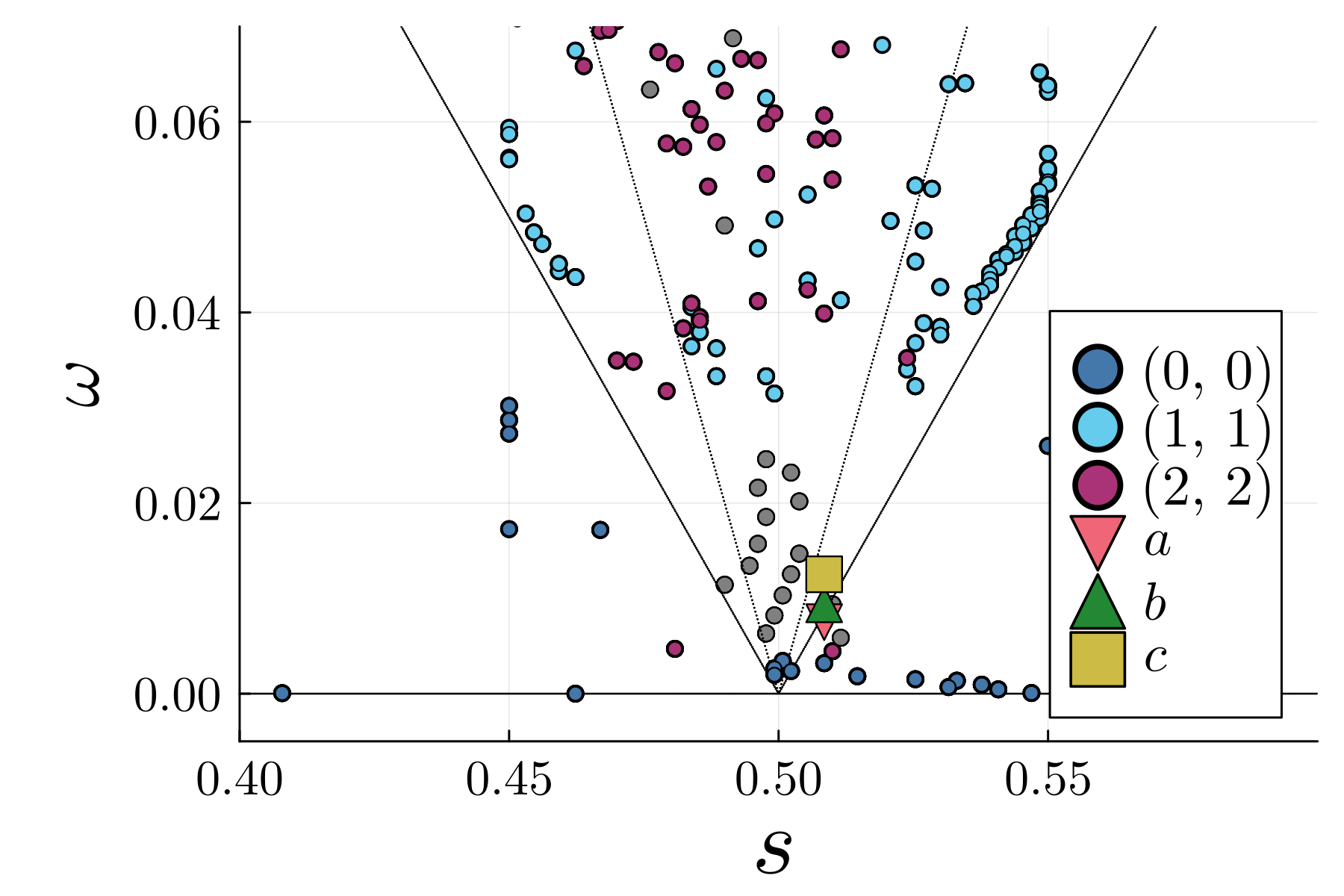}
    \caption{Continuum computed in QFM coordinates, $(s, \vartheta, \zeta)$, with a single magnetic island chain. Highlighted are $3$ likely global modes.}
    \label{fig:isl_cont}
\end{figure}

\begin{figure}[htp]
    \centering
    \subfloat[Global mode $a$]{\includegraphics[width=0.45\textwidth]{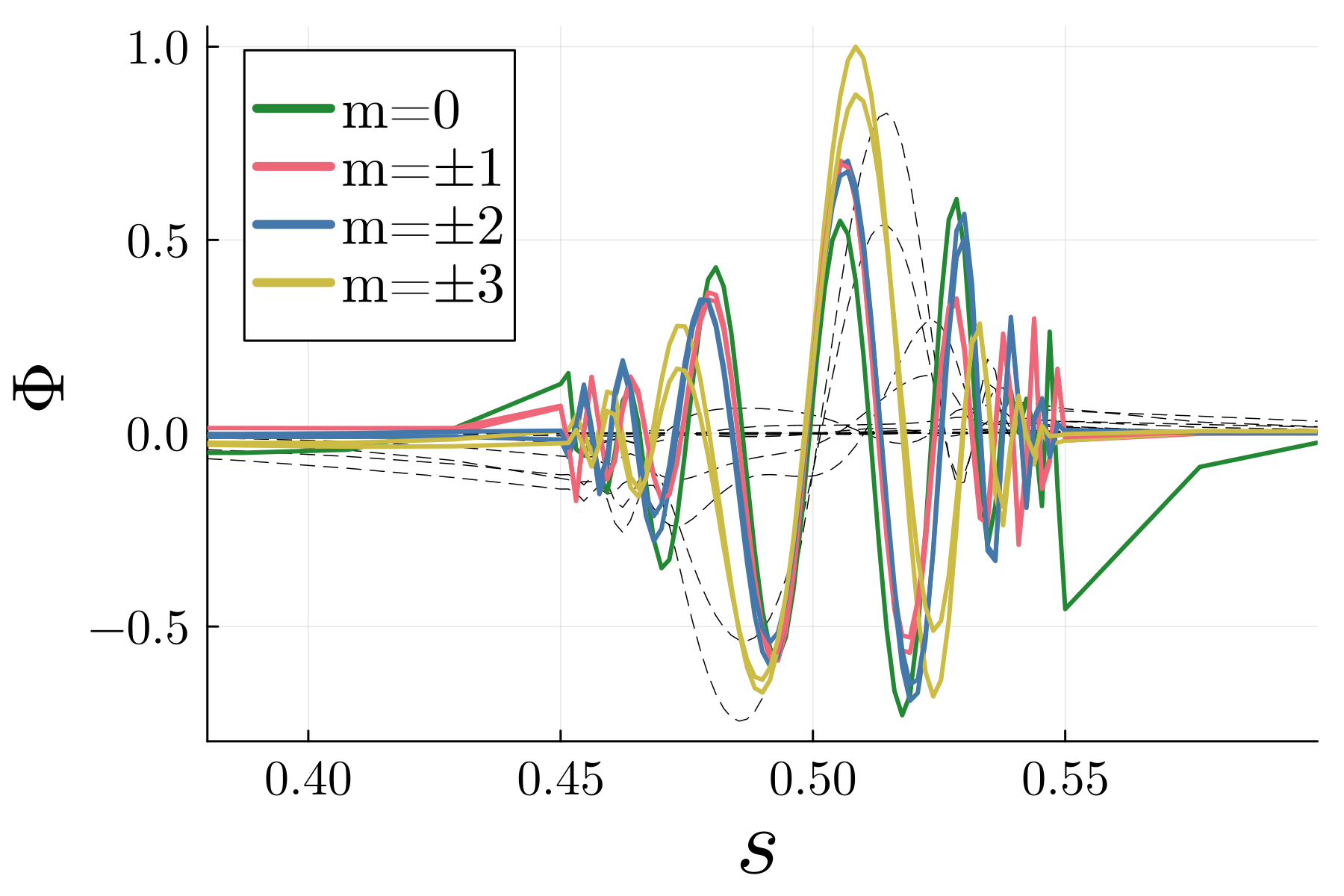}\label{subfig:isl_a}}
    \qquad
    \subfloat[Global mode $a$]{\includegraphics[width=0.45\textwidth]{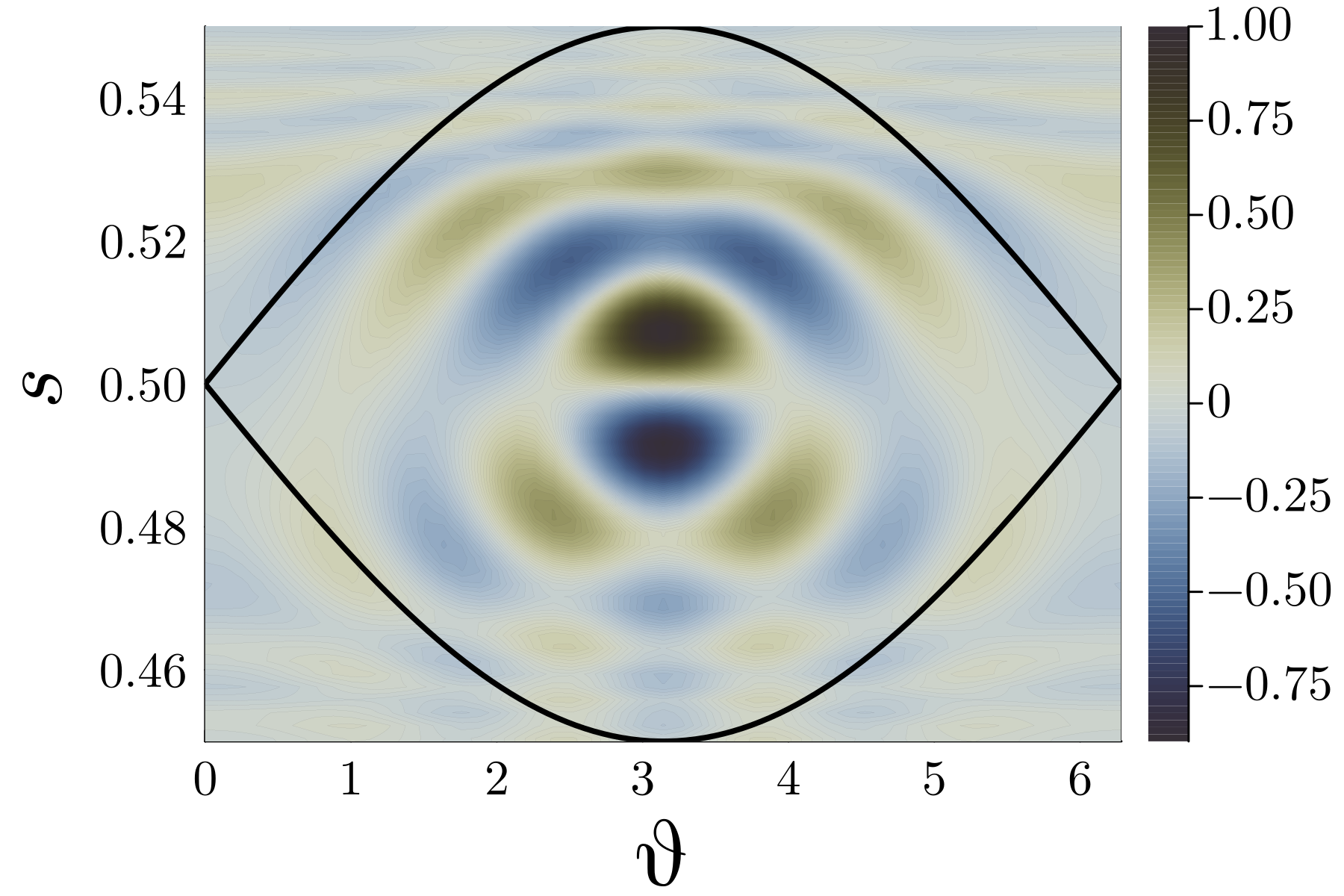}\label{subfig:isl_a_cont}}

    \subfloat[Global mode $b$]{\includegraphics[width=0.45\textwidth]{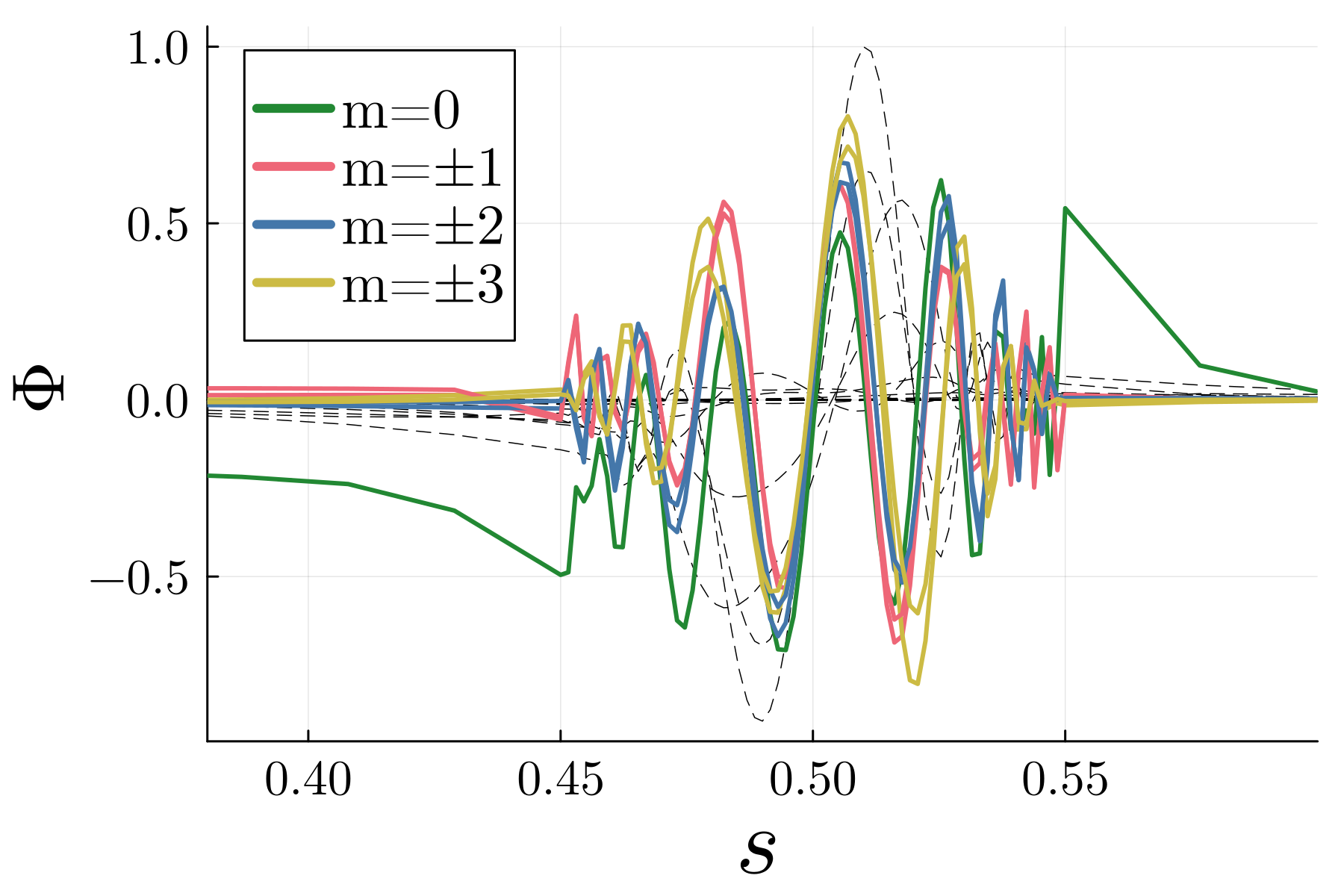}\label{subfig:isl_b}}
    \qquad
    \subfloat[Global mode $b$]{\includegraphics[width=0.45\textwidth]{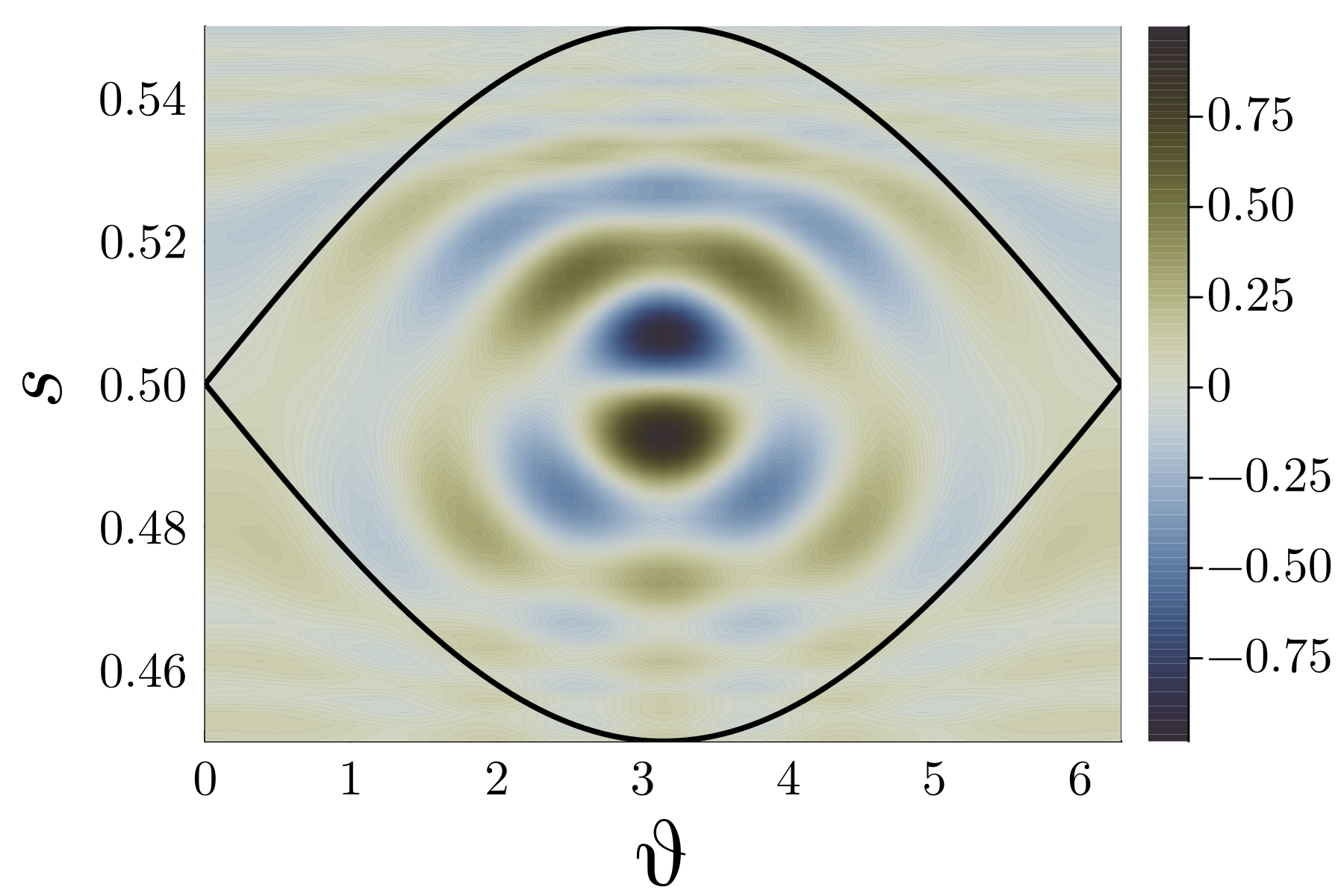}\label{subfig:isl_b_cont}}

    \subfloat[Global mode $c$]{\includegraphics[width=0.45\textwidth]{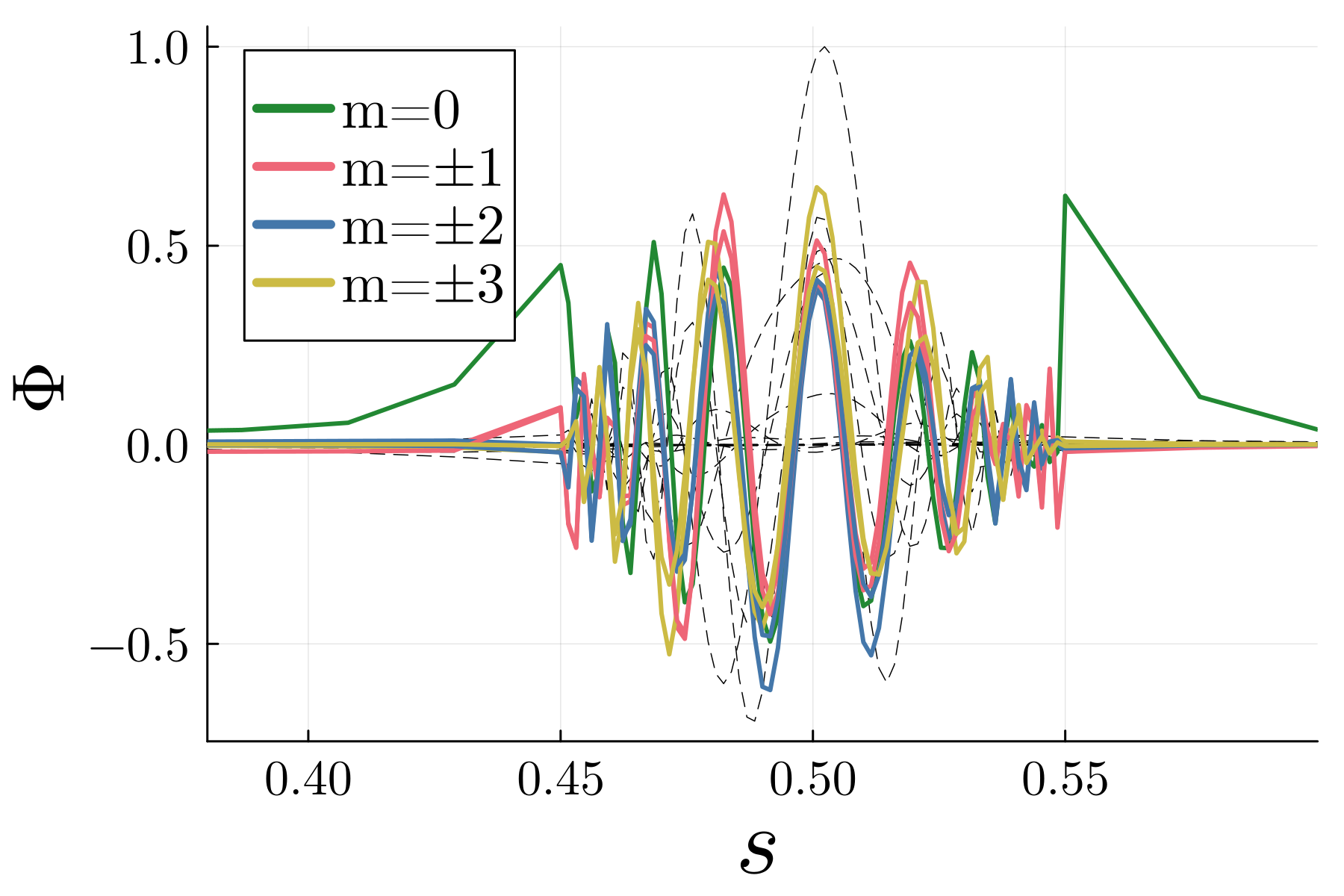}\label{subfig:isl_c}}
    \qquad
    \subfloat[Global mode $c$]{\includegraphics[width=0.45\textwidth]{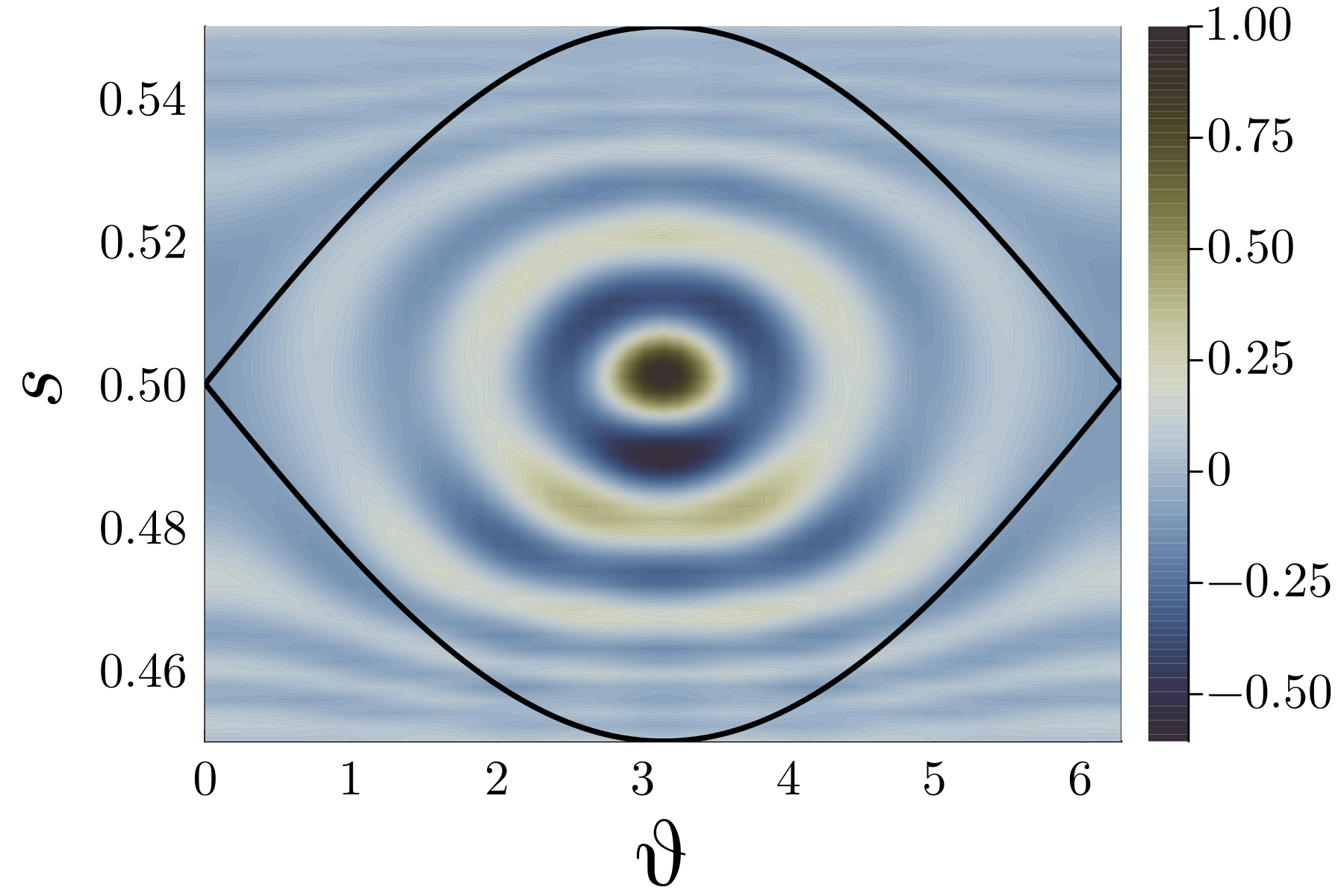}\label{subfig:isl_c_cont}}
    \caption{Three likely island global modes computed using QFM coordinates, $(s, \vartheta, \zeta)$. Left column shows Fourier harmonics and right column shows contour plot at $\zeta=0$, with the island separatrix in black.}
    \label{fig:isl_solutions}
\end{figure}

One of the most notable features of the spectrum inside a magnetic island is the new frequency gap and discrete magnetic island induced Alfvén eigenmodes present in the gap.
These both occur at low frequencies \cite{biancalani_continuous_2010, qu_shear_2023, konies_numerical_2022, konies_shear_2024}, so we will focus on this region.
The spectrum produced under this perturbation is shown in figure \ref{fig:isl_cont}, where the black lines show the unperturbed continuum.
Notably, we can see branches that previously go down to zero instead flatten out at $\omega \approx 0.03$.
We have also labelled $3$ specific solutions, $a,b,c$.
The Fourier harmonics of these are shown on the left column of figure \ref{fig:isl_solutions}, with the corresponding contour plots shown in the right column.

The Fourier harmonics shown here resemble the harmonic structure of the island induced mode found by Cook \textit{et al}. \cite{cook_identification_2016}.
Looking at the contour plots, we can see that these modes are clearly conforming to the island structure and are highly localised inside the separatrix, shown in black.

For a more quantitative comparison, we employ the straight field line coordinates used in previous work \cite{qu_shear_2023} to recompute the same spectrum, focusing on the inside of the island.
Additionally, we use the continuum calculation of Qu and Hole \cite{qu_shear_2023} to verify this result.

\begin{figure}[ht]
    \centering
    \subfloat[]{\includegraphics[width=0.45\textwidth]{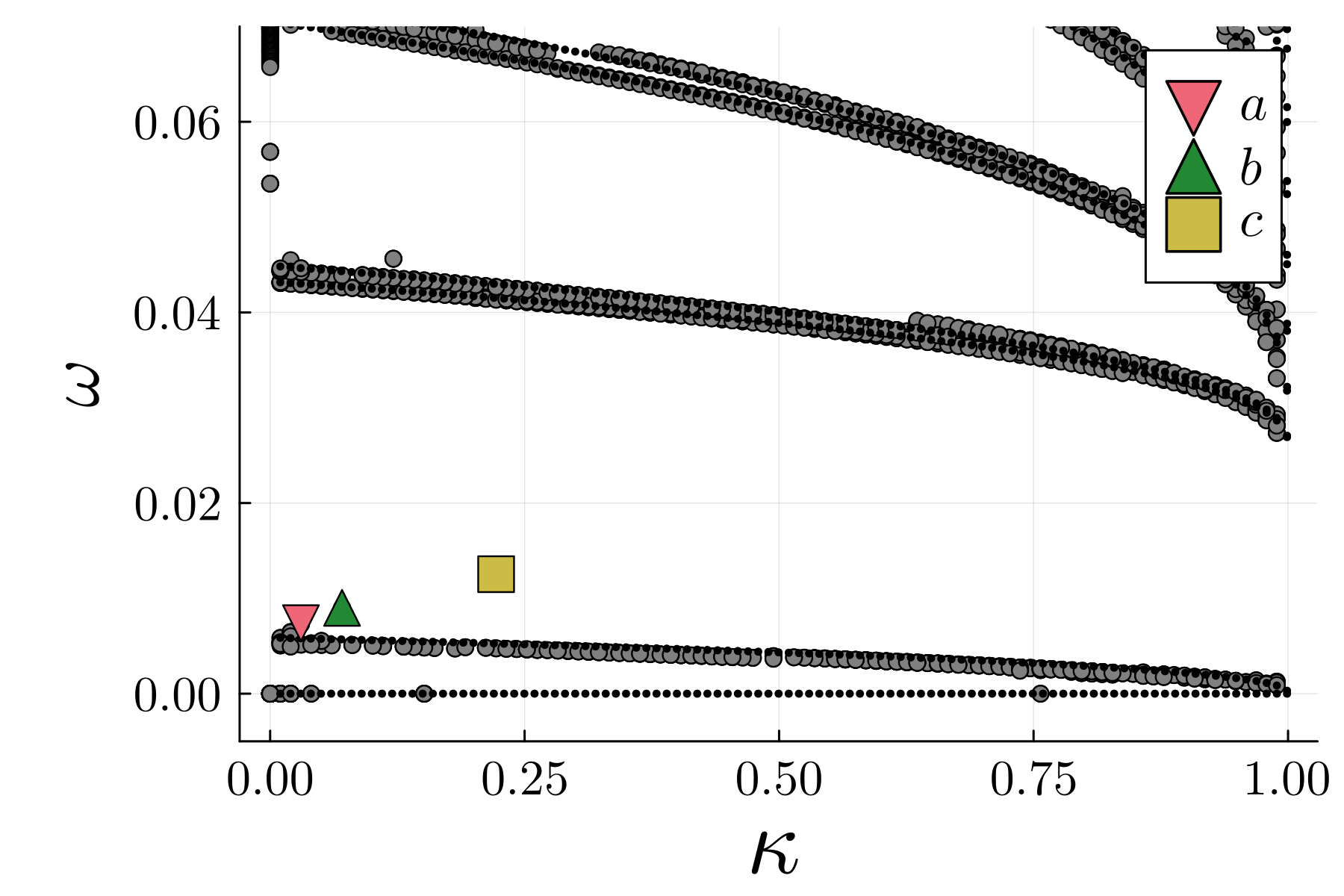}\label{subfig:in_isl_cont}}
    \qquad
    \subfloat[]{\includegraphics[width=0.45\textwidth]{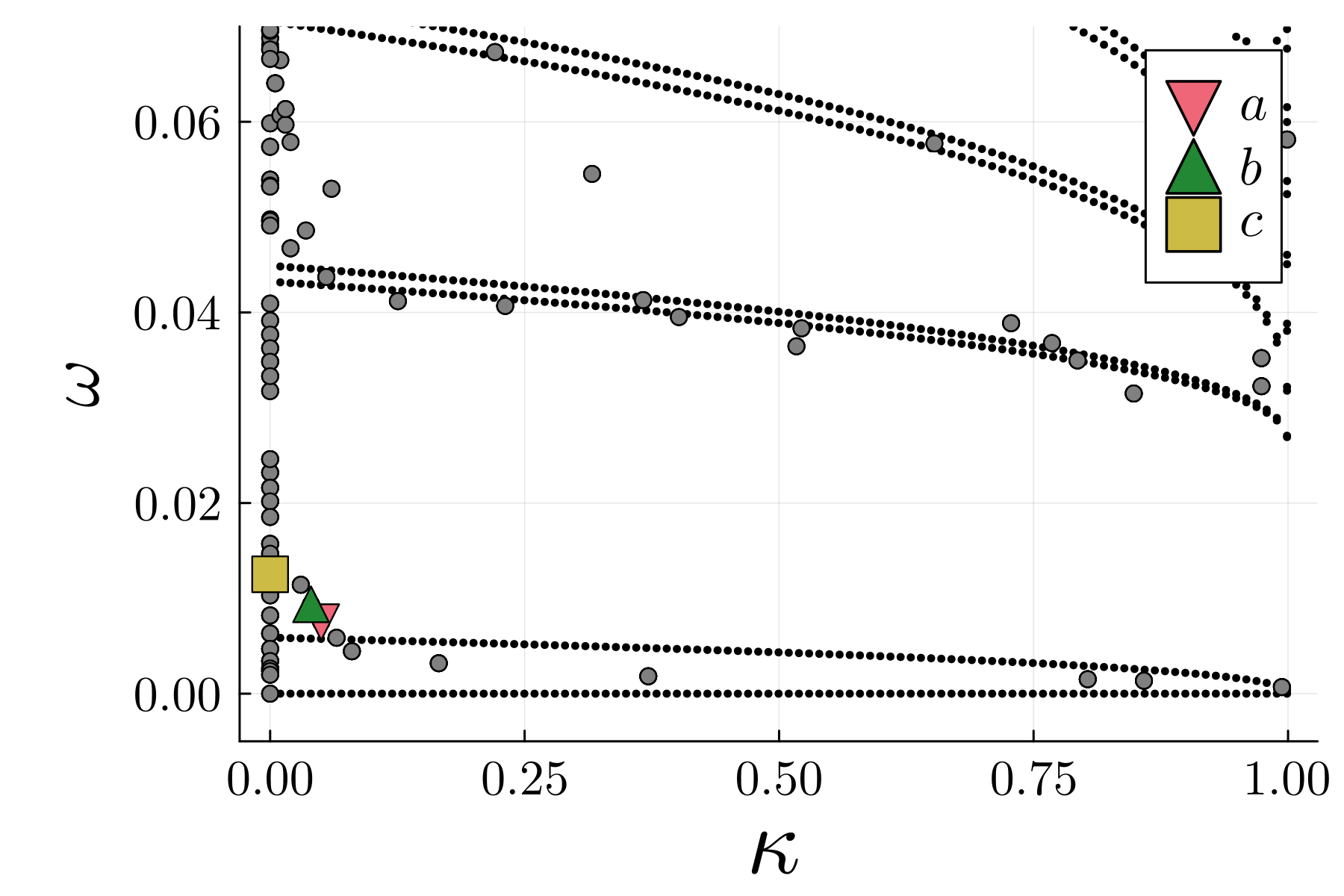}\label{subfig:mapped_isl_cont}}
    \caption{Shear Alfvén spectrum inside a magnetic island. Computed with island coordinates, $(\kappa, \ab, \tau)$ (a) and computed in QFM coordinates, $(s, \vartheta, \zeta)$, then mapped to island coordinates for comparison (b).}
    \label{fig:in_isl_cont}
\end{figure}

The spectrum computed with these coordinates is shown in figure \ref{subfig:in_isl_cont}, where we have highlighted $3$ global modes found, which provided the frequencies for choosing the modes selected earlier.
This figure also shows the previous computation of the continuum in black, showing good agreement.

To compare directly, we have mapped the calculation performed in QFM coordinates, $(s, \vartheta, \zeta)$, into the coordinates used for the island calculation, $(\kappa, \ab, \tau)$, noting that only solutions that peak inside the island are mapped.
This mapping is shown in figure \ref{subfig:mapped_isl_cont}, again with the previous continuum calculation shown in black.
Here we can see some small aspects of the continuum predicted with straight field line island coordinates, there is a small gap in frequency around $\omega=0.03$, however, it is much smaller than expected. We also see that gap modes at similar frequencies, but there are many more than predicted by the island coordinates. Finally, the continuum branch at $\omega\approx0.04$ appears to have a few matching solutions but it is clear the continuum is not adequately computed.

\begin{figure}[htp]
    \centering
    \subfloat[Global mode $a$]{\includegraphics[width=0.45\textwidth]{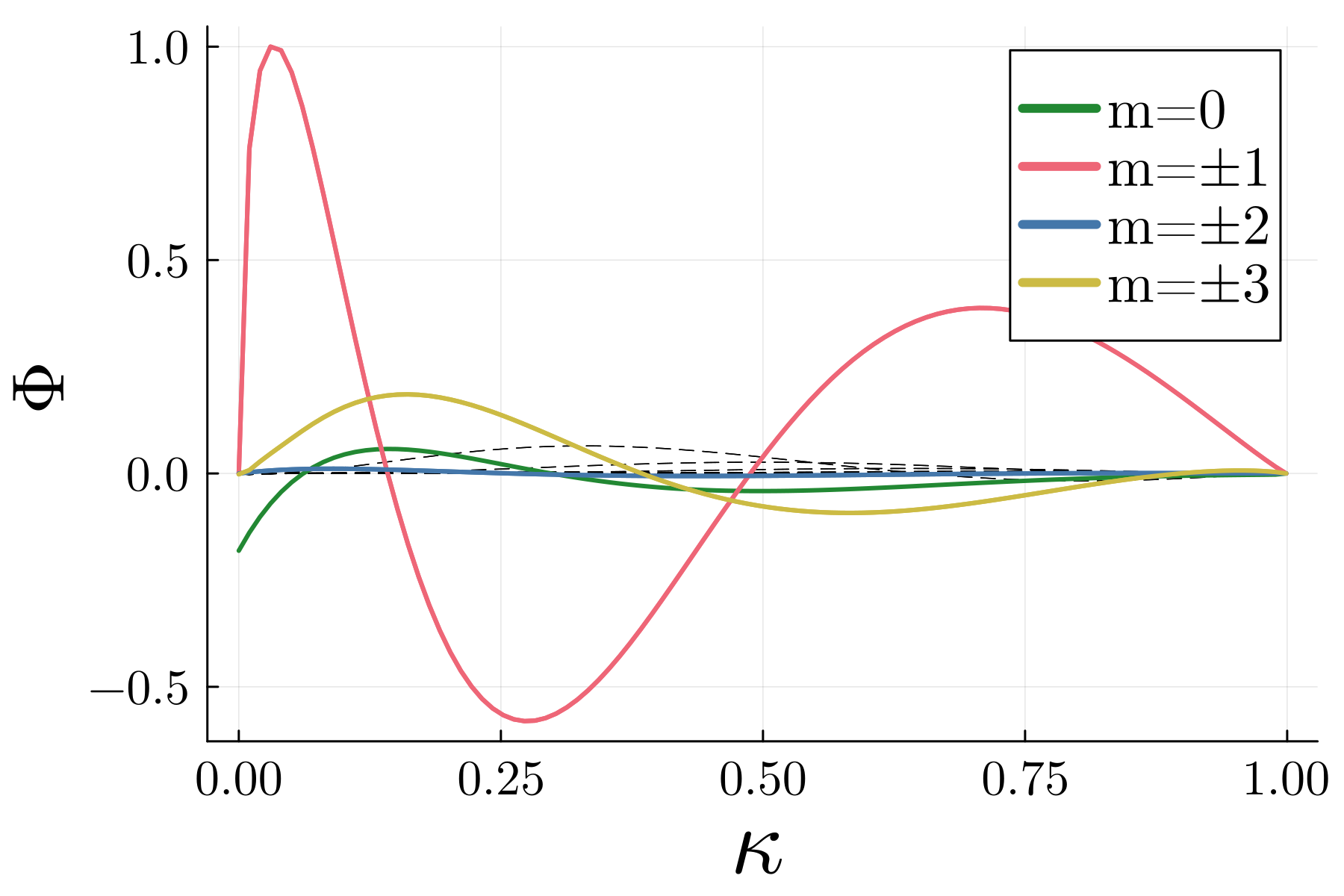}\label{subfig:in_isl_a}}
    \qquad
    \subfloat[Global mode $a$]{\includegraphics[width=0.45\textwidth]{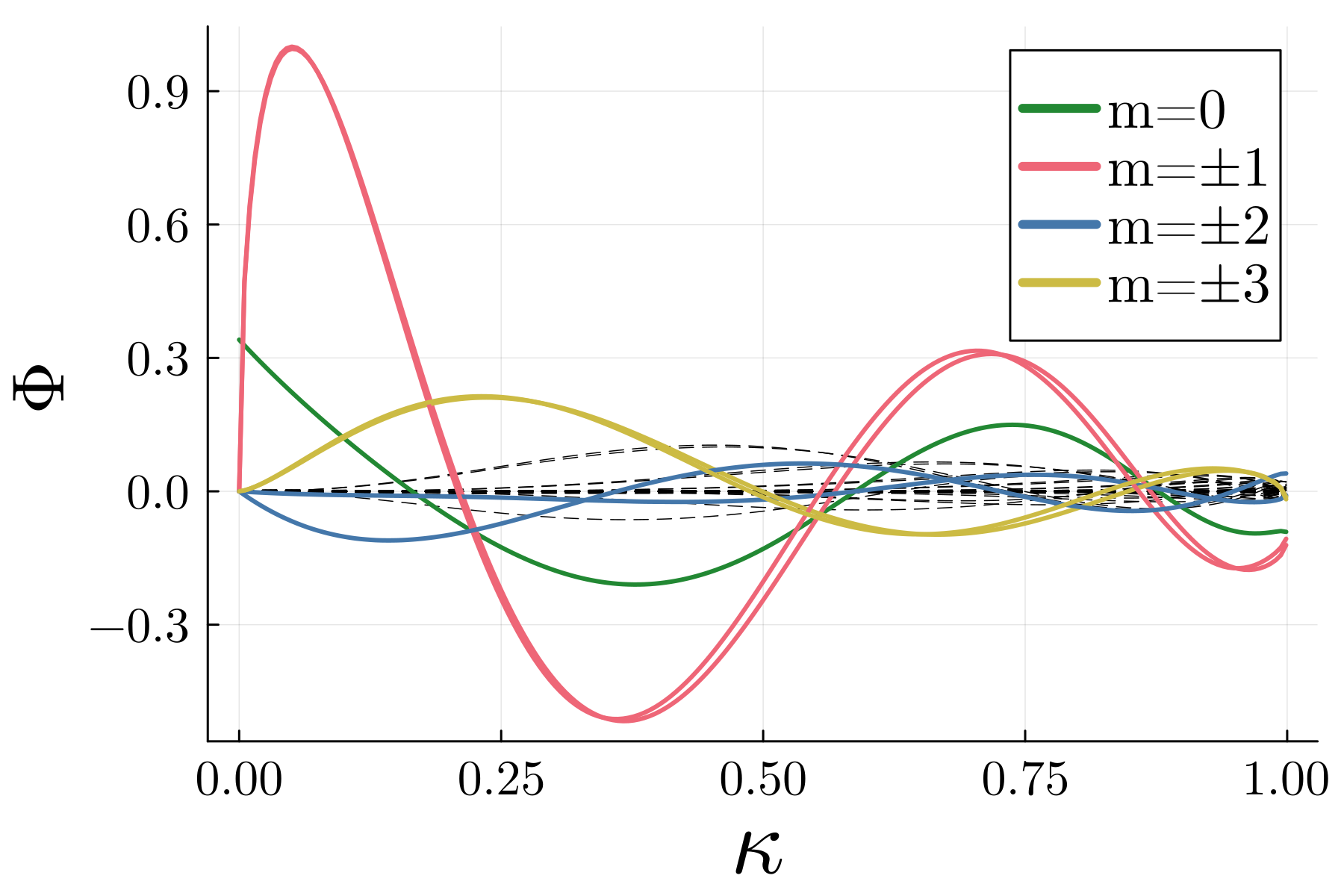}\label{subfig:mapped_isl_a}}

    \subfloat[Global mode $b$]{\includegraphics[width=0.45\textwidth]{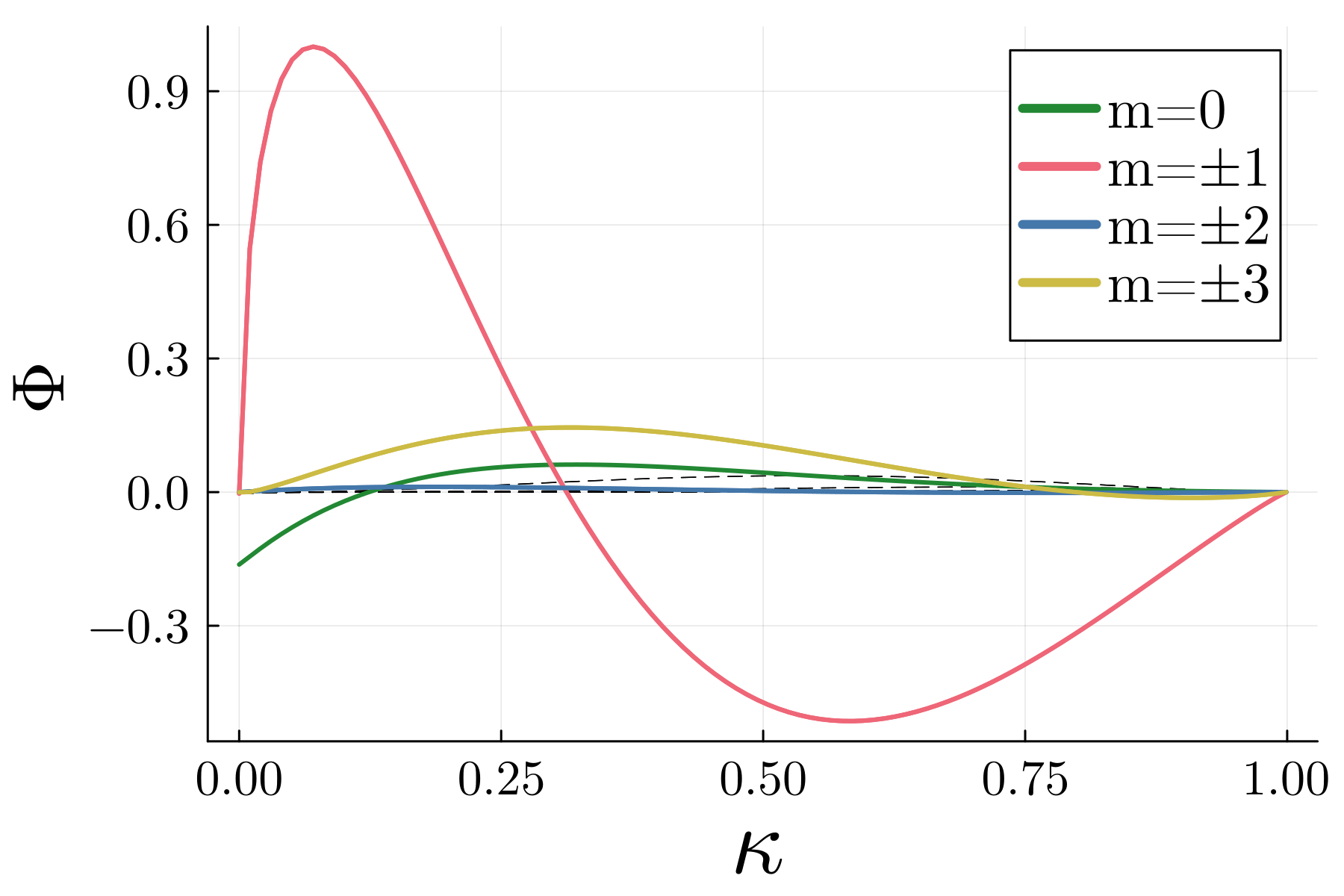}\label{subfig:in_isl_b}}
    \qquad
    \subfloat[Global mode $b$]{\includegraphics[width=0.45\textwidth]{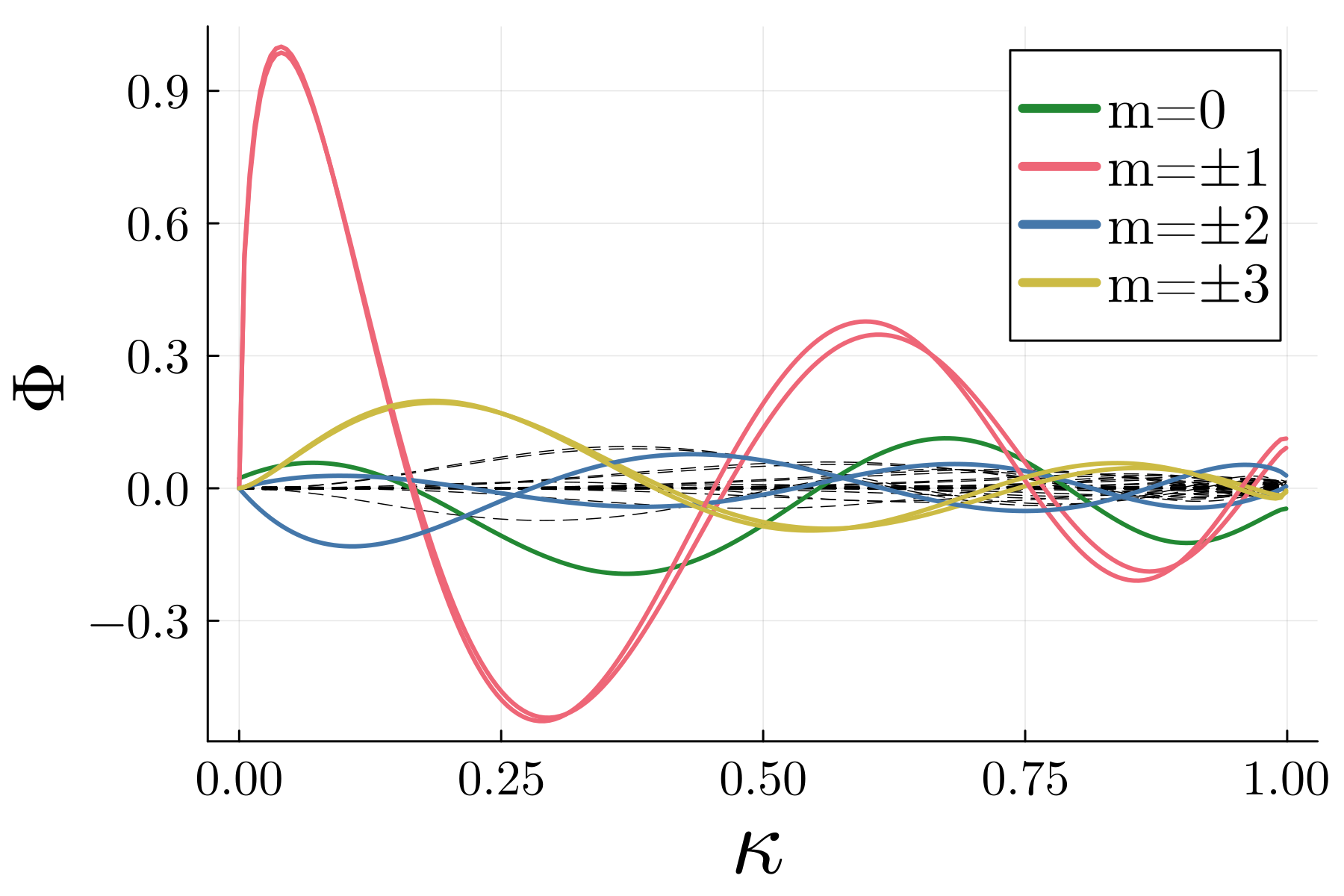}\label{subfig:mapped_isl_b}}

    \subfloat[Global mode $c$]{\includegraphics[width=0.45\textwidth]{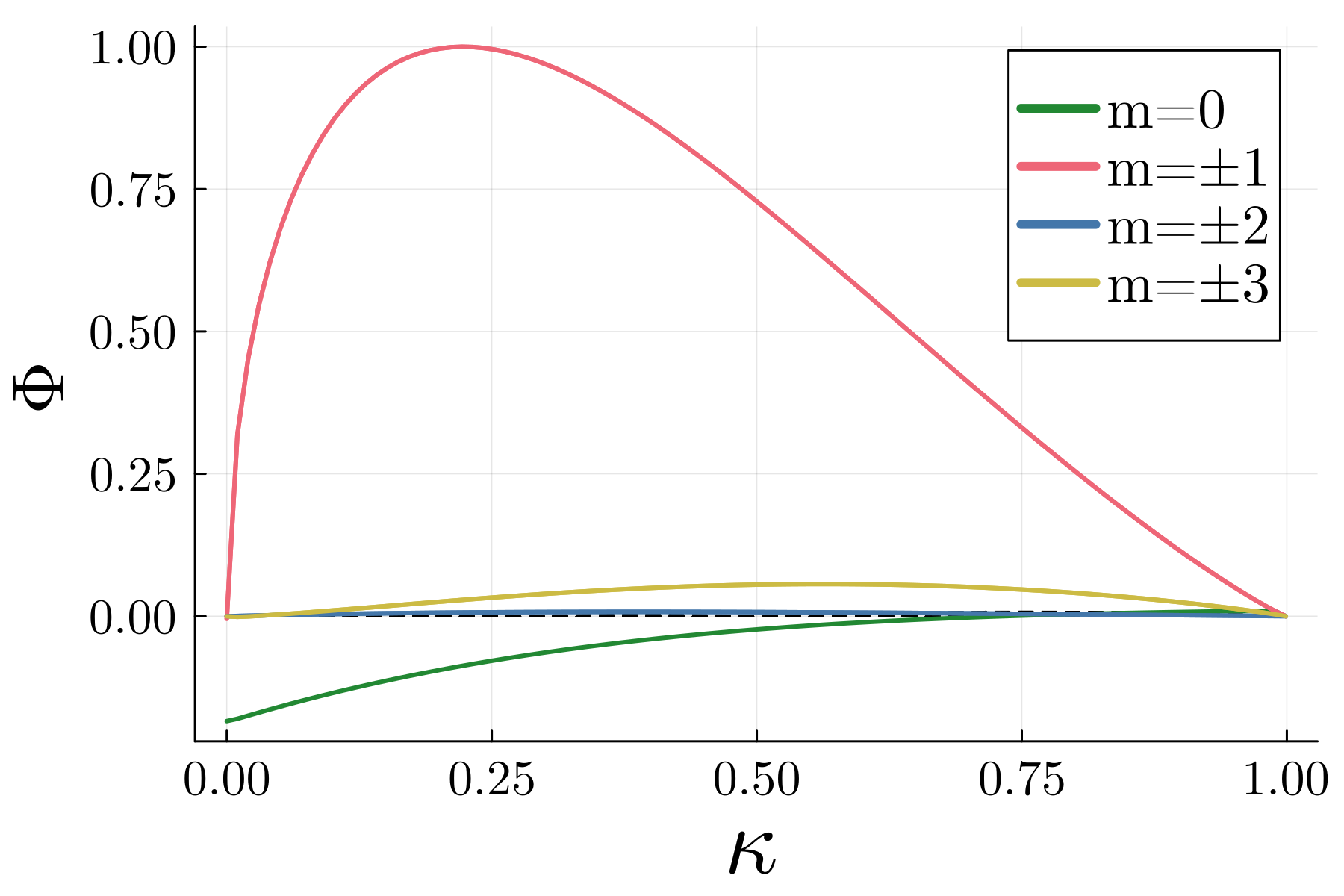}\label{subfig:in_isl_c}}
    \qquad
    \subfloat[Global mode $c$]{\includegraphics[width=0.45\textwidth]{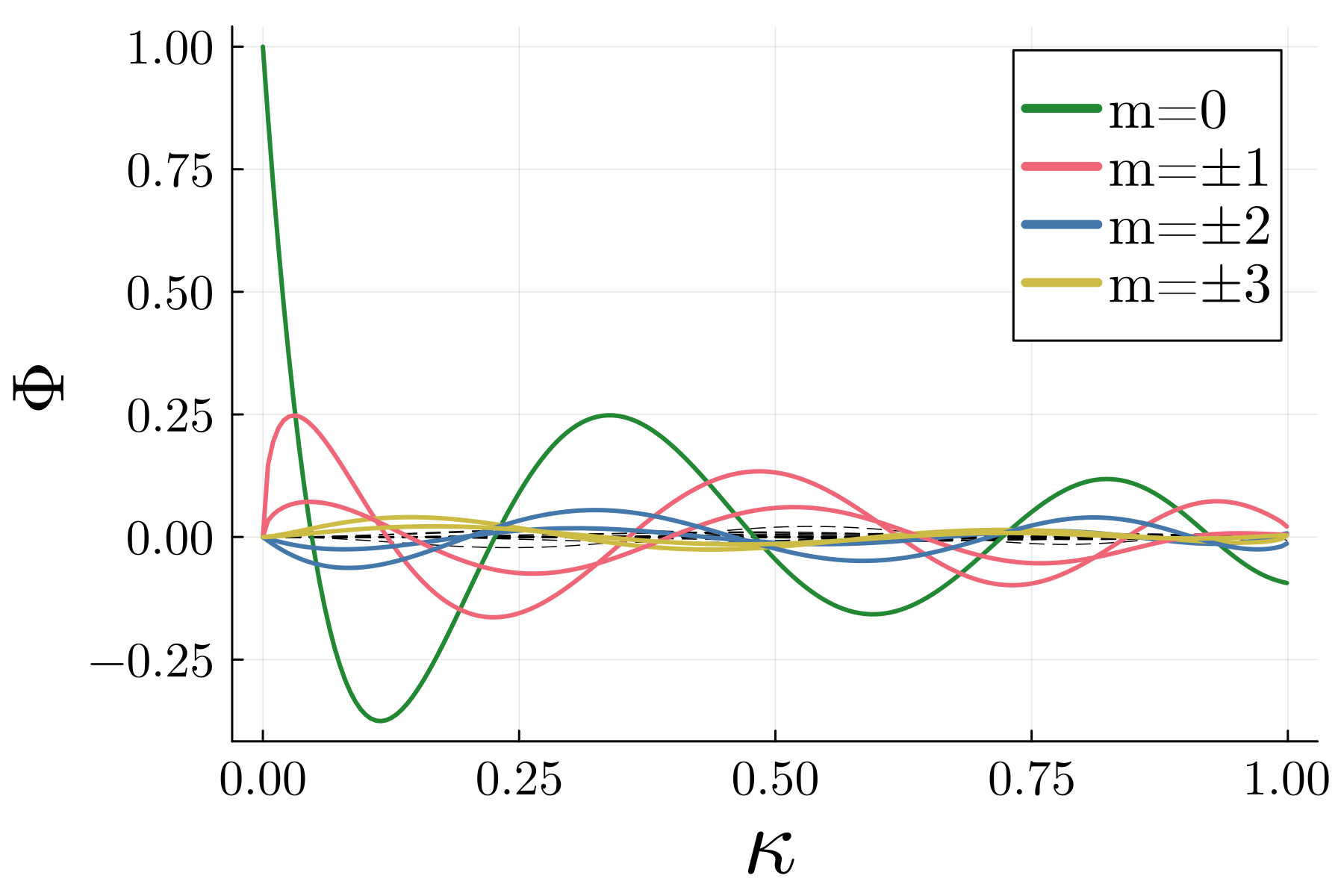}\label{subfig:mapped_isl_c}}
    \caption{Global island eigenmodes. Computed with island coordinates, $(\kappa, \ab, \tau)$, (left column) and with QFM coordinates, $(s, \vartheta, \zeta)$, then mapped to island coordinates for comparison (right column).}
    \label{fig:in_isl_solutions}
\end{figure}

Looking closer at the specific gap modes, we can compare the harmonic structure when computed with island coordinates, shown in the left column of figure \ref{fig:in_isl_solutions}, against those computed in QFM coordinates, then mapped to island coordinates, shown in the right column.
These three eigenmodes computed in island coordinates qualitatively match the solutions $a,b,c$ shown in figure 2 of Könies \textit{et al}. \cite{konies_shear_2024}, noting that the $x$-axis used here is the square of the $x$-axis used by Könies \textit{et al}.

The first of the mapped solutions, figures \ref{subfig:in_isl_a} and \ref{subfig:mapped_isl_a}, show excellent agreement, however, the other two show less, despite all solutions existing at very similar frequencies.
One notably difference is the prevalence of the $(0, 0)$ harmonic.
This may just be an issue of numerical resolution, however, it could be hinting that these island modes, existing at low frequencies, are interacting with the $(0, 0)$ harmonic outside the island chain.
This behaviour could not possibly be predicted with the island coordinates due to their limitations at the separatrix.

These results clarify the observations seen when investigating the chaotic spectrum.
Without island coordinates, the intricate structures inside the islands are not adequately resolved.
We can understand this further by consider the effect of our QFM coordinates.
Most notable in figure \ref{fig:qfm_poincare}, the QFM coordinates straighten intact flux surfaces and provide structure to the magnetic field, however, they have minimal effect on the magnetic island chains.

Because the magnetic island chains still exist as nested flux surfaces in the QFM coordinates, the rectangular grid we apply is poorly aligned with the elliptical flux surfaces of the islands.
The numerical benefit we are gaining with the QFM coordinates in the chaotic region is not true for the magnetic island chains.
We can then assert that the QFM coordinates and our current numerical approach are not sufficient for resolving the shear Alfvén waves inside magnetic island chains.

\subsection{Numerical Demand}

One technique to increase the accuracy of the results is to increase the grid resolution used.
Here we will consider a rough estimate of the numerical demand for resolving the solutions inside the island chains present in the cases covered in section \ref{sec:chaotic_spectrum}.
We used a grid resolution of $150 \times 40 \times 30$, with multiple shift and inverts each targeting $1000$ eigenvalues.
At this resolution, the two sparse matrices we construct, $\bm P$ and $\bm Q$, require approximately $5$GB to store, however, in the process of solving, the Krylov-Schur algorithm \cite{stewart_krylov--schur_2002} requires a partial factorisation of the matrix system.
This factorisation is performed by \ttt{MUMPS} \cite{goos_mumps_2001}, requiring approximately $500$GB for the factorisation.
As this step has the highest memory demand we will neglect the remaining use of memory in this discussion.

Above, we have shown that a grid resolution of roughly $66\times 50 \times 30$ per island node is capable of resolving some global modes while not being sufficient for the continuum structure.
If we naively take this as an adequate resolution for island modes, we can estimate the expected demand to fully resolve the spectrum in the chaotic case.
In order to have a sufficient resolution for the second order $(7, -5)$ island chain, we would require at least $350$ poloidal and $150$ toroidal grid points to have the same number of points per node.
Approximating the $(7, -5)$ island chain as extending across $20\%$ of the chaotic region, this same resolution for the entire chaotic region would require at least $300$ radial points, then a much smaller amount outside the chaotic region.

Taking the total resolution as $350\times 350\times 150$, assuming the same density of non-zeros in the sparse matrices $\bm P, \bm Q$ of approximately $1.5\times10^{-4}$, these sparse matrices would require over $500$ TB to store.
Then, to actually solve the problem, assuming the same memory scaling for the factorisation, we would require memory of the order $50$ petabytes.

Naturally, these calculations are very approximate, but we can see that even for an underestimate of the required grid resolution, the memory demands are completely infeasible.
This highlights the need for an improved approach for handling the island chains that form as part of a symmetry breaking perturbation, as this brute force approach will not scale adequately.

This is especially impactful for our solutions residing on broken flux surfaces, which smear across the chaotic region.
The broadening of these solutions increases the interaction with the magnetic islands.
Because the island structures are not fully resolved, we cannot make conclusive statements on the specific harmonics produced or their specific shape in this smearing process.

This also limits our ability to model global modes such as the TAE, as this will also interact strongly with the magnetic islands.
For specific modelling of a TAE with an RMP-like field, we will therefore need a different method.
This may require a serious modification to the numerical approach taken here, for example, instead of solving the equilibrium eigenvalue problem it may be more practical to consider the time dependent evolution of a TAE.
Alternatively, we may want to modify the grid used, for example, it may be possible to partially use the island coordinates and `stitch' them together with a different coordinate system outside the island chains, avoiding the coordinate singularity at the separatrix.
Another option is to introduce an unstructured mesh grid that aligns with the magnetic field inside the islands, which the finite element method is well equipped to handle.
So far we have relied on magnetic coordinates to capture the geometry of the problem, but perhaps a combination of a mesh grid and magnetic coordinates would reduce the numerical demand inside the island chains.

Regardless of any future modifications made, we can see that the QFM coordinates will likely be an essential aspect of future solutions.
Outside the islands, QFM coordinates provide a clear benefit to the numerical scheme.
Additionally, as shown in figure \ref{fig:qfm_poincare}, the QFM coordinates add structure and regularity to the magnetic island chains, meaning that any future modifications used are likely to be much simpler when using QFM coordinates.

This highlights a key aspect of this study, QFM coordinates provide pseudo straight field line coordinates for chaotic perturbations, and may well be an essential tool for this kind of work.
However, because they do not make significant difference to the alignment of the numerical grid with the magnetic field inside the magnetic islands, QFM coordinates alone are not sufficient for computing the entire spectrum of shear Alfvén waves.


\section{Conclusion and Future Work}\label{sec:conclusion}

Pseudo straight field line quadratic flux minimising coordinates have provided a method for computing the shear Alfvén spectrum under symmetry breaking perturbations with chaotic magnetic field trajectories.
We have shown that while a given flux surface remains intact, shear Alfvén waves localised to that surface remain relatively unchanged compared to the unperturbed case.
Additionally, once the flux surface is broken, we find that the original Fourier harmonic structure persists with the addition of a smearing across the chaotic region.

We have also highlighted the limited effect QFM coordinates provide for resolving the spectrum inside the magnetic island chains and shown that additional tools will be required for accurately computing the full spectrum, and eventually the interaction of a TAE with an RMP field.
While this does limit the practical usage of our current method, it does clarify where QFM coordinates excel and where they offer limited benefit.
We also have a clear problem to solve in future work; the magnetic island chains are the obvious bottleneck for this research.
If we can accurately and practically determine the spectrum inside a magnetic island, combining this method with QFM coordinates should allow accurate computation of the full spectrum and further verification of the trends we have observed.

Provided such methods are possible, it will then become practical to study the interaction of a TAE with a chaotic region providing an understanding of the influence of an RMP field.
Of particular interest will be the computation of changes to continuum damping of a TAE, requiring a simple, already implemented, non-ideal addition to equation \ref{eqn:SAW_gov} \cite{poedts_damping_1992}.

We also note that there are other physically interesting additions that could be made to this problem.
One example is that near the island separatrix, especially at the $X$-points, the length scale of the problem becomes comparable to the ion Larmor radius, invalidating some of the assumptions present in ideal MHD.
Inclusion of additional non-ideal effects may provide a more realistic picture of the behaviour of the spectrum near the islands, in particular, it would include the addition of kinetic Alfvén waves \cite{hasegawa_kinetic_1975}, which may provide a physical connection between the inside and outside of the magnetic islands.

Throughout this work we have also assumed the $q$-profile is unchanged under the perturbation, which is imposed on top of the equilibrium field.
Instead, we may consider a self consistent magnetic field, which would modify the $q$-profile and may have a large effect on the spectrum \cite{hole_identifying_2011}.
Our code is set up to handle generic geometry for both the construction of QFM coordinates and the solution to the eigenvalue problem for shear Alfvén waves.
It should therefore be straightforward to adapt our code use a more realistic equilibrium generated by equilibrium codes such as \ttt{SPEC} \cite{hudson_computation_2012}, and allow us to compute the spectrum in more realistic scenarios.

%
%

\ack{We would like thank Axel Könies, Jinjia Cao and Ralf Kleiber for useful discussions regarding the shear Alfvén spectrum inside magnetic island chains.
We also thank Stuart Hudson for helpful discussions on quadratic-flux-minimising surfaces.
This research was undertaken with the assistance of resources from the National Computational Infrastructure (NCI Australia), an NCRIS enabled capability supported by the Australian Government.}

\funding{This work was partially funded by the Australian Government through the Australian Government Research Training Program (RTP) Scholarship.
This work is partly funded by Ministry of Education (MOE) AcRF Tier 1 grants RS02/23 and RG156/23, and National Research Foundation Singapore (NRF) core funding ``Fusion Science for Clean Energy''.}

\appendix
\counterwithin*{equation}{section}
\renewcommand\theequation{\thesection\arabic{equation}}

\section{Cylindrical metric tensor and Jacobian}\label{app:cyl_metric}

The metric tensor in cylindrical geometry is given in the limit $R_0\rightarrow \infty$ of the usual toroidal metric \cite{sharapov_energetic_2021}, expressed in terms of the toroidal flux $\psi$,
\begin{subequations}
\begin{align}
    g_{\psi\psi} &= \oo{r^2},\\
    g_{\theta\theta} &= r^2,\\
    g_{\varphi\varphi} &= R_0^2,
\end{align}
\end{subequations}
where $r=\sqrt{2\psi}$. The off-diagonal terms are zero and the Jacobian is given by
\begin{align}
    \jac = R_0.
\end{align}

\section{Weak Form}\label{app:weakform}
To convert the weak form, equation \ref{eqn:weak_form}, into a matrix equation we define a vector containing the unique derivatives of the basis functions, 
\begin{align}\label{eqn:basis_vector}
    \partial^2\bm\Phi = (\partial_\psi\Phi, \partial_\theta\Phi, \partial_\varphi\Phi, \partial_{\psi\psi}\Phi, \partial_{\psi\theta}\Phi, \partial_{\psi\varphi}\Phi, \partial_{\theta\theta}\Phi, \partial_{\theta\varphi}\Phi, \partial_{\varphi\varphi}\Phi).
\end{align}
This is then contracted with the two differential operators, $\hat{P}$ and $\hat{Q}$.
The term containing the eigenvalue, $\omega^2$, and shown on the left hand side of equation \ref{eqn:weak_form}, is define by $\hat{Q}$, which is computed via
\begin{align}
    \hat{Q}^{ij} = \jac \frac{\mu_0\rho}{B^2}\hat{D}^{ij}
\end{align}
with 
\begin{align}
    D^{ij} = g^{i j} - B^i B^j / |B|^2.
\end{align}
where Latin indices run from $1$ to $3$, as this term is only contracting with the first derivative terms of the vector defined in equation \ref{eqn:basis_vector}.
The second operator, $\hat{P}$, contains the right hand side of equation \ref{eqn:weak_form}, which we split up into two terms; the Laplacian-like first term, $T_l$, and the second term containing the parallel current, $T_j$.
The first term is given by
\begin{align}
    T_l^{\mu\nu} = \jac C^{\mu}_{i} \hat{D}^{ij}C_j^{\nu},
\end{align}
where Greek indices run from $1$ to $9$ and the non-square $C$ matrix is given by,
\begin{align}
    C^\mu_i = 
    \left(
    \begin{array}{c c c c c c c c c}
        \left(\frac{B^\psi}{B^2}\right)_\psi & \left(\frac{B^\theta}{B^2}\right)_\psi & \left(\frac{B^\varphi}{B^2}\right)_\psi & \frac{B^\psi}{B^2} & \frac{B^\theta}{B^2} & \frac{B^\varphi}{B^2} & 0 & 0 & 0 \\
        \left(\frac{B^\psi}{B^2}\right)_\theta & \left(\frac{B^\theta}{B^2}\right)_\theta & \left(\frac{B^\varphi}{B^2}\right)_\theta & 0 & \frac{B^\psi}{B^2} & 0  & \frac{B^\theta}{B^2} & \frac{B^\varphi}{B^2} & 0 \\
        \left(\frac{B^\psi}{B^2}\right)_\varphi  & \left(\frac{B^\theta}{B^2}\right)_\varphi  & \left(\frac{B^\varphi}{B^2}\right)_\varphi  & 0 & 0 & \frac{B^\psi}{B^2} & 0 & \frac{B^\theta}{B^2} & \frac{B^\varphi}{B^2}
        \end{array}
    \right),
\end{align}
where the subscript denotes differentiation and $B^2 = B^iB^jg_{ij}$.
The final term is split into two parts which are identical when swapping the test and trial functions. Each of these is given by two matrices,
\begin{align}
    \hat{T}_1 = \left(
    \begin{array}{c c c}
        \Gamma^\psi_i\oo{\sqrt{g}}\epsilon^{ijk}\partial_j(\Gamma^\psi_k) & \Gamma^\psi_i\oo{\sqrt{g}}\epsilon^{ijk}\partial_j(\Gamma^\theta_k) & \Gamma^\psi_i\oo{\sqrt{g}}\epsilon^{ijk}\partial_j(\Gamma^\varphi_k) \\
        \Gamma^\theta_i\oo{\sqrt{g}}\epsilon^{ijk}\partial_j(\Gamma^\psi_k) & \Gamma^\theta_i\oo{\sqrt{g}}\epsilon^{ijk}\partial_j(\Gamma^\theta_k) & \Gamma^\theta_i\oo{\sqrt{g}}\epsilon^{ijk}\partial_j(\Gamma^\varphi_k) \\
        \Gamma^\varphi_i\oo{\sqrt{g}}\epsilon^{ijk}\partial_j(\Gamma^\psi_k) & \Gamma^\varphi_i\oo{\sqrt{g}}\epsilon^{ijk}\partial_j(\Gamma^\theta_k) & \Gamma^\varphi_i\oo{\sqrt{g}}\epsilon^{ijk}\partial_j(\Gamma^\varphi_k),
    \end{array}
    \right)
\end{align}
and 
\begin{align}
(\hat{T}_2)^T = \oo{\jac}\left(
    \begin{array}{c c c c c c}
         \Gamma^\psi_i\epsilon^{i\psi k}\Gamma^\psi_k & \Gamma^\theta_i\epsilon^{i\psi k}\Gamma^\psi_k & \Gamma^\varphi_i\epsilon^{i\psi k}\Gamma^\psi_k \\
         \Gamma^\psi_i\epsilon^{i\psi k}\Gamma^\theta_k + \Gamma^\psi_i\epsilon^{i\theta k}\Gamma^\psi_k & \Gamma^\theta_i\epsilon^{i\psi k}\Gamma^\theta_k + \Gamma^\theta_i\epsilon^{i\theta k}\Gamma^\psi_k & \Gamma^\varphi_i\epsilon^{i\psi k}\Gamma^\theta_k + \Gamma^\varphi_i\epsilon^{i\theta k}\Gamma^\psi_k\\
         \Gamma^s_i\epsilon^{isk}\Gamma^\varphi_k + \Gamma^s_i\epsilon^{i\varphi k}\Gamma^s_k & \Gamma^\theta_i\epsilon^{i\psi k}\Gamma^\varphi_k + \Gamma^\theta_i\epsilon^{i\varphi k}\Gamma^\psi_k & \Gamma^\varphi_i\epsilon^{i\psi k}\Gamma^\varphi_k + \Gamma^\varphi_i\epsilon^{i\varphi k}\Gamma^\psi_k\\
         \Gamma^\psi_i\epsilon^{i\theta k}\Gamma^\theta_k & \Gamma^\theta_i\epsilon^{i\theta k}\Gamma^\theta_k & \Gamma^\varphi_i\epsilon^{i\theta k}\Gamma^\theta_k\\
         \Gamma^\psi_i\epsilon^{i\theta k}\Gamma^\varphi_k + \Gamma^\psi_i\epsilon^{i\varphi k}\Gamma^\theta_k & \Gamma^\theta_i\epsilon^{i\theta k}\Gamma^\varphi_k + \Gamma^\theta_i\epsilon^{i\varphi k}\Gamma^\theta_k & \Gamma^\varphi_i\epsilon^{i\theta k}\Gamma^\varphi_k + \Gamma^\varphi_i\epsilon^{i\varphi k}\Gamma^\theta_k \\\Gamma^\psi_i\epsilon^{i\varphi k}\Gamma^\varphi_k & \Gamma^\theta_i\epsilon^{i\varphi k}\Gamma^\varphi_k & \Gamma^\varphi_i\epsilon^{i\varphi k}\Gamma^\varphi_k
    \end{array}
    \right).
\end{align}
Where $(\psi, \theta, \varphi)\equiv (1, 2, 3)$, $\epsilon^{ijk}$ is the Levi-Civita symbol, and we have defined
\begin{align}
    \Gamma_i^j = \left(\delta_i^j-g_{ik}b^kb^j\right).
\end{align}
The current term, $T_j$, is given by the sum of $T_1$, $T_2$ and the transpose of this to reflect the symmetry of the two terms in equation \ref{eqn:weak_form}.

\section{Island Metric Tensor}\label{app:island_metric}

The metric tensor for the straight field line coordinates is computed following the method of Könies \textit{et al.} \cite{konies_shear_2024}, using the notation of Qu and Hole \cite{qu_shear_2023}. 
Notably, we start with the toroidal flux, $\psi$, rather than the geometric radius squared, $r^2$. 
Additionally, the resulting straight field line radial coordinate, $\kappa$, used here is the square of the coordinate used by Könies \textit{et al.}

These coordinates are defined only for $\kappa < 1$ and require a linear rotational transform to be analytical. 
The original coordinates are the cylindrical coordinates defined in appendix \ref{app:cyl_metric}.

From \cite{qu_shear_2023}, we have 
\begin{align}
    \kappa = \frac{4}{w^2}(\psi-\psi_0)^2 + \sin^2(m\alpha/2),
\end{align}
where $\kappa$ acts like a radial coordinate inside the island, with $\kappa=1$ defining the separatrix and $\kappa=0$ the $O$-point.
We have also defined the helical angle $\alpha=\theta-\varphi/q_0$, with $q_0=-m_0/n_0$.
We next introduce an intermediate coordinate transform satisfying
\begin{align}
    \sqrt{\kappa}\sin\beta=\sin(m\alpha/2).
\end{align}
The metric tensor for this intermediate coordinate system, $(\kappa, \beta, \tau)$, is given by
\begin{subequations}\label{eqn:intermediate_island_metric}
\begin{align}
    \nabla\kappa\cdot\nabla\kappa &= \frac{32\psi\kappa}{w^2}\cos^2\beta + \hat{b}\kappa\sin^2\beta,\\
    \nabla\kappa\cdot\nabla\beta &= \left(-\frac{16\psi}{w^2} + \frac{\hat{b}}{2}\right)\cos\beta\sin\beta,\\
    \nabla\kappa\cdot\nabla\tau &= \frac{n}{R_0^2}\sqrt{\kappa} \sin\beta\sqrt{1-\kappa\sin^2\beta},\\
    \nabla\beta\cdot\nabla\beta &= \frac{8\psi}{w^2\kappa}\sin^2\beta + \frac{\hat{b}}{2\kappa}\cos^2\beta,\\
    \nabla\beta\cdot\nabla\tau &= \frac{n}{2\sqrt{\kappa}R_0^2} \cos\beta\sqrt{1-\kappa\sin^2\beta},\\
    \nabla\tau\cdot\nabla\tau &= \frac{1}{R_0^2},
\end{align}
\end{subequations}
where 
\begin{align}
    \hat{b} &= \left(\frac{m^2}{2\psi} + \frac{n^2}{R_0^2}\right)(1-\kappa\sin^2\beta),\\
    \psi &= \psi_0 + \frac{w\sqrt{\kappa}}{2}\cos\beta.
\end{align}
The coordinate system is then straightened via the transformation
\begin{align}
    \ab = \frac{\pi}{2 K(\kappa)} F(\beta, \kappa),
\end{align}
where $K(\kappa), F(\beta, \kappa)$ are the complete and incomplete elliptical integrals of the first kind respectively.
The two derivatives of this are
\begin{subequations}
\begin{align}
    \p{\ab}{\kappa} &= \frac{\pi}{4(1-\kappa)\kappa K(\kappa)} \left( Z(\beta, \kappa) - \frac{\kappa\sin\beta\cos\beta}{\sqrt{1-\kappa\sin^2\beta}}\right),\\
    \p{\ab}{\beta} &= \frac{\pi}{2K(\kappa)}\sqrt{1-\kappa\sin^2\beta},
\end{align}
\end{subequations}
where $Z(\beta, \kappa) = E(\beta, \kappa) - E(\kappa)F(\beta, \kappa) / K(\kappa)$, is the Jacobi Zeta function.
Inverting the transformation,
\begin{align}
    \beta = \text{am}(2K(\kappa)\ab /\pi, \kappa),
\end{align}
we can then write the trigonometric functions as,
\begin{subequations}\label{eqn:isl_met_trig}
\begin{align}
    \sin\beta &= \text{sn}(2K(\kappa)\ab/\pi, \kappa),\\
    \cos\beta &= \text{cn}(2K(\kappa)\ab/\pi, \kappa),\\
    \sqrt{1-\kappa\sin^2\beta} &= \text{dn}(2K(\kappa)\ab/\pi, \kappa).
\end{align}
\end{subequations}
Finally, we can define the metric tensor, in terms of the straightened coordinate system $(\kappa, \ab, \tau)$.
\begin{subequations}
\begin{align}
    g^{\kappa\kappa} &= \frac{32\psi\kappa}{w^2} \text{cn}^2(2K(\kappa)\ab/\pi, \kappa) + \hat{b}\kappa \text{sn}^2(2K(\kappa)\ab/\pi),\\
    g^{\kappa\ab} &= \p{\ab}{\kappa} \nabla\kappa^2 + \p{\ab}{\beta}\nabla\kappa\cdot\nabla\beta,\\
    g^{\kappa\tau} &= \frac{n}{R_0^2}\sqrt{\kappa} \text{sn}(2K(\kappa)\ab/\pi)\text{dn}(2K(\kappa)\ab/\pi),\\
    g^{\ab\ab} &= \left(\p{\ab}{\kappa}\right)^2 \nabla\kappa^2 + 2 \p{\ab}{\kappa}\p{\ab}{\beta} \nabla\kappa\cdot\nabla\beta + \left(\p{\ab}{\beta}\right)^2\nabla\beta^2,\\
    g^{\ab\tau} &= \p{\ab}{\kappa} \nabla\kappa\cdot\nabla\tau + \p{\ab}{\beta}\nabla\beta\cdot\nabla\tau,\\
    g^{\tau\tau} &= \frac{1}{R_0^2},
\end{align}
\end{subequations}
where the unspecified expressions are given in equation \ref{eqn:intermediate_island_metric}, using the expressions defined in equation \ref{eqn:isl_met_trig}.

\printbibliography

\end{document}